\documentclass[iop]{emulateapj}
\usepackage{amsthm}
\usepackage{amsmath}
\usepackage{bm}
\usepackage[printonlyused]{acronym}
\usepackage{xcolor}

\defcitealias{Skinner:2013}{SO13}
\defcitealias{Ostriker:2011}{OS11}

\newcommand\gram{\mbox{ g}}
\newcommand\pcc{\mbox{ cm}^{-3}}

\newcommand\Msun{\; M_\odot}
\newcommand\kms{\mbox{ km}\mbox{ s}^{-1}}

\newcommand\pc{\mbox{ pc}}

\newcommand\Surf{\Msun\mbox{ pc}^{-2}}
\newcommand\cmsg{\mbox{ cm}^{2}\mbox{ g}^{-1}}
\newcommand\gcms{\mbox{ g}\mbox{ cm}^{-2}}

\newcommand\simgt{\lower.5ex\hbox{$\; \buildrel > \over \sim \;$}}
\newcommand\simlt{\lower.5ex\hbox{$\; \buildrel < \over \sim \;$}}

\newcommand\cs{c_s}

\newcommand\tff{\; t_\mathrm{ff}}
\newcommand\Mcloud{{M_\mathrm{cloud}}}
\newcommand\Rcloud{R_\mathrm{cloud}}

\shorttitle{RHD Simulations of Turbulent, Star-Forming Clouds}
\shortauthors{Skinner and Ostriker}

\begin{document}

\title{Numerical Simulations of Turbulent Molecular Clouds 
Regulated by Reprocessed Radiation Feedback from Nascent Super Star Clusters}

\author{M. Aaron Skinner and Eve C. Ostriker}
\affil{Department of Astrophysical Sciences, Princeton University, Princeton, NJ 08544-1001;
askinner@astro.princeton.edu,eco@astro.princeton.edu}

\begin{abstract}
Radiation feedback from young star clusters embedded in \acp{GMC}
is believed to be important to the control of
star formation.  For the most massive and dense clouds, including
those in which \acp{SSC} are born, pressure from
reprocessed radiation exerted on dust grains may disperse a
significant portion of the cloud mass back into the \ac{ISM}.  
Using our \ac{RHD} code, \textit{Hyperion},
we conduct a series of numerical simulations to test this idea.  Our
models follow the evolution of self-gravitating, strongly turbulent
clouds in which collapsing regions are replaced by radiating sink
particles representing stellar clusters.  We evaluate the dependence
of the \ac{SFE} on the size and mass of the
cloud and $\kappa$, the opacity of the gas to \ac{IR} radiation.
We find that the single most important parameter determining the
evolutionary outcome is $\kappa$, with $\kappa \simgt 15 \cmsg$ needed
to disrupt clouds.  For $\kappa = 20-40 \cmsg$, the resulting
\ac{SFE} $=50-70\%$ is similar to empirical estimates for some \ac{SSC}-forming
clouds.  The opacities required for \ac{GMC} disruption likely apply only
in dust-enriched environments.  We find that the subgrid model
approach of boosting the direct radiation force $L/c$ by a ``trapping
factor'' equal to a cloud's mean \ac{IR} optical depth can overestimate the
true radiation force by factors of $\sim 4-5$.  We conclude that
feedback from reprocessed \ac{IR} radiation alone is unlikely to
significantly reduce star formation within \acp{GMC} unless their dust
abundances or cluster light-to-mass ratios are enhanced.  
\end{abstract}


\section{Introduction} \label{intro}
\acresetall

Giant molecular clouds (\acsp{GMC}\acused{GMC}), the sites of star formation, form out
of the diffuse \ac{ISM} due to some combination of
self-gravity, the gravity of the stellar disk and bulge (which compresses the
\ac{ISM} vertically everywhere and horizontally in spiral arms), and
large-scale gas motions associated with turbulence, supernova remnant
or superbubble expansion, and other \ac{ISM} flows
\citep[e.g.,][]{McKee:2007,Dobbs:2014}.  
However, \acp{GMC} are not believed
to be permanent structures, and based on age-dating of associated star 
clusters (in the Milky Way and other galaxies)
are thought to survive for at most several internal free-fall
times \citep{Leisawitz:1989,Kawamura:2009,Miura:2012,Whitmore:2014}.
It is widely believed that ``feedback'' from massive stars is
responsible for the demise of \acp{GMC}, but exactly how this works is
still poorly understood.  Among the most basic uncertainties is which
among the possible feedback effects predominate for different regimes
of \ac{GMC} properties and surrounding environment.  Currently, the
most-discussed candidate effects \citep[see, e.g., the review of][]{Krumholz:2014}
include (1) (magneto)hydrodynamic
forces from overpressured regions produced by ionizing radiation,
shocked stellar winds, or supernova blasts; and (2) radiation forces
from the primary absorption of stellar optical and \ac{UV}, and from the
secondary absorption of reprocessed \ac{IR}.

Both theoretical and observational motivations have led to an
increased interest in the effects of radiation forces. Although
supernovae inject an order of magnitude more momentum per stellar mass
to their surroundings than other forms of feedback
\citep{Ostriker:2011,Kim:2015,Iffrig:2015,Walch:2014,Martizzi:2015,
Geen:2015}, there is a significant 
time delay between the advent of star formation in a \ac{GMC} and the
explosion of the first supernova (and further delay before the last massive stars die).
If other agents are able to destroy the \ac{GMC} in this interval, much of the 
momentum and energy from \acp{SN} may be delivered to the diffuse \ac{ISM} rather 
than the progenitor's birth cloud.  

In \acp{GMC} with low escape speeds hosting clusters containing massive stars, 
the combination of photoevaporation 
and the pressure force from the expanding \ion{H}{2} region can unbind much 
of a cloud's initial mass
\citep[e.g.,][and references therein]{Matzner:2002,Dale:2012,Dale:2013a}. 
However, analytic spherical models
\citep{Krumholz:2009a,Murray:2010,Fall:2010} of clusters within clouds 
suggest that
the effects from radiation pressure will exceed that from ionized gas
pressure at high values of the total cloud mass and surface density.
Furthermore, the relative importance of reprocessed \ac{IR} compared to
direct optical/\ac{UV} is expected to increase as the cloud surface density
increases.  Observational evidence suggests that radiation pressure
exceeds ionized gas pressure close to the centers of \ion{H}{2}
regions around massive clusters (consistent with the theory of
\citealt{Draine:2011}), and for younger systems
\citep{Lopez:2011,Lopez:2014,Pellegrini:2007}.  Although
historically the gas pressure from shocked stellar winds was expected
to drive dynamics around massive clusters at early times, 
this has been called into
question for some systems due to the lack of X-ray emission, with the
suggestion that the primary wind and/or hot shocked gas largely
escapes \citep{Townsley:2003,Harper-Clark:2009,Rogers:2013}.

Embedded \acp{SSC} in dense \acp{GMC} represent the systems for which radiation
pressure is expected to play the greatest role.  Considering that they are
still deeply buried in \acp{GMC}, the primary signature of these \acp{SSC}
is thermal radio emission, indicating clusters of masses $\sim 10^4-10^6 \Msun$
powering \ion{H}{2} regions only a few pc in size 
\citep{Turner:1998,Turner:2000,Kobulnicky:1999,Johnson:2001,
Johnson:2003,Johnson:2015,Reines:2008,Tsai:2009,Kepley:2014}.  The molecular
clouds associated with these \acp{SSC} are far denser than typical Milky 
Way \acp{GMC}.  For example, ``Cloud D'' in NGC 5253 has a mass 
$\sim 2\times10^6 \Msun$ and diameter $\sim 40\pc$ based on the SMA 
observations of \citet{Turner:2015}, giving a mean density 
of hydrogen nuclei $n_H=1800\pcc$ and mean surface density 
$\Sigma =1600 \Surf = 0.33 \gram \pcc$.  The ``pre-\ac{SSC}'' cloud in the 
Antennae system studied with ALMA by \citet{Johnson:2015} has an estimated 
mass $\sim 0.3-1.5 \times 10^7 \Msun$ and diameter $< 40 \pc$, yielding 
$n_H \simgt 0.26 - 1.3 \times 10^4 \pcc$ and $\Sigma \simgt 0.2 -1.2 \times 
10^4 \Surf$.  In the central starburst region of NGC 253, ALMA 
observations of high critical density tracers by  \citet{Leroy:2015}
identified clouds with radii $10-50 \pc$ and masses $0.2 - 6 \times 10^7 \Msun$,
implying typical $n_H \sim 2000 \pcc$ and $\Sigma \sim 6000 \Surf$.

The deeply embedded nature of \acp{SSC} makes it difficult to constrain 
the relative importance of different feedback mechanisms empirically.  
However, the SFEs appear to be at least an order of magnitude higher 
than in typical Milky Way \acp{GMC} \citep[e.g.,][]{Tsai:2009}.  The 
young, isolated Cloud D in NGC 5253 has an estimated \ac{SFE} $\sim 0.6$, 
well above the range $\sim 0.01 -0.2$ estimated for the most luminous 
Milky Way star-forming complexes \citep{Murray:2011}.  It is 
clearly of interest to develop theoretical models that explore the effects
of feedback in controlling star formation and setting \acp{SFE} in 
\acp{GMC} comparable to \ac{SSC} hosts.  This may also inform our 
understanding of globular cluster formation, which presumably occurred 
in similarly dense and massive clouds.

Because forces from reprocessed \ac{IR} radiation become increasingly
important as gas surface densities increase,\footnote{For the spherical
  case, the total force imparted to the gas is $(1+\tau_\mathrm{IR})L_*/c$ 
  where $L_*$ is the stellar luminosity and $\tau_\mathrm{IR}$ 
  is the center-to-edge \ac{IR} optical depth, proportional to the
  cloud's surface density.} the pressure from trapped \ac{IR} has been
argued to dominate the regulation of star formation in the most
extreme systems \citep{Thompson:2005,Murray:2010}.  However, analytic
models typically require major simplifications to be tractable 
(including having all the gas mass collected in a thin,
uniform-density spherical shell, and having a single stellar cluster that is
centrally located with a luminosity that is fixed in time), and it is
unclear the extent to which these may affect the conclusions.  It is
therefore useful to employ time-dependent numerical simulations to
investigate systems with more structural and temporal complexity.

In this paper, we shall consider numerical models of massive, compact, 
turbulent \acp{GMC} that fragment gravitationally to form massive star
clusters. Our model clouds have surface density $1600 - 6200 \Surf$,  comparable to 
the \acp{GMC} in starburst nuclei within which \acp{SSC} are born.
Optical and \ac{UV} radiation from young, hot stars dominate the
luminosity of massive clusters, but these primary photons are likely to
be absorbed by dust very near their source.  This warm dust then
radiates isotropically into the lower-frequency \ac{IR} band, which
has a much smaller absorption cross-section but can still be absorbed and
re-emitted multiple times in high-column \acp{GMC}.  

In our
simulations, we apply \ac{RHD} methods to focus on the
effects of long-wavelength radiation.  We consider the 
regime of clouds at relatively 
high mass and density, such that the optical depth to \ac{IR} exceeds unity and 
simple spherical models would predict that the radiation 
force from reprocessed \ac{IR} exceeds that of the direct optical/\ac{UV}.  
We follow the co-evolution of the gas and radiation in the
system, under the assumption that radiation forces applied to dust are
transferred to the gas (i.e., assuming perfect gas-dust collisional
coupling), and that all radiation that is absorbed is locally
re-emitted.  Over time, the fraction of a cloud's mass converted to
stars increases as more and more gas collapses.  However, as the stellar 
luminosity grows, the radiation field exerts increasing outward forces on the
gas.  Under certain conditions, we show that a significant portion of
the cloud's initial gas mass is expelled from the system.

Our models are highly idealized, in that we consider exclusively the
effects of long-wavelength radiation, with spatially-uniform opacity.
As our numerical code presently allows only a single opacity, here we do not
consider the direct effects of either ionizing or non-ionizing \ac{UV}; effectively,
we treat radiation as being degraded to \ac{IR} close to each stellar 
cluster source. In reality, several feedback effects operate simultaneously, and
complete models must eventually include (at least) non-ionizing as well as
ionizing \ac{UV}, and stellar winds.  The present set of simulations
provides a baseline for more comprehensive studies, similar to the
baseline provided by simulations that focus exclusively on the effects
of ionizing radiation and cloud disruption from \ion{H}{2} region
expansion \citep[e.g.,][]{Dale:2012,Dale:2013a,Walch:2012}. 

We note that all of our clouds have escape speeds exceeding $20\kms$.  
In this regime, the simulations of \citet{Dale:2012,Dale:2013a} suggest 
that photoevaporation combined with the pressure from ionized gas would be 
able to unbind no more than $1\%$ of the cloud's mass prior to the advent of 
supernovae.  Additionally, although realistic \ac{RHD} models focused on 
non-ionizing \ac{UV} radiation have not yet been completed for this extreme regime 
(very high $\Sigma$ and large $v_{\rm esc}$),
analytic models have been developed that account for the effects of 
turbulence-driven internal structure in the radiation/gas interaction 
(\citealt{Thompson:2014}; Raskutti, Ostriker, \& Skinner 2015, in preparation); 
the Raskutti et al model has also been 
verified via numerical \ac{RHD} simulations in the lower-$\Sigma$ regime.    
These models predict that it would be difficult for the direct \ac{UV} radiation 
to expell substantial material from large, dense \acp{GMC}, since 
the already-high mean surface density is increased in the filamentary 
structures that comprise most of the cloud's mass. Thus,
as prior studies suggest that neither ionizing or non-ionizing \ac{UV} will 
be effective in destroying \acp{GMC} for the cloud regime studied here, we are 
motivated to turn the focus to the effects of reprocessed \ac{IR}.  
To our knowledge, this work represents the first direct \ac{RHD} 
study of the dynamical effects of reprocessed radiation in turbulent, 
self-gravitating, star-forming clouds.  

The plan of this paper is as follows.  We begin with a summary of our
numerical methods, model specification, and model parameters, together
with tests to verify code performance (Section
\ref{methodsandmodels}).  We then consider evolution of a fiducial
model (Section~\ref{evolution}), which also serves to illustrate the
radiation structure and differences in the radiation/matter
interaction compared to simple spherical systems (Section~\ref{radstructure}).  
In Section~\ref{varyingopacity}, we analyze the
effects of varying the opacity, demonstrating that quite high opacity
($\kappa > 10 \cmsg$) is required for substantial mass to be ejected
by radiation forces; we also quantify the gas-radiation
anticorrelation and the level of radiation trapping.  Our set of
simulations allows for varying cloud mass and radius as well as a
range of opacity; Section~\ref{parameterstudy} compares simulation
outcomes (including net \ac{SFE} and net momentum
ejected) for different model parameters.  Section~\ref{conclusions}
summarizes and discusses our conclusions.

\section{Numerical Methods \& Model Descriptions}  \label{methodsandmodels}

\subsection{Numerical Methods}  \label{methods}

We evolve the equations of \ac{RHD} using our Godunov code
\textit{Hyperion} \citep[][hereafter \citetalias{Skinner:2013}]{Skinner:2013}.  
\textit{Hyperion} is an
extension of the \textit{Athena} code \citep{Stone:2008} for
computational hydrodynamics and \ac{MHD}; we employ
the van Leer algorithm of \citet{Stone:2009} to integrate the gas
equations, utilizing the HLLC Riemann solver and a piecewise-linear
spatial reconstruction scheme.  For simplicity in this first study, we
neglect magnetic fields and adopt an isothermal \ac{EOS}; as the sound
speed is small compared to other speeds in the problem, the dynamics
are insensitive to its exact value.  Gravity of the gas as well as the
``star particles'' (representing collapsed regions that have formed
star clusters) is computed via Fourier methods, as described below.

In \textit{Hyperion}, the radiation energy density and flux, i.e., the zeroth
and first moments of the radiation intensity, are advanced in time using a
piecewise-linear spatial reconstruction and an HLL-type Riemann
solver.  For the radiation energy equation, we adopt the limit of
radiative equilibrium, such that all absorbed radiation is re-emitted locally.  
For long wavelength radiation, which we consider,
scattering is small compared to true absorption and is neglected in
our treatment.  As velocities are small compared to the speed of light
and optical depths are moderate, we adopt the static diffusion limit
in which terms of $\mathcal{O}(v/c)$ and $\mathcal{O}(\tau v/c)$ are
neglected.  Finally, we employ the \ac{RSLA}
\citep{Gnedin:2001}, which allows us to solve the radiation subsystem
explicitly rather than implicitly \citep[see also][]{Rosdahl:2013}.  
The only sources of radiation are
the star particles.

The system of equations to be solved for the gas and radiation is given by
\begin{subequations} \label{graysystem}
\begin{eqnarray}
  \partial_t \rho + \nabla \cdot (\rho \mathbf{v}) &=& 0, \label{graysystem:density} \\
  \partial_t (\rho \mathbf{v}) + \nabla \cdot (\rho \mathbf{v} \mathbf{v} + P\mathbb{I}) &=& -\rho\nabla\Phi + \rho\kappa\frac{\mathbf{F}}{c},  \label{graysystem:momentum} \\
  \frac{1}{\hat{c}} \,\partial_t \mathcal{E} + \nabla \cdot \left(\frac{\mathbf{F}}{c}\right) &=& \frac{j_*}{c},  \label{graysystem:radenergy} \\
  \frac{1}{\hat{c}} \,\partial_t \left(\frac{\mathbf{F}}{c}\right) + \nabla \cdot \mathbb{P} &=& -\rho\kappa\frac{\mathbf{F}}{c},  \label{graysystem:radflux}
\end{eqnarray}
\end{subequations}
where $\rho$, $\mathbf{v}$, and $P$ are the gas density, velocity, and
pressure, respectively, and $\Phi$ is the gravitational
potential.  In Equations~\eqref{graysystem}, $\mathcal{E}$,
$\mathbf{F}$, and $\mathbb{P}$ are the frequency-integrated radiation
energy density, flux vector, and pressure tensor, respectively, which
are measured in the inertial frame, and $\hat{c}$ is the reduced speed
of light.  The absorption opacity, $\kappa$, is taken to be a spatial
constant, although we vary this parameter in different models.
Finally, the term $j_*$ in Equation~\eqref{graysystem:radenergy}
represents emission of radiation from star particles.

The \textit{Hyperion} code uses the $M_1$ closure relation
\citep{Levermore:1981,Gonzalez:2007} to express $\mathbb{P}$ as a
function of $\mathcal{E}$ and $\mathbf{F}$ in
Equations~\eqref{graysystem}.  The radiation subsystem in
Equations~\eqref{graysystem:radenergy} and~\eqref{graysystem:radflux}
is operator-split from the gas subsystem in
Equations~\eqref{graysystem:density} and~\eqref{graysystem:momentum}
and is explicitly evolved by subcycling on a time step determined by
the radiation propagation speed, $\hat{c}$.  Using the \ac{RSLA}, we
choose $\hat{c}$ satisfying the \ac{RSLA} static diffusion criterion
\citepalias{Skinner:2013},
\begin{equation}
	\hat{c} \gg v_\mathrm{max}\max\{1,\tau_\mathrm{max}\}, \label{chat}
\end{equation}
where $v_\mathrm{max}$ is the maximum velocity of the gas and
$\tau_\mathrm{max}$ is the maximum optical depth in the simulation.
This allows us to reduce the radiation-to-gas time step ratio to a
reasonable level, but has a negligible effect on the system's dynamics
as the gas evolves much more slowly than the radiation field.  Our
simulations typically use a value of $\hat{c}$ such that
$\hat{c}/v_\mathrm{max} \sim 10-100$, which is substantially less than
realistic values of $c/v_\mathrm{max} \sim 10^4-10^5$, making explicit
time integration computationally feasible.

In \textit{Hyperion}, star particles are treated using the algorithm
developed by \cite{Gong:2013}, with added functionality to provide for
the luminosity of the particles.\footnote{At the resolution of our
  simulations, each ``star particle'' actually represents a
  fully-sampled star cluster.}  The $M_1$ closure cannot resolve the
behavior of a streaming radiation field too near to a true point
source, since the the radiation flux from such a
source varies rapidly in angle.  Therefore, radiation sources in our algorithm
must be resolved over some minimum number of grid zones.  We have
found it convenient to add radiation energy density to the grid using
a Gaussian source function given by
\begin{equation}
	j_*(\mathbf{x}) = \frac{L_*}{(2\pi \sigma_*^2)^{3/2}} \exp \left(-\frac{|\mathbf{x} - \mathbf{x}_*|^2}{2\sigma_*^2}\right),  
\label{sourceprofile}
\end{equation}
where $L_*$ is the star particle's luminosity, $\mathbf{x}_*$ is the
star particle's position, and $\sigma_* \equiv R_*/\sqrt{2\ln 2}$ is
set such that the \ac{HWHM} of the distribution is equal to the star
particle's effective size, $R_*$.  Note that $\int_0^r j_*(r')
\,4\pi r'^2\,dr' \to L_*$ for $r/R_*\gg 1$.  In practice, we have found that
sources with $R_*/\Delta x \gtrsim 8$ are sufficiently well-resolved
that angular variations in the radiation flux at radii $r \gg R_*$ are
negligible (see Section~\ref{models}).  

To compute gravitational forces including contributions from both gas
and star particles, we first use the \ac{PM} method to assign the star
particle masses to a discrete grid via the \ac{TSC} method, as described in
\citet{Gong:2013}.  We then apply the ``zero-padding'' method of
\cite{Hockney:1988} to obtain the potential, $\Phi(\mathbf{x})$, of an
isolated source distribution subject to open (vacuum) boundary
conditions via \acp{FFT}.  This potential is given by the solution of
Poisson's equation,
\begin{equation}
\nabla^2 \Phi = 4\pi G \rho(\mathbf{x}), \label{poisson}
\end{equation}
for a given density field, $\rho(\mathbf{x})$; $\rho$ includes both contributions from the gas and the star particles.  Solutions of Equation~\eqref{poisson} can be expressed as the convolution
\begin{equation}
	\Phi(\mathbf{x}) = G \int \mathcal{G}(\mathbf{x},\mathbf{x}')\rho(\mathbf{x}') \,d^3\mathbf{x}', \label{gravity:phi}
\end{equation}
where $\mathcal{G}(\mathbf{x},\mathbf{x}') = \mathcal{G}(|\mathbf{x}-\mathbf{x}'|)=-|\mathbf{x}-\mathbf{x}'|^{-1}$ 
is a Green function solution of the equation 
$\nabla^2 \mathcal{G} = 4\pi\delta^3(\mathbf{x}-\mathbf{x}')$.  Using the Fourier convolution theorem, 
Equation~\eqref{gravity:phi} may be re-expressed as the convolution of the 
Fourier transforms of the density distribution and Green function; 
details of this are given in the Appendix.

\subsection{Model Description}  \label{models}

Each model cloud is initiated as a uniform-density sphere of radius
$\Rcloud$ and mass $\Mcloud$.  The cloud is centered in a
computational box of side length 
$L_\mathrm{box} = 4\Rcloud$ with background density
set to $1\%$ of the initial cloud density.\footnote{Realistic clouds 
may have lower density contrast relative to their
surroundings, which could lead to additional late-time accretion.  However, 
for this first study we consider isolated clouds for simplicity.}  A highly supersonic turbulent
velocity field is applied initially, which rapidly creates density
structure within the cloud.  Over time, this initial velocity field
decays, although collapse and radiation forces drive further turbulent
motions.

We initialize the turbulent velocity field using Gaussian random
perturbations with power spectrum $|\delta \mathbf{v}| \propto
k^{-4}$, for $k/dk \in [2,64]$, where $dk = 2\pi / L_\mathrm{box}$, as
described in \cite{Stone:1998}.  The velocity perturbations are
normalized such that $\alpha_\mathrm{vir, init} = 2$, corresponding to
a just-bound state, where
\begin{equation}
	\alpha_\mathrm{vir} \equiv 2 \frac{E_\mathrm{kin}}{E_\mathrm{grav}}
\end{equation}
is the virial parameter, and where $E_\mathrm{kin} = 
(1/2)\Mcloud \sigma^2$ and $E_\mathrm{grav} = (3/5) G
\Mcloud^2/\Rcloud$ are the initial kinetic and gravitational energies of the gas,
respectively; $\sigma=\left[\alpha_\mathrm{vir, init} 3 G \Mcloud/(5 \Rcloud)\right]$ 
is the initial turbulent
velocity dispersion.\footnote{Note that for a uniformly dense sphere, such as our
initial configuration, $\alpha_\mathrm{vir}$ is equivalent to the often-used 
definition of the virial parameter from \citet[][see Equation~2.8a]{Bertoldi:1992}.}  
The velocities of the perturbations are
recentered such that they add no net momentum to the computational
domain, i.e., such that $\int \rho \delta \mathbf{v} \,dV = 0$.

For all simulations, the isothermal sound speed is set to $c_s =
2\mbox{ km s}^{-1}$.  We adopt this value, somewhat larger than true
thermal sound speeds in cold clouds, in part to limit extreme shock
compression, as would occur if we had included magnetic fields.  We
have found that results are insensitive to the exact choice of the
sound speed, provided the Mach number is large.  The initial turbulent
Mach number is $\sim 11$ for our fiducial parameters (see below), and
more generally the turbulent flows in all of our models are highly
supersonic. Tests show that there is little variation in
simulation outcomes for different realizations of the initial random
perturbed velocity field
(see Section~\ref{evolution}), so we use a single realization for most sets
of model parameters.

Following \citet{Gong:2013}, the density threshold for star particle creation is set using the Larson-Penston criterion, $\rho_\mathrm{thr} = \rho_\mathrm{LP}(r=\Delta x/2)$, where
\begin{equation}
	\rho_\mathrm{LP}(r) = \frac{8.86 c_s^2}{4\pi G r^2}
\end{equation}
describes the density profile of the asymptotic state produced by
gravitational collapse of an isothermal sphere
\citep{Larson:1969,Penston:1969}; this profile is approached for
collapse initiated from a wide variety of initial conditions,
including supersonic turbulent flows (see
\citealt{Gong:2009,Gong:2011}).  As discussed in \cite{Gong:2013}, the
Larson-Penston density criterion is a factor of $\sim 14$ times larger
than the Truelove criterion \citep{Truelove:1997} used in many other
star particle creation implementations \citep[see,
  e.g.,][]{Krumholz:2004,Federrath:2012}, but the star particles
produced in a turbulent flow follow essentially the same histories.

We assume the star clusters that form in the cloud (represented as
star particles) have a bolometric luminosity per unit mass of 
$\Psi =1700 \mbox{ erg s}^{-1} \mbox{ g}^{-1}$, as obtained from a
Starburst99 model \citep{Leitherer:1999} for the total luminosity of a
young cluster 
that fully samples a Kroupa \ac{IMF}, averaged over a lifetime of 5 Myr.\footnote{For random 
sampling of a fixed \ac{IMF}, $\Psi$ saturates for clusters of mass 
$M_\mathrm{cl} \gtrsim 10^4 \Msun$.  In our simulations, we find that star particles, which represent
star clusters on the scales we consider, may occasionally form with masses of order a few times less than 
$10^4 \Msun$.  However, these somewhat under-sampled clusters typically comprise a trivial fraction 
of the total mass in stars, which is dominated by fully-sampled clusters.  Furthermore, these lower-mass star 
particles tend to accrete sufficient gas over a short enough time scale to justify the use of a constant 
specific luminosity.}  Since the total luminosity of a cluster would not
change substantially over the cloud's lifetime, for simplicity 
we set $L_* = \Psi M_*$, independent of the star particle's age.

We adopt a constant effective physical size for each star particle,
which we take to be $R_* = 1\pc$.  Sources of this size are
generally consistent with observations of young, embedded super-star
clusters \citep[e.g.,][]{Johnson:2003} as well as older clusters
\citep{Murray:2009}.  Note that here, we do not attempt to model the \ion{H}{2} 
region that those star clusters would create.  For high luminosity
sources, \ac{UV} radiation pressure on dust significantly alters the
conditions within \ion{H}{2}  regions compared to the classical Str\"omgren
solution \citep{Binette:1997,Dopita:2002,Draine:2011}, 
with radiation forces strongly compressing the photoionized gas.  In 
principle, high pressure from a shocked stellar wind could further compress 
the photoionized gas, but observed ionization parameter measures suggest 
that this does not occur in practice  \citep{Yeh:2012}.  Given the uncertainties about the
effective size of the emission region in which the cluster's radiation
is reprocessed into \ac{IR}, we opt to keep $R_*$ fixed rather than
having it depend on the mass and age of each star particle.
Tests indicate that varying the physical size $R_*$ of each star particle  
radiation source by a factor of
two in either direction at fixed grid resolution 
only affects the final total mass 
in stars by about 5\% (see Section~\ref{evolution}).

\begin{figure}
  \centering
  \epsscale{1}
  \plotone{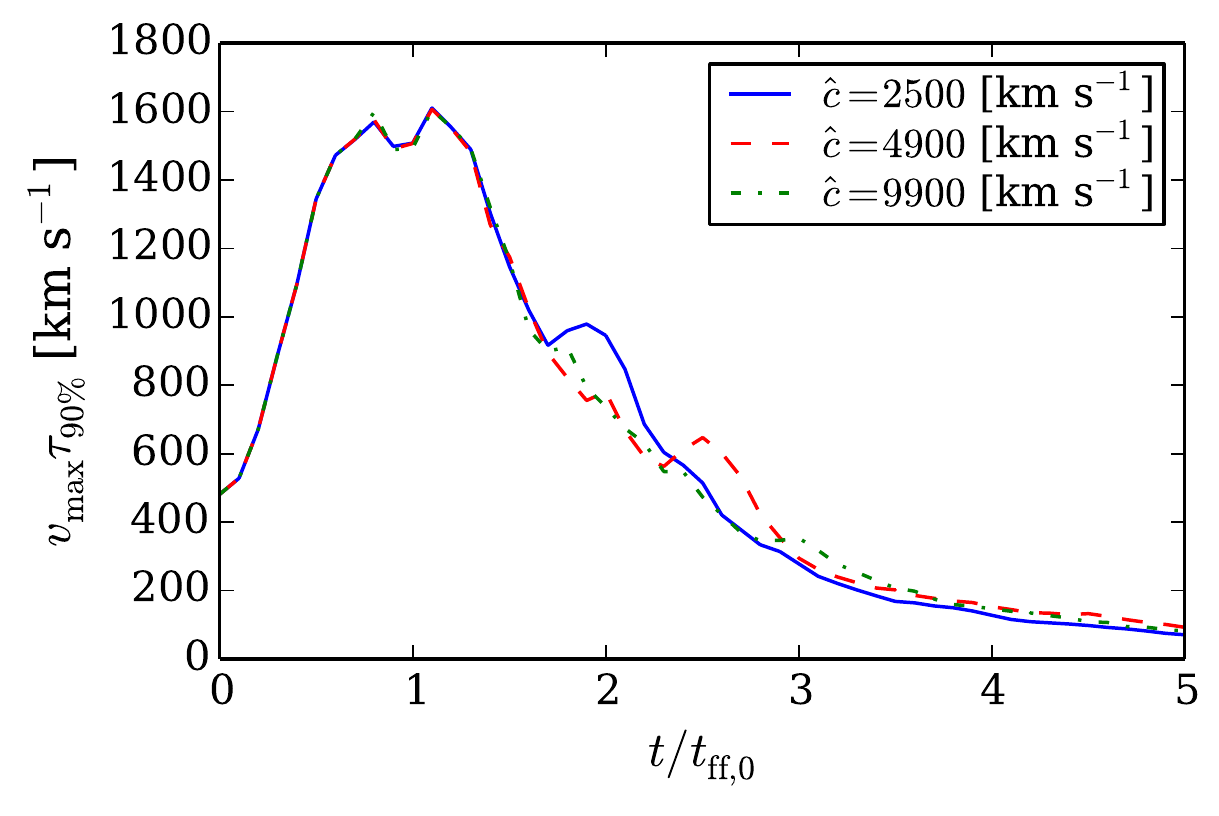}
\caption{
Time evolution of $v_\mathrm{max}\tau_{90\%}$ for three different values
of the reduced speed of light, $\hat{c}$, where $v_\mathrm{max} \equiv
\sigma + c_s$ approximates the maximum hydrodynamic signal
speed and $\tau_{90\%}$ is the optical depth at the 90th percentile
over all projections of the grid in all directions.  All runs were
performed with $N=128^3$ zones.  This demonstrates that the time
evolution of the optical depth for the bulk of the gas is insensitive
to the actual value of the reduced speed of light used, provided the
\ac{RSLA} static diffusion criterion of Equation~\eqref{chat} is 
satisfied.  The value of $\hat{c}=4900$ km s$^{-1}$ is the reduced
speed of light in our fiducial run.}
	\label{methods:fig:tau90_vs_crad}
\end{figure}

From Equation~\eqref{chat}, the required value of $\hat{c}$ depends on
the maximum optical depth and velocity in the model.  If we define
$\tau_{90\%}$ at a given time as the optical depth at the 90th
percentile over all projections of the grid in all directions, our
tests show that the maximum of $\tau_{90\%}$ over a simulation run is
comparable to $\tau$, the initial \ac{IR} optical depth across
the \ac{GMC}, within a factor of a few, and that this value is
well-converged if we use $\hat{c} = 5 v_\mathrm{max}
\tau$.  
Figure~\ref{methods:fig:tau90_vs_crad} shows the
time evolution of $v_\mathrm{max}\tau_{90\%}$ for three different values
of $\hat{c}$, where $v_\mathrm{max} = \sigma + c_s$
approximates the maximum hydrodynamic signal speed.  Each of these
runs was performed at a reduced resolution of $N=128^3$ zones 
for parameters otherwise the same as the fiducial model (K20), 
and the choices of 2500, 4900, and 9900 km s$^{-1}$ for $\hat{c}$ correspond
to 5, 10, and 20 times 
$(\sigma+c_s) \tau$,
respectively.  The data clearly indicate that the optical depth that
is ``seen'' by the majority of the radiation is insensitive to the specific
value of $\hat{c}$, provided the \ac{RSLA} static diffusion criterion of
Equation~\eqref{chat} is satisfied.  In all of our
simulations, we therefore choose $\hat{c}$ such that $\hat{c} \ge 5
(\sigma+\cs) \tau$; in most simulations we have
$\hat{c} = 10 (\sigma+c_s)\tau$.

In our current investigation, which focuses on effects of reprocessed
radiation that is absorbed and re-emitted by dust, the absorption
opacity of the medium, $\kappa$, is taken to be a constant in space and time.  
More realistically, the opacity law would depend on the
frequency of the radiation as well as on local dust abundance and
properties, such that the frequency-averaged opacity would depend on
these properties and also the particular radiation regime.  For
example, consider the energy-, Planck-, flux-, and Rosseland-mean
opacities, which are defined, respectively, by
\begin{subequations}
\begin{eqnarray}
	\kappa_{\mathcal{E}} &\equiv& \frac{\int \kappa_\nu \mathcal{E}_\nu \,d\nu}{\int \mathcal{E}_\nu \,d\nu}, \\
	\kappa_\mathrm{P}       &\equiv& \frac{\int \kappa_\nu B_\nu \,d\nu}{\int B_\nu \,d\nu}, \\
	\kappa_F             &\equiv& \frac{\int \kappa_\nu F_\nu \,d\nu}{\int F_\nu \,d\nu}, \\
	\kappa_\mathrm{R}       &\equiv& \frac{\int \partial B_\nu/\partial T \,d\nu}{\int  \kappa_\nu^{-1}(\partial B_\nu/\partial T) \,d\nu}.
\end{eqnarray}
\end{subequations}
For an optically thick flow, if we assume the radiation field is that
of a blackbody, then $\kappa_{\mathcal{E}}=\kappa_\mathrm{P}$ and
$\kappa_F=\kappa_\mathrm{R}$.  In contrast, for an optically thin
flow, since $F_\nu \propto \mathcal{E}_\nu$, it follows that
$\kappa_F=\kappa_{\mathcal{E}}$.  However, the assumption that the
radiation field is a blackbody is dubious in this regime, hence it is
unclear whether or not $\kappa_{\mathcal{E}}$ and $\kappa_\mathrm{P}$
are even related.  A full frequency-dependent treatment of radiation
would require some transition between optically thick and thin
regimes.  Here, the primary interest is investigating the effects of
radiation forces that develop under optically thick conditions, such
that the most relevant single value of the opacity is
$\kappa_F=\kappa_\mathrm{ R}$.  For temperatures $\lesssim 100$ K, the frequency dependence of
$\kappa_\nu\propto \nu^2$ leads to an approximate dependence $\kappa_R
\propto T^2$ \citep{Draine:2011a}, and some recent simulations
investigating radiation in dusty starburst disks have included this
dependence \citep{Krumholz:2012,Davis:2014}. For the moderate optical
depth as would apply in molecular clouds of surface density $\sim 
1\mathrm{ g\, cm^{-2}}$, the opacity would
increase by less than a 
factor $\sim 2$ from the edge to the center of the cloud, and 
for simplicity we adopt a constant $\kappa$.  We allow, however, for a range 
of values of $\kappa$, as described below.

We consider four series of runs over a wide range of \ac{GMC} conditions,
which we summarize in Table~\ref{table:modelparams}.  Our model parameters
are the adopted opacity, $\kappa$, and the initial cloud
radius and mass, $\Rcloud$ and $\Mcloud$, respectively.
  In addition to the basic input parameters, 
for each model Table~\ref{table:modelparams} lists the initial gas surface 
density, $\Sigma \equiv \Mcloud/(\pi \Rcloud^2)$, the 
initial turbulent velocity dispersion, 
$\sigma$, the initial optical depth through the cloud,
and the initial free-fall time of the cloud, $\tau$
$\tff \equiv [\pi^2 \Rcloud^3/(8G \Mcloud)]^{1/2}$, 
We run each simulation for 8 times the initial free-fall time, i.e.,
to $t_\mathrm{final} = 8 \tff$.

In the K series, we vary the opacity while holding $\Rcloud=10$
pc and $\Mcloud=10^6 \Msun$ fixed, which allows
$\tau$ to vary while $\Sigma$ and $\sigma$ do
not.  As we shall discuss 
in Section~\ref{radstructure}, simple scaling arguments
suggest that the most important parameter in determining the fate of a
cloud with feedback dominated by \ac{IR} radiation 
is $f_\mathrm{Edd,*} \equiv \kappa \Psi/(4 \pi cG)$; 
the K series allows us to focus on this
dependence. For gas-to-dust ratio $\sim 100$,
realistic values of the Rosseland mean opacity for
reprocessed \ac{IR} radiation are likely in the range $\kappa \sim 1-5 \cmsg$ 
\citep{Semenov:2003}, but
we include a much larger range $\kappa = 1-40 \cmsg$ in the K series to explore
the physical dependence of the results on $\kappa$.  
The values at the upper end of the opacity range in our models 
may apply for \ac{ISM} regions 
that have high dust abundances.  Potentially, \acp{GMC} with high \ac{SFE} 
may have dust abundances enhanced by self-enrichment.
As discussed in Section~\ref{intro}, with $\tau \ge 1$ in all
runs we expect the reprocessed \ac{IR} radiation force to dominate over
the direct radiation force.  Also, since $\sigma \ge 16 \kms$, which
exceeds the maximum expansion velocity of $\sim 10 \kms$ for
\ion{H}{2} regions, we would not expect significant contribution from
\ion{H}{2} pressure for clouds in the regime under investigation.

In the R series, we vary $\Rcloud$ while holding
$\Mcloud=10^6 \Msun$ fixed, while in the M series, we vary
the cloud mass, $\Mcloud$, while holding the cloud radius
$\Rcloud=10$ pc fixed.  In both the R and M series, we hold
$\kappa$ fixed, but $\Sigma$, $\sigma$, and
$\tau$ all depend on both the cloud radius and mass,
hence vary with either of them, albeit in different ways.  Finally, in
the RM series, we vary both $\Rcloud$ and $\Mcloud$
together such that $\Mcloud \propto \Rcloud^2$, while
holding $\kappa$ fixed.  Thus, in the RM series 
the gravitational potential well depth and 
$\sigma$ vary while
$\Sigma$ and $\tau$ do not.  For the R, M,
and RM series, we set $\kappa=20$ cm$^2$ g$^{-1}$.  This value is
likely larger than realistic values for Solar metallicity.  
However, we find from the K
series that radiation does not have significant effects for lower
values of $\kappa$, and in the interest of exploring how the dynamical
outcomes depend on cloud size and mass we must choose a larger value
of $\kappa$.

For computational expediency, we use $\hat{c} = 5(\sigma+\cs)
\tau$ in runs K30, K40, R7.1, and M2, instead of the larger value $\hat{c} =
10(\sigma+\cs) \tau$ used in all other runs.  Our studies have shown
that the typical values of $\tau_{90\%}$ are in fact converged with respect to
this somewhat relaxed criterion (see Figure~\ref{methods:fig:tau90_vs_crad}).

\newcommand\tna{\,\tablenotemark{a}}
\newcommand\tnb{\,\tablenotemark{b}}
\newcommand\tnc{\tablenotemark{c}}
\newcommand\tnd{\tablenotemark{d}}
\newcommand\tne{\tablenotemark{e}}
\newcommand\tnf{\tablenotemark{f}}
\newcommand\tng{\tablenotemark{g}}
\newcommand{\bftab}{\fontseries{b}\selectfont}
\newcommand{\phsq}{\phantom{$/\sqrt{2}$}}
\begin{deluxetable*}{cccccccccccc}
  \tabletypesize{\footnotesize}
  \tablecaption{Initial Model Parameters}
  \tablewidth{0pt}
  \tablecolumns{11}
  \centering
  \tablehead{
    \colhead{\phantom{\quad}} &
    \colhead{Model} &
    \colhead{$\kappa$} &
    \colhead{$\Rcloud$} &
    \colhead{$\Mcloud$} &
    \colhead{$\Sigma$} &
    \colhead{$\sigma$} &
    \colhead{$f_\mathrm{Edd,*}$} &
    \colhead{$\tau$\tna} &
    \colhead{$t_\mathrm{ff}$} &
    \colhead{$\hat{c}$} &
    \colhead{\phantom{\quad}} \\
    \colhead{} &
    \colhead{} &
    \colhead{(cm$^2$ g$^{-1}$)} &
    \colhead{(pc)} &
    \colhead{($10^6 \Msun$)} &
    \colhead{(g cm$^{-2}$)} &
    \colhead{(km s$^{-1}$)} &
    \colhead{} &
    \colhead{} &
    \colhead{(Myr)} &
    \colhead{($10^3$ km s$^{-1}$)} &
    \colhead{}
  }
  \startdata
	&        K1      &  1 & 10 & 1 & 0.67 & 23 & 0.068 &  1 & 0.54 & 0.25 & \\
	&        K5      &  5 & 10 & 1 & 0.67 & 23 & 0.34  &  5 & 0.54 & 1.2  & \\
	&        K10     & 10 & 10 & 1 & 0.67 & 23 & 0.68  & 10 & 0.54 & 2.5  & \\
	& \bftab K20\tnb & 20 & 10 & 1 & 0.67 & 23 & 1.4   & 20 & 0.54 & 4.9  & \\
	&        K30     & 30 & 10 & 1 & 0.67 & 23 & 2.0   & 30 & 0.54 & 4.9  & \\
	&        K40     & 40 & 10 & 1 & 0.67 & 23 & 2.7   & 40 & 0.54 & 4.9  & \\
	\vspace{-0.2cm} \\
	\tableline
	\vspace{-0.2cm} \\
	&  R7.1           & 20 & 10$/\sqrt{2}$ & 1 & 1.30 & 27 & 1.4 & 40 & 0.32 & 5.8 & \\
	&  \bftab R10\tnb & 20 & 10            & 1 & 0.67 & 23 & 1.4 & 20 & 0.54 & 4.9 & \\
	&  R14.1          & 20 & 10$\sqrt{2}$  & 1 & 0.33 & 19 & 1.4 & 10 & 0.90 & 2.1 & \\
	\vspace{-0.2cm} \\
	\tableline
	\vspace{-0.2cm} \\
	&  M0.5          & 20 & 10 & 0.5 & 0.33 & 16 & 1.4 & 10 & 0.76 & 1.8 & \\
	&  \bftab M1\tnb & 20 & 10 & 1   & 0.67 & 23 & 1.4 & 20 & 0.54 & 4.9 & \\
	&  M2            & 20 & 10 & 2   & 1.30 & 32 & 1.4 & 40 & 0.38 & 6.8 & \\
	\vspace{-0.2cm} \\
	\tableline
	\vspace{-0.2cm} \\
	&  R5M0.25          & 20 & 5             & 0.25 & 0.67 & 16 & 1.4 & 20 & 0.38 & 3.6 & \\
	&  R7.1M0.5         & 20 & 10$/\sqrt{2}$ & 0.5  & 0.67 & 19 & 1.4 & 20 & 0.32 & 4.2 & \\
	&  \bftab R10M1\tnb & 20 & 10            & 1    & 0.67 & 23 & 1.4 & 20 & 0.54 & 4.9 & \\
	&  R14.1M2          & 20 & 10$\sqrt{2}$  & 2    & 0.67 & 27 & 1.4 & 20 & 0.90 & 5.8 & \\
	&  R20M4            & 20 & 20            & 4    & 0.67 & 32 & 1.4 & 20 & 0.76 & 6.8 &
  \enddata
  \label{table:modelparams}
\noindent
\tablenotetext{a}{The optical depth through the center of the initial uniform 
cloud, 
  \mbox{$\tau \equiv 2\kappa \Rcloud \Mcloud/(\case{4}{3} \pi \Rcloud^3) 
\equiv \case{3}{2} \kappa \Sigma$}.}  
\tablenotetext{b}{The parameters for these runs (in bold face) are identical and refer to our fiducial model.}  

\end{deluxetable*}

\section{Results}  \label{results}

\subsection{Evolution of a Fiducial Model} \label{evolution}

\begin{figure}
\epsscale{1}
\plotone{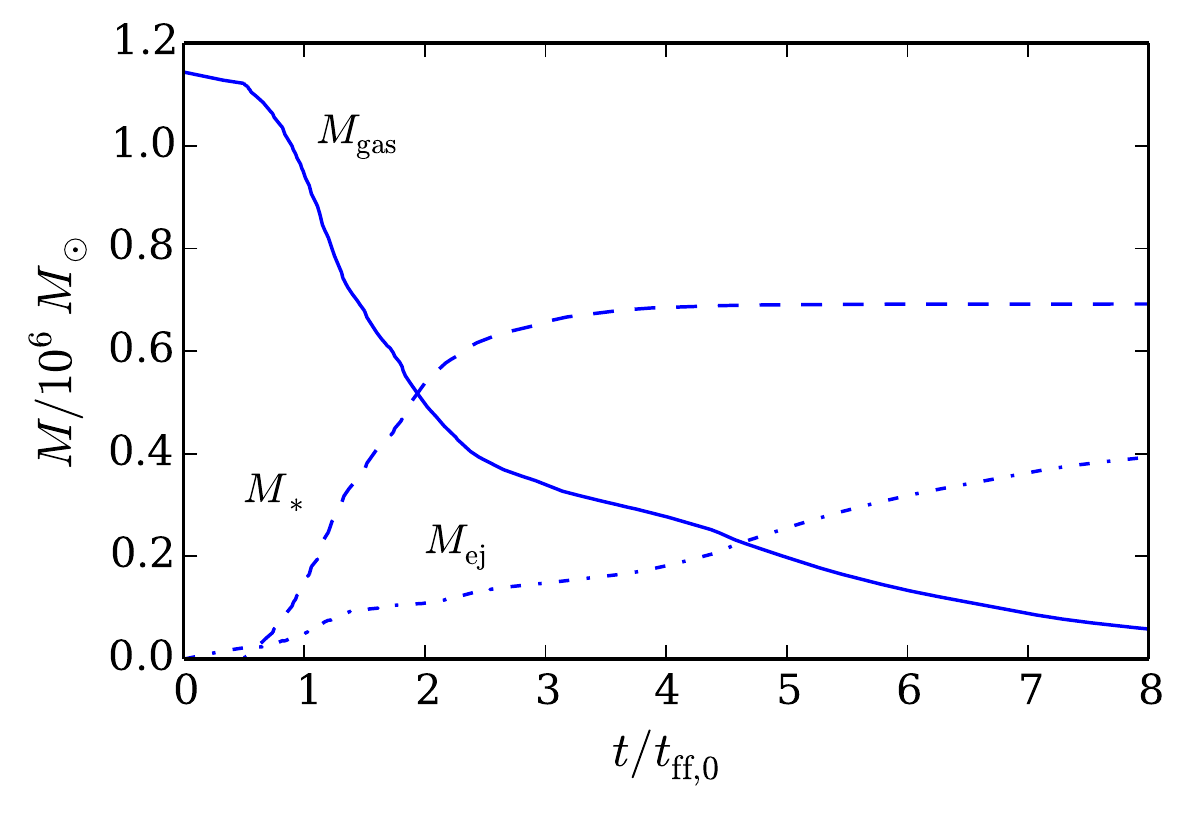}
\caption{Evolutionary history of the fiducial model, with 
$\kappa = 20\mbox{ cm}^2 \mbox{ g}^{-1}$,
$\Rcloud = 10\mbox{ pc}$, 
and $\Mcloud = 10^6 \Msun$ (see K20 in Table~\ref{table:modelparams}).
Shown are the gas mass within the domain (including the diffuse background
gas, which is initially set to 1\% of the initial cloud density), the mass that has been 
ejected from the box, and the total mass in star particles.  Time is in 
units of the initial free-fall time within the cloud.  
\label{evolution:fig:masshist}
}
\end{figure}

We begin by describing the overall evolution of a fiducial case.  
To illustrate key features of the interaction between the 
turbulent gas of the cloud and 
the diffuse radiation field that permeates it, we select the model with 
$\kappa = 20\mbox{ cm}^2 \mbox{ g}^{-1}$,
$\Rcloud = 10\mbox{ pc}$, and $\Mcloud = 10^6 \Msun$. This run 
has initial properties as listed, e.g., under K20 in Table~\ref{table:modelparams}.  
Figure~\ref{evolution:fig:masshist} shows the history of the mass of 
gas within the box, the mass ejected, and the total mass of all star 
particles.

Over time, the gas mass steadily declines (slowing after $2\tff$), the
ejected mass steadily increases, and the mass in star particles
increases rapidly between $\sim 1 - 2\tff$, and then reaches a
plateau.  Accretion onto the star particles effectively ceases after
$\sim 3 \tff$, because the strong outward radiation force exceeds the
inward gravitational force. Over the duration of the simulations,
$\sim 60\%$ of the initial mass in the box (or $70\%$ of the initial cloud mass) 
is accreted onto the star
particles.  Although a small amount of gas ($\sim 10\%$ of the total) is ejected
from the box at early times (because a portion of the turbulent cloud
had sufficiently large initial velocities to escape), the gas ejection
at $t\simgt 3\tff$ is driven by the radiation force.

\begin{figure}
\epsscale{1}
\plotone{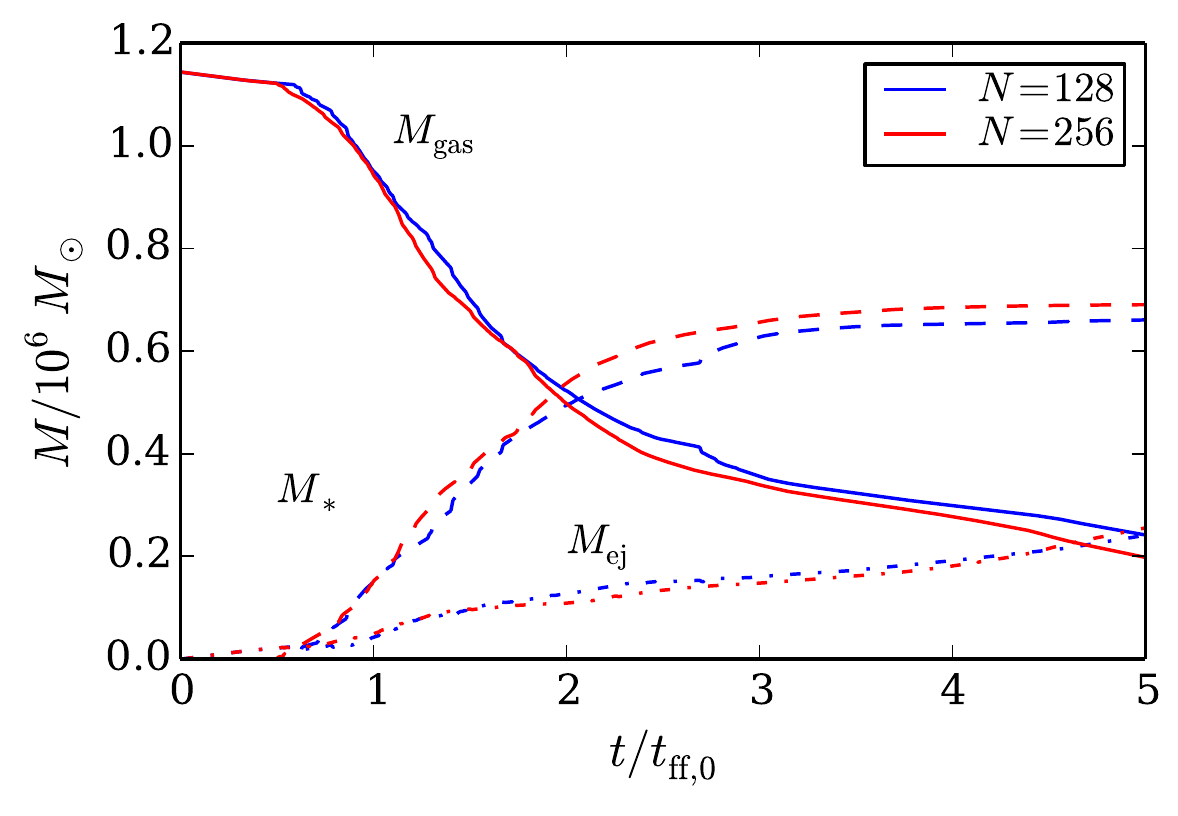}
\caption{Effect of resolution on the mass history of the fiducial model 
(see K20 in Table~\ref{table:modelparams}).  After 5 free-fall times, the 
variation of the final stellar mass is of order 5\%, indicating that the 
evolution of the simulations is acceptably converged at a resolution of $N=256$.
\label{evolution:fig:masshist_res}}
\end{figure}

\begin{figure}
\epsscale{1}
\plotone{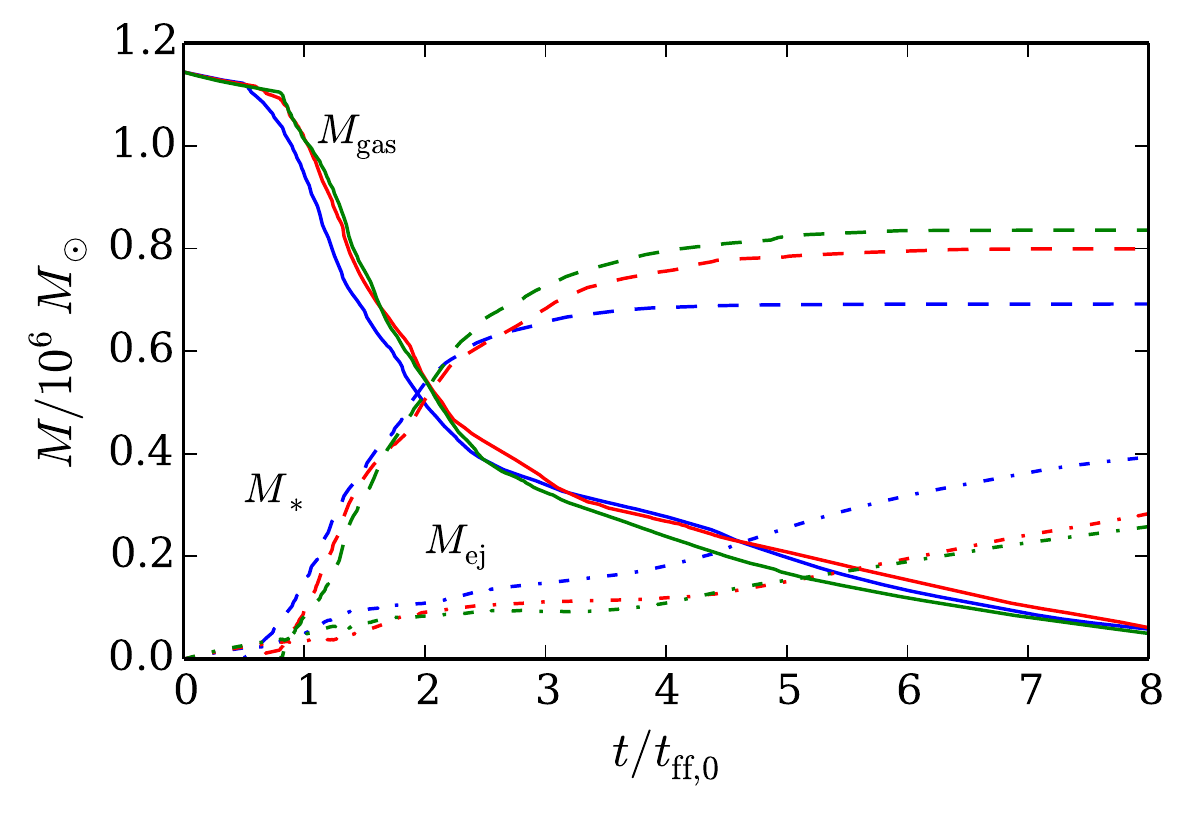}
\caption{Effect of the detailed 
initial turbulent velocity field on the mass history of the fiducial model 
(see K20 in Table~\ref{table:modelparams}).
Red, green, and blue curves represent 
three distinct realizations of the initial turbulence
spectrum (i.e., produced by three differently seeded random number sequences), 
all with the same initial kinetic energy.  
After 8 free-fall times, the variation of the final stellar and ejected masses is of order 10\%.}
\label{evolution:fig:masshist_rseed}
\end{figure}

\begin{figure}
\epsscale{1}
\plotone{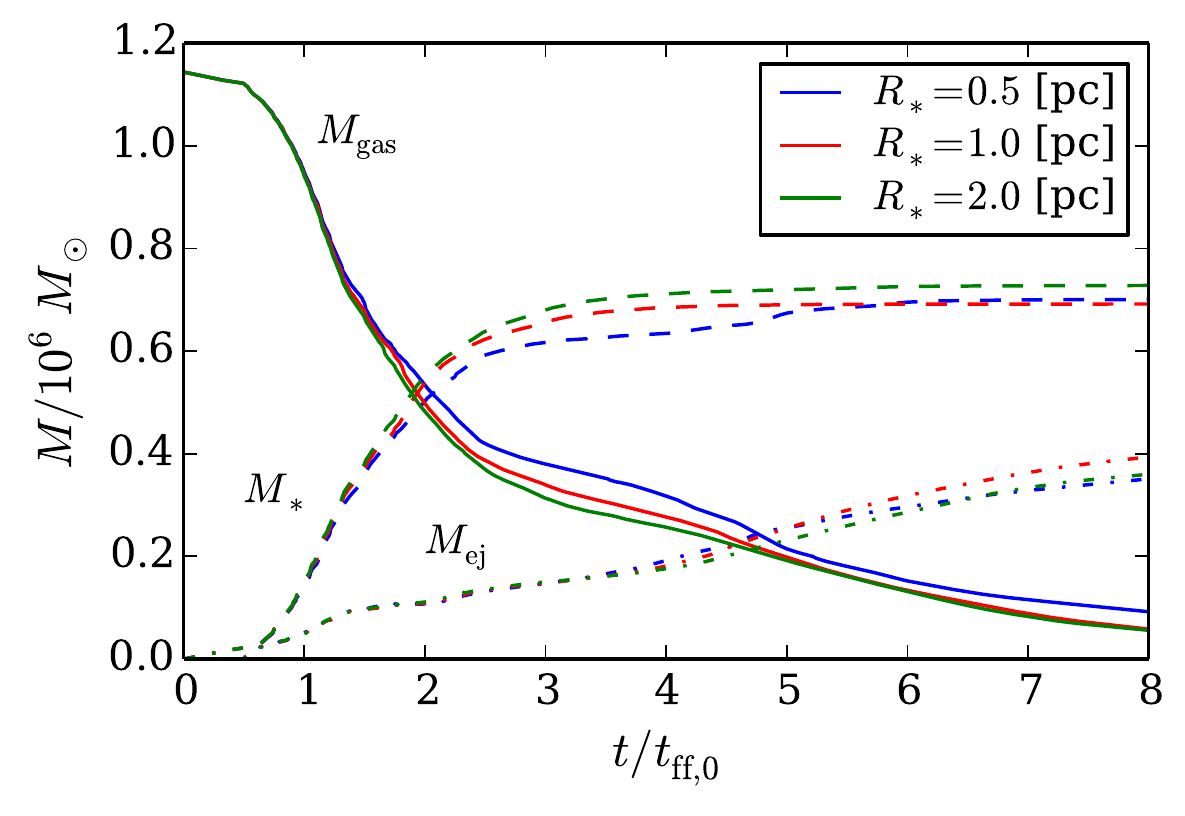}
\caption{Effect of the source size, $R_*$, on the mass history of the fiducial model 
(see K20 in Table~\ref{table:modelparams}).  After 5 free-fall times, the 
variation of the final stellar and ejected masses is of order 5\% for runs with $R_*$
a factor of 2 larger and smaller than the fiducial size $R_*=1$ pc.
\label{evolution:fig:masshist_src_res}}
\end{figure}

The standard resolution of our simulations, $N=256$, is dictated by the 
combined constraints of computational cost steeply increasing 
with resolution, and the desire to explore a range of parameters.  Although 
the details of turbulence models are always subject to resolution, we have 
confirmed that the overall evolution of our simulations is acceptably 
converged.  
Figure~\ref{evolution:fig:masshist_res} shows the mass histories analogous to 
Figure~\ref{evolution:fig:masshist}, comparing results for resolutions $N=128$
and $N=256$ for the first 5 free-fall times.  Evidently, the variations are only 
of order 5\%.
We note that different seeds for the random number generator used to form the 
initial turbulent velocity field (with a given total kinetic energy) can 
also lead to of order 10\% variation in the mass history of a cloud, 
as shown in Figure~\ref{evolution:fig:masshist_rseed}.
Finally, we note that the physical size of the source, $R_*$, used in
Equation~\ref{sourceprofile}, leads to of order 5\% variation in the mass histories for 
values a factor of 2 above and below the fiducial source size of $R_*=1$ pc, as shown
in Figure~\ref{evolution:fig:masshist_src_res}.

\begin{figure*}

\epsscale{0.57}
\plotone{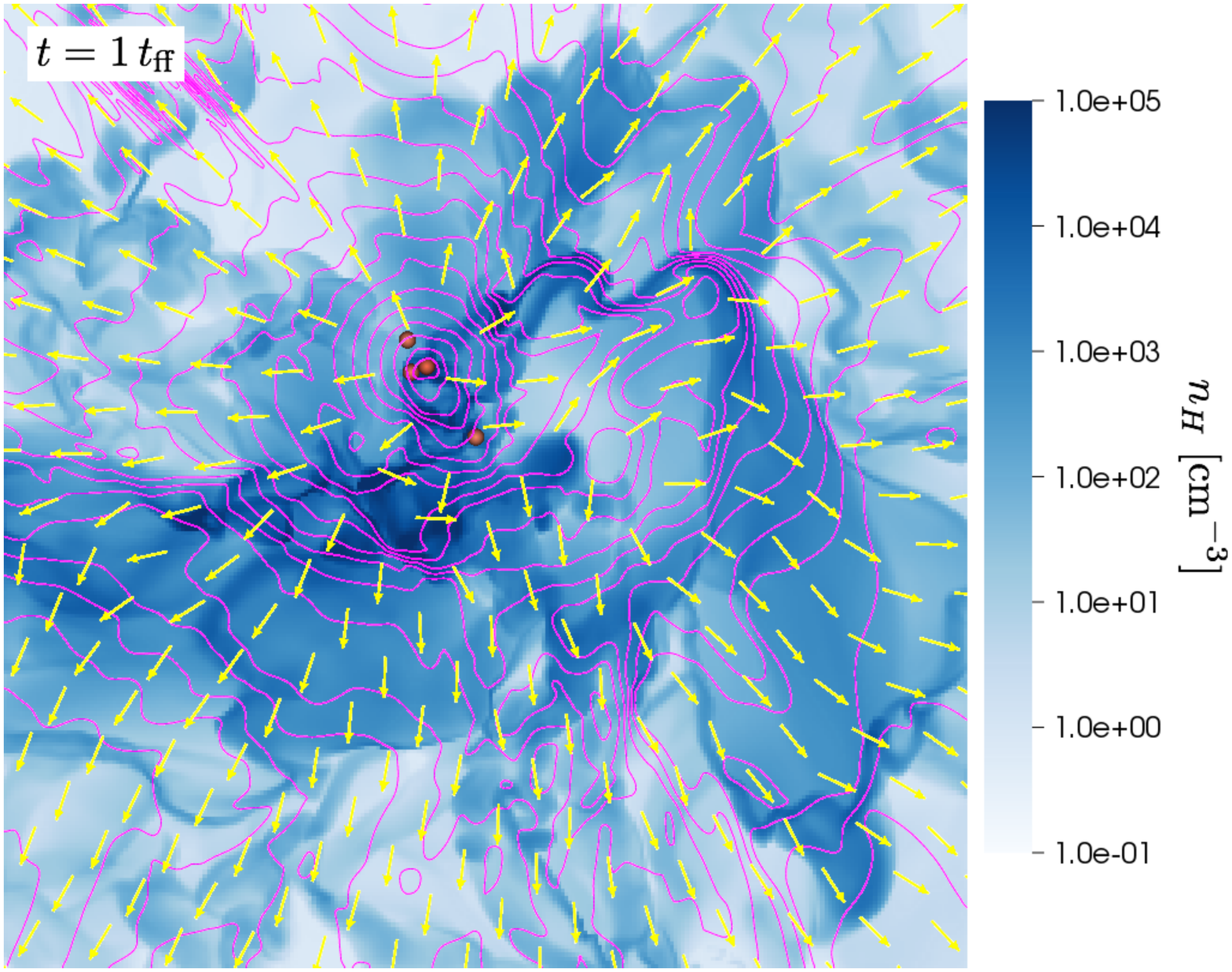}
\plotone{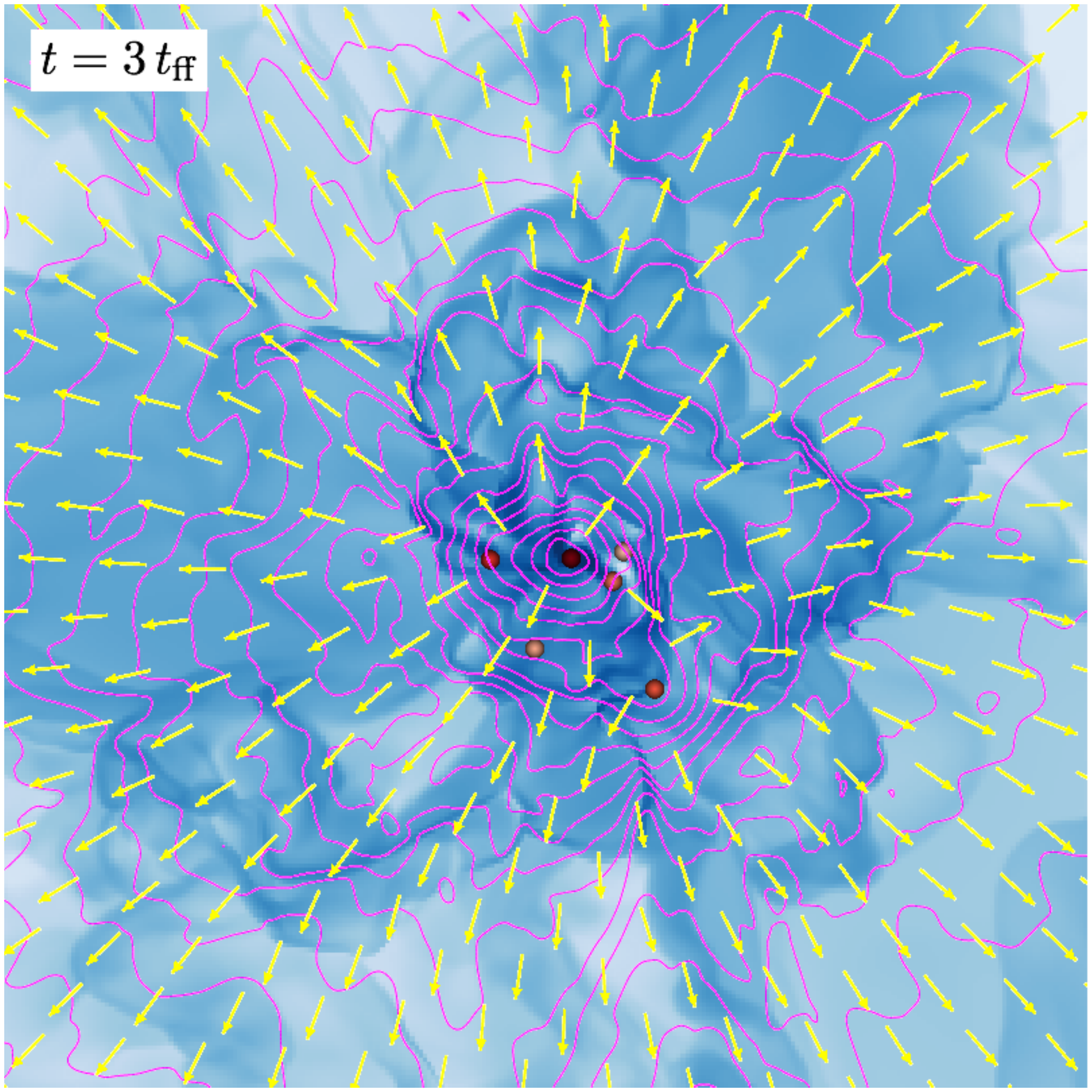}
\plotone{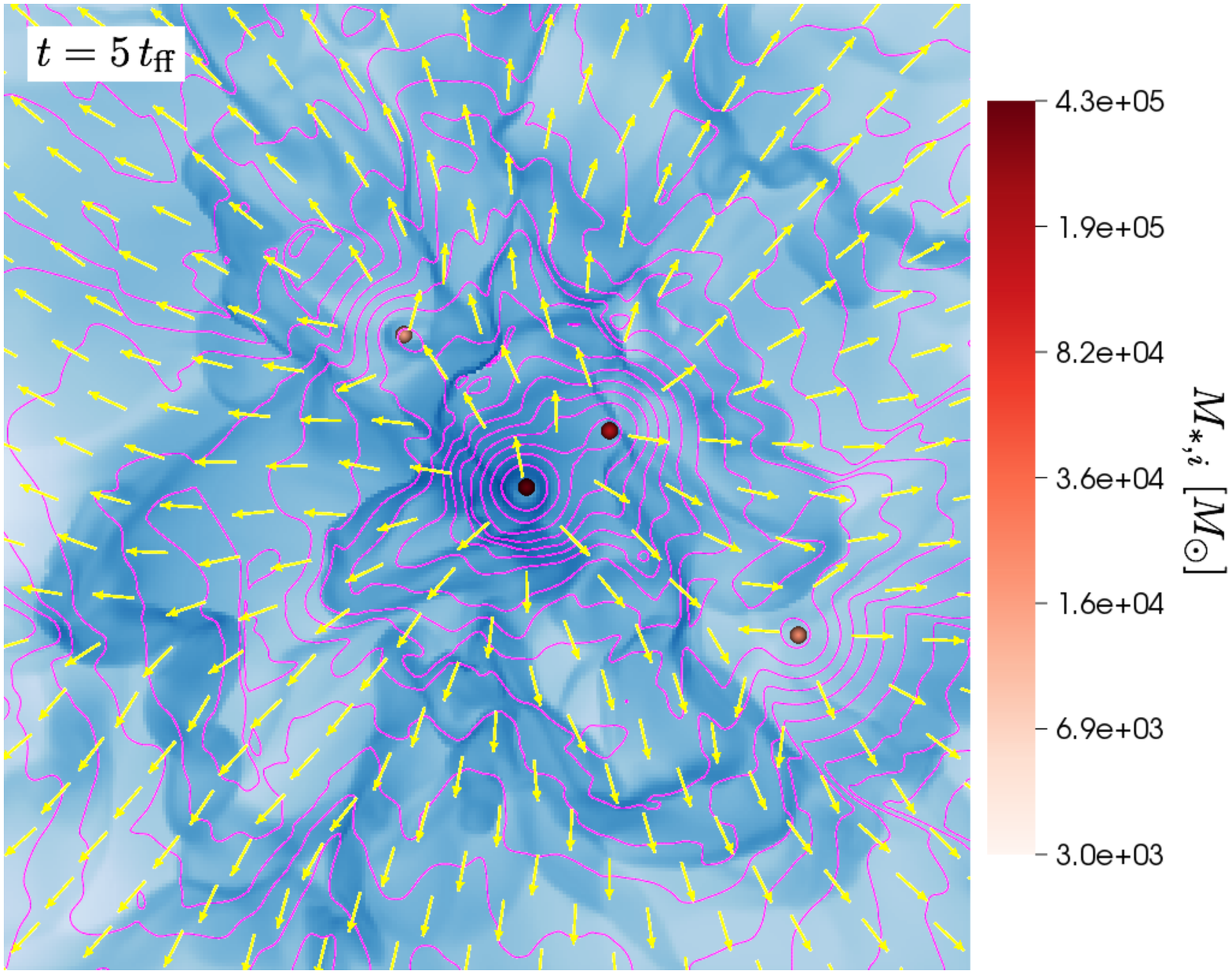}
\plotone{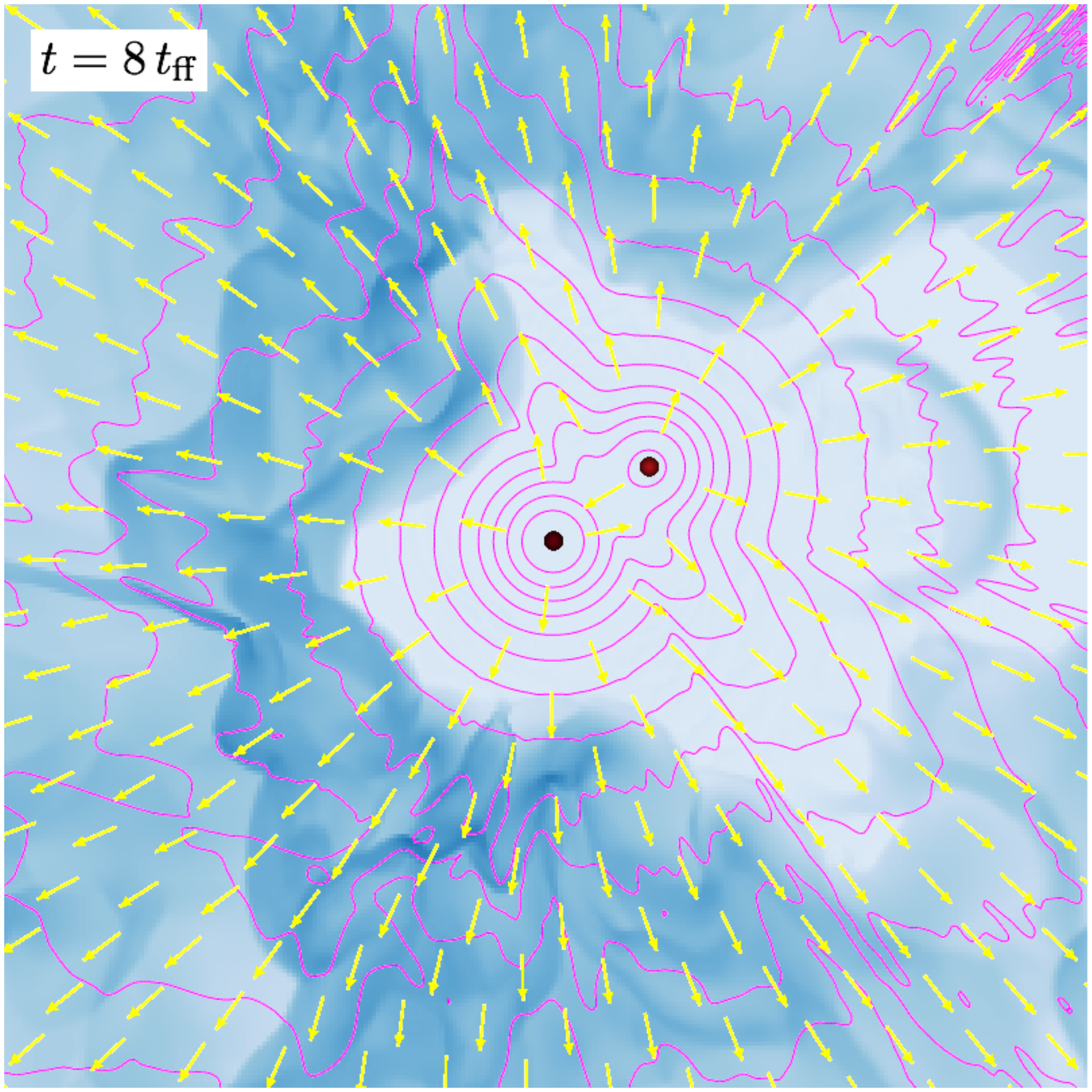}
\caption{Snapshots of slices through the fiducial model, K20, at
  successive stages in its evolution.  Each plot shows the gas number 
  density of hydrogen nuclei $n_H$
  (logarithmic color scale), radiation energy density (contours),
  direction of radiation flux (vectors), and star particles (spheres with
  logarithmic color scale indicating mass).  The slices are through
  the most massive star particle in each snapshot, at (top left)
  $t=1\tff$ and $z=2.8\pc$, (top right) $t=3\tff$ and $z=1.3\pc$,
  (bottom left) $t=5\tff$ and $z=-0.78\pc$, and (bottom right)
  $t=8\tff$ and $z=-2.3\pc$.  Star particles within 
  $\Delta z = \pm 2\pc$ of the slice are plotted.  
The 20 energy density contours are logarithmically spaced over the 
data range in each slice.  
The color scale for the gas density (blue, top) 
is in units of $\pcc$,
and the color scale for the star particle mass (red, bottom) 
is in units of $\Msun$.  The slice dimensions are $40\times 40\pc^2$.
\label{evolution:fig:rho_snapshots}
}
\end{figure*}

\begin{figure*}

\epsscale{0.57}
\plotone{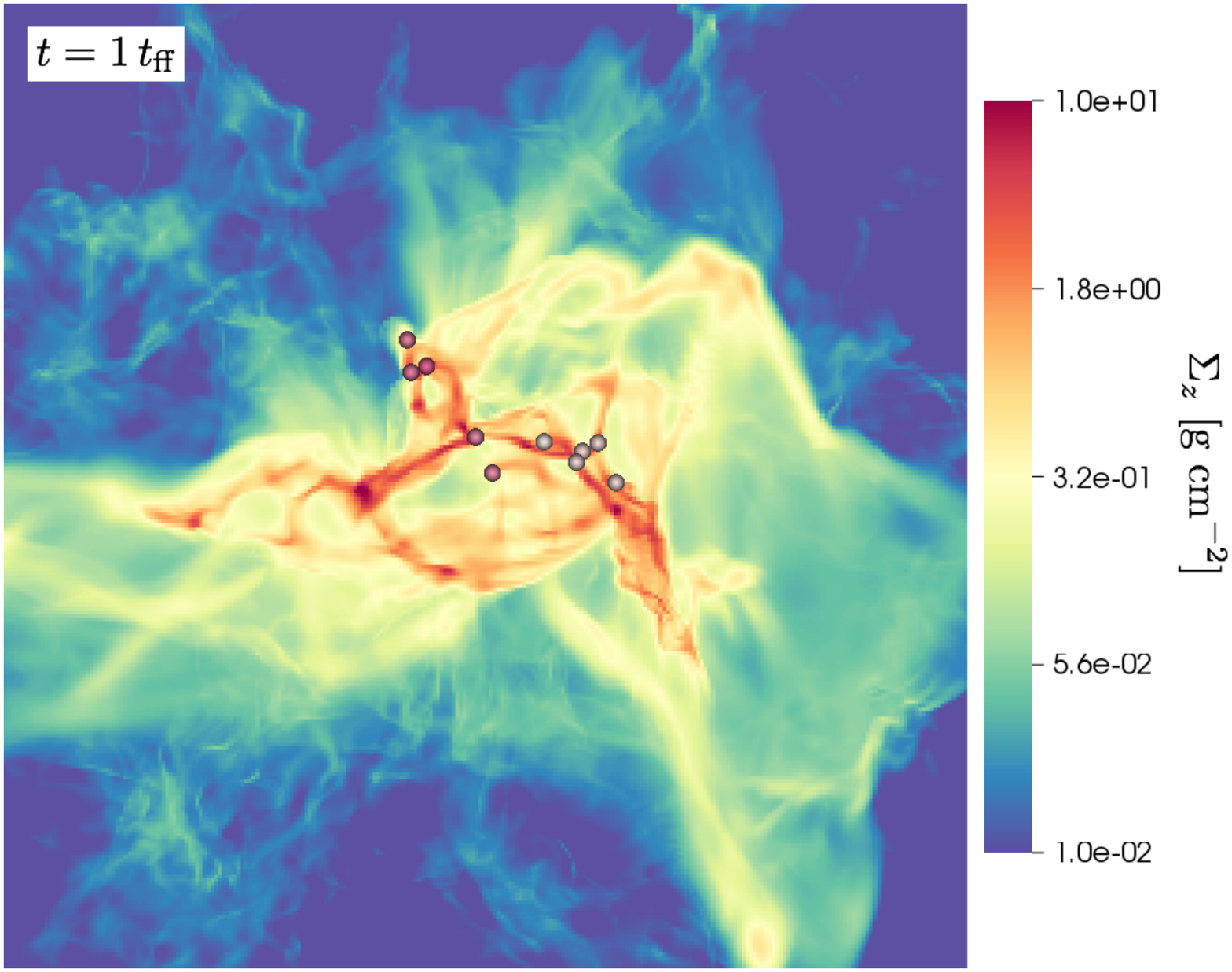}
\plotone{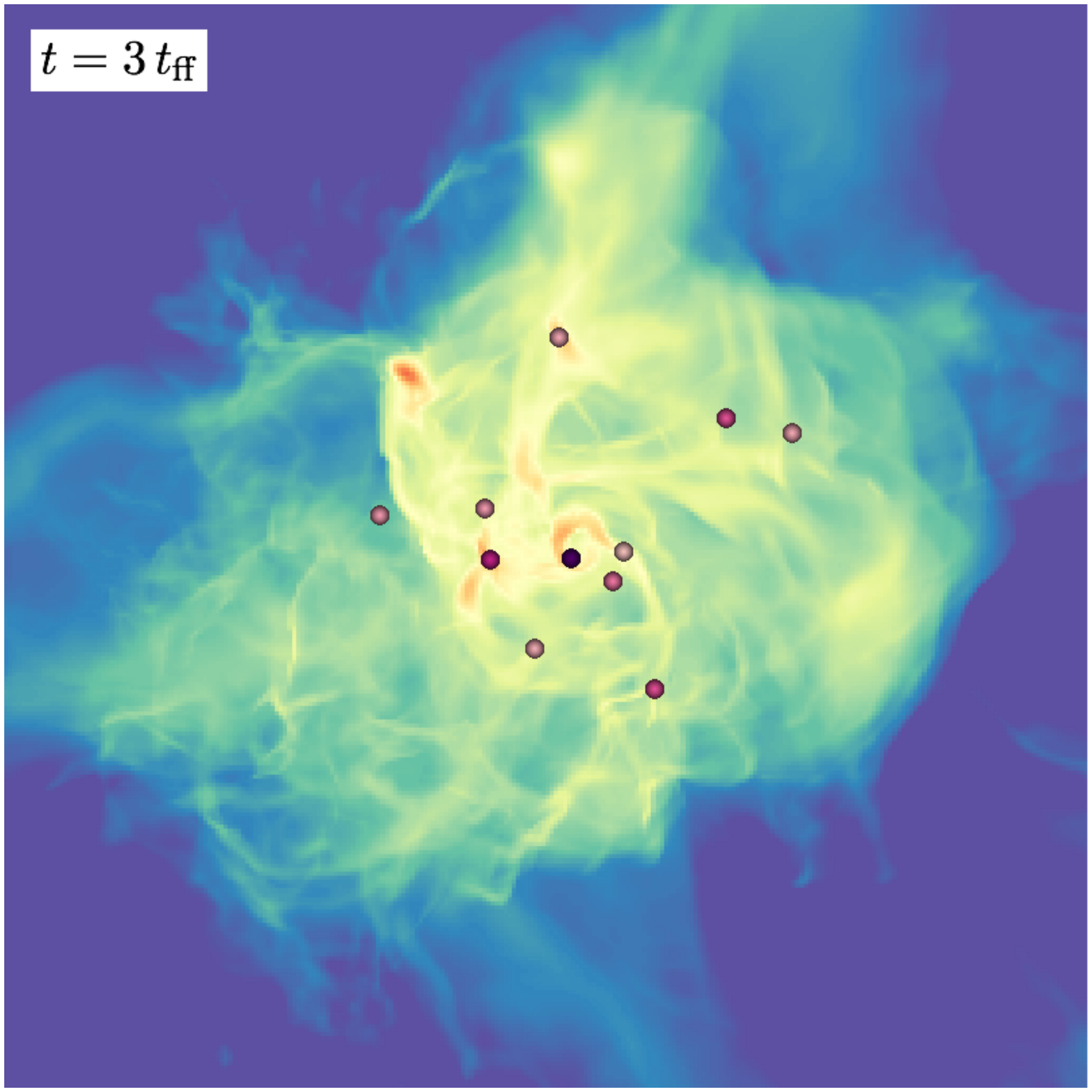}
\plotone{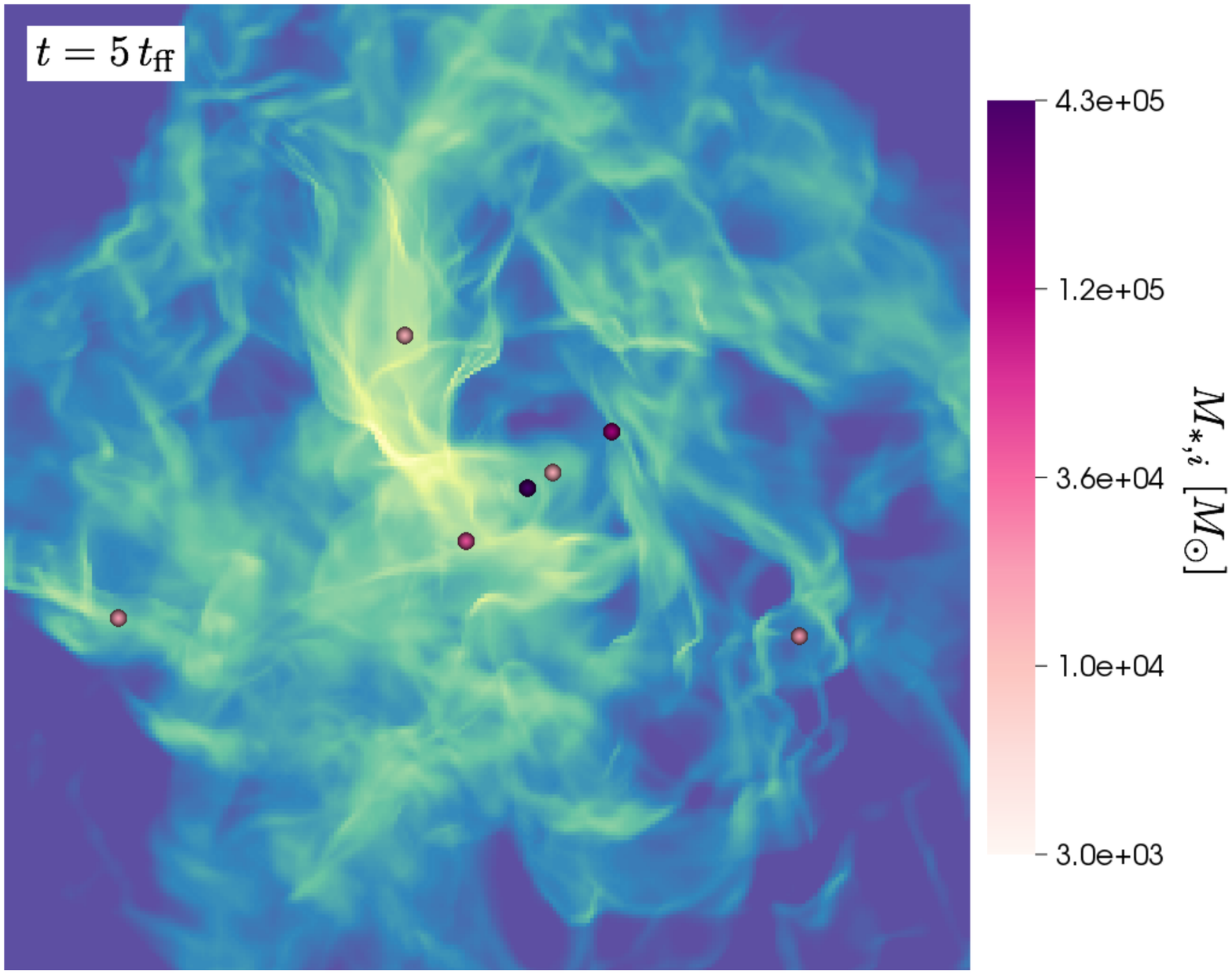}
\plotone{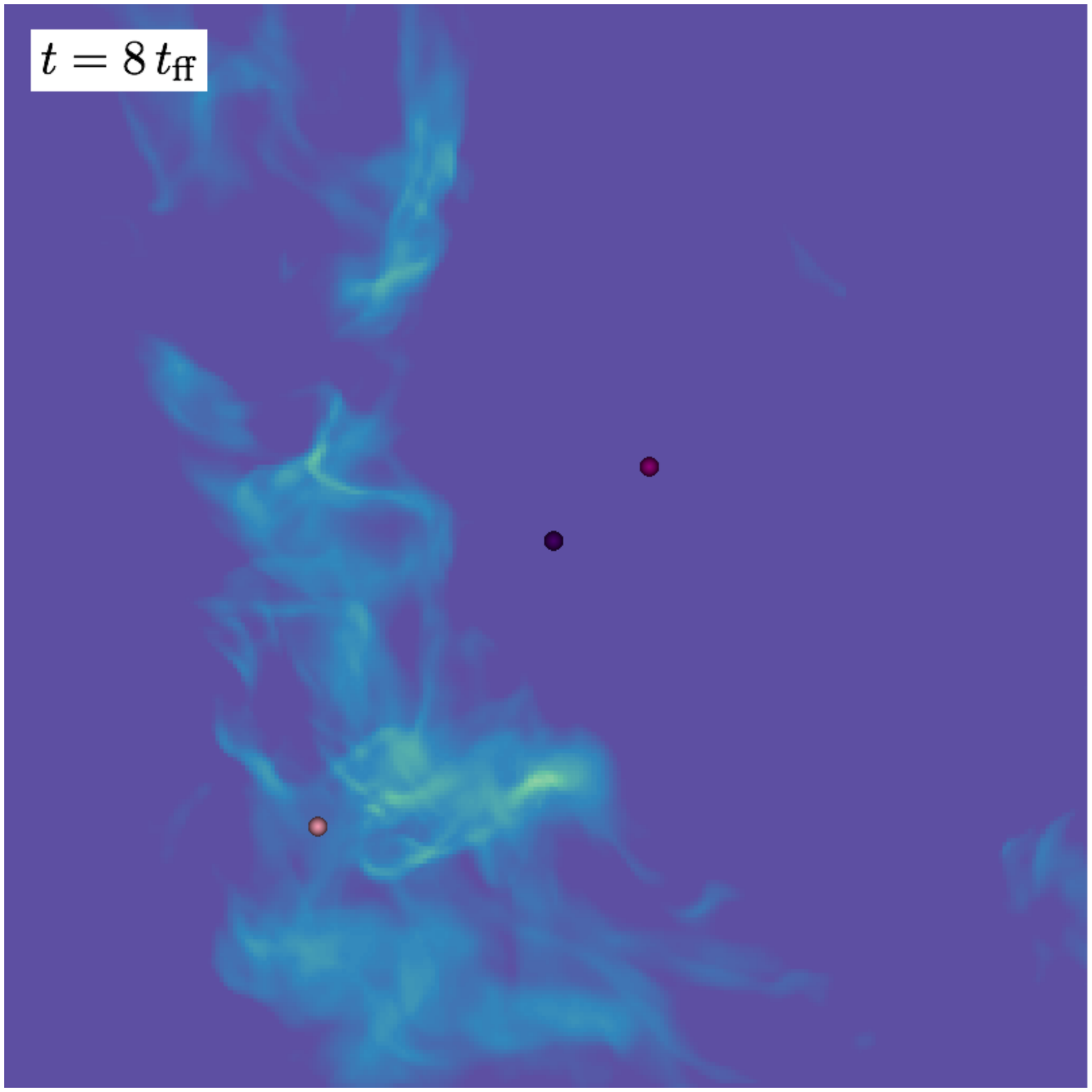}

\caption{ Same as Figure~\ref{evolution:fig:rho_snapshots}, but for 
snapshots of $\Sigma_z$, the gas surface density integrated in the $z$-direction 
(logarithmic color scale), along with a two-dimensional projection of all star 
particles (spheres with logarithmic color scale indicating particle mass $M_{i,*}$).  
The snapshots are at (top left) $t=1\tff$, (top right) $t=3\tff$, (bottom left) $t=5\tff$, 
and (bottom right) $t=8\tff$.  The color scale for the gas surface density (spectrum, top) 
is in units of g cm$^{-2}$, and the color scale for the star particle mass (violet, bottom) 
is in units of $\Msun$.  The image dimensions are $40\times 40\pc^2$.
\label{evolution:fig:Sigma_snapshots}
}
\end{figure*}

The action of the radiation force on the gas can be seen in snapshots of 
the structure within the cloud at different stages of its evolution, as 
shown in Figure~\ref{evolution:fig:rho_snapshots}.
In all of the snapshots, the gas density is highly clumpy and
filamentary due to the turbulence.  The radiation energy density is
highest immediately surrounding star particles, and has other local
variations due to interaction with the gas.  Overall, the radiation
energy density decreases outward, as does the gas density.  The
radiation flux points primarily radially away from near the center of
box, where the most luminous star particles are found, except in the
immediate vicinity of other star particles. Inspection of
Figure~\ref{evolution:fig:rho_snapshots} shows that unlike in
implementations of \ac{RHD} that adopt \ac{FLD} methods, the radiation flux
vectors (unscaled by magnitude) point independently from the gradient in the radiation energy
density.  Thus, the radiation force is not necessarily co-aligned with the 
negative gradient of the radiation energy density, as it is by construction
in \ac{FLD}-based implementations.  For comparison, Figure~\ref{evolution:fig:Sigma_snapshots}
shows snapshots of $\Sigma_z \equiv \int \rho dz$, the gas surface density in the
$z$-direction at the same times shown in Figure~\ref{evolution:fig:rho_snapshots}.

\begin{figure}
\epsscale{1.0}
\plotone{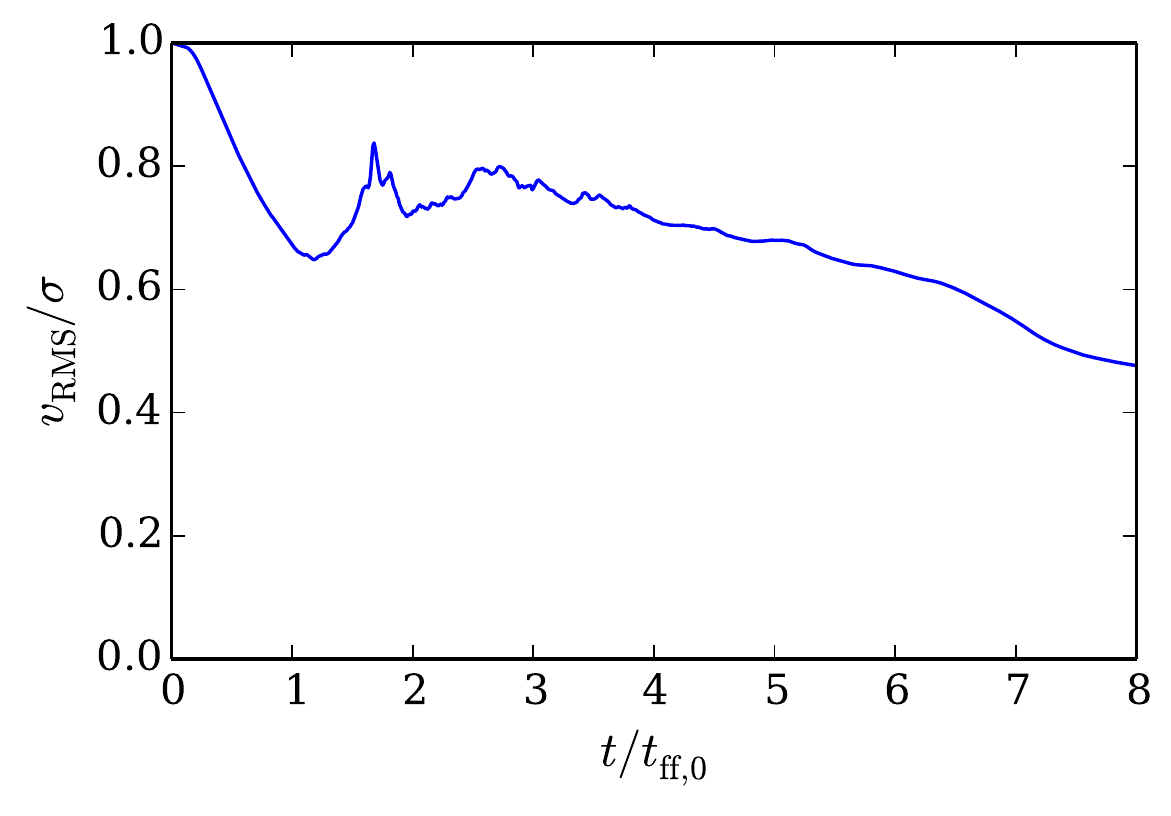}
\caption{
Time evolution of the \ac{RMS} gas velocity in units of the initial turbulent velocity, $\sigma$, for the fiducial model (K20). 
\label{evolution:fig:vturb}
}
\end{figure}

Over time, the mean density decreases as gas is expelled from the volume.  The 
gas remains turbulent and filamentary throughout the evolution, and 
expansion does not lead to the kind of simple shell-like structure that 
results for non-turbulent clouds (as seen, e.g., for the simple test 
shown in Fig. 23 and 26 of 
\citetalias{Skinner:2013}).
Figure~\ref{evolution:fig:vturb} shows the mass-weighted
\ac{RMS} gas velocity, 
$v_\mathrm{RMS} \equiv (2 E_\mathrm{kin}/M_\mathrm{gas})^{1/2}$, 
where $E_\mathrm{kin}$ and $M_\mathrm{gas}$ 
are respectively the total kinetic energy 
and mass of the gas remaining on the grid, over the evolution of the fiducial 
model.
This velocity declines over $\sim \tff$ as the initial turbulence dissipates, 
rises as radiation feedback becomes important, and then 
declines slowly for the remainder of the evolution.

\begin{figure}
\epsscale{1.0}
\plotone{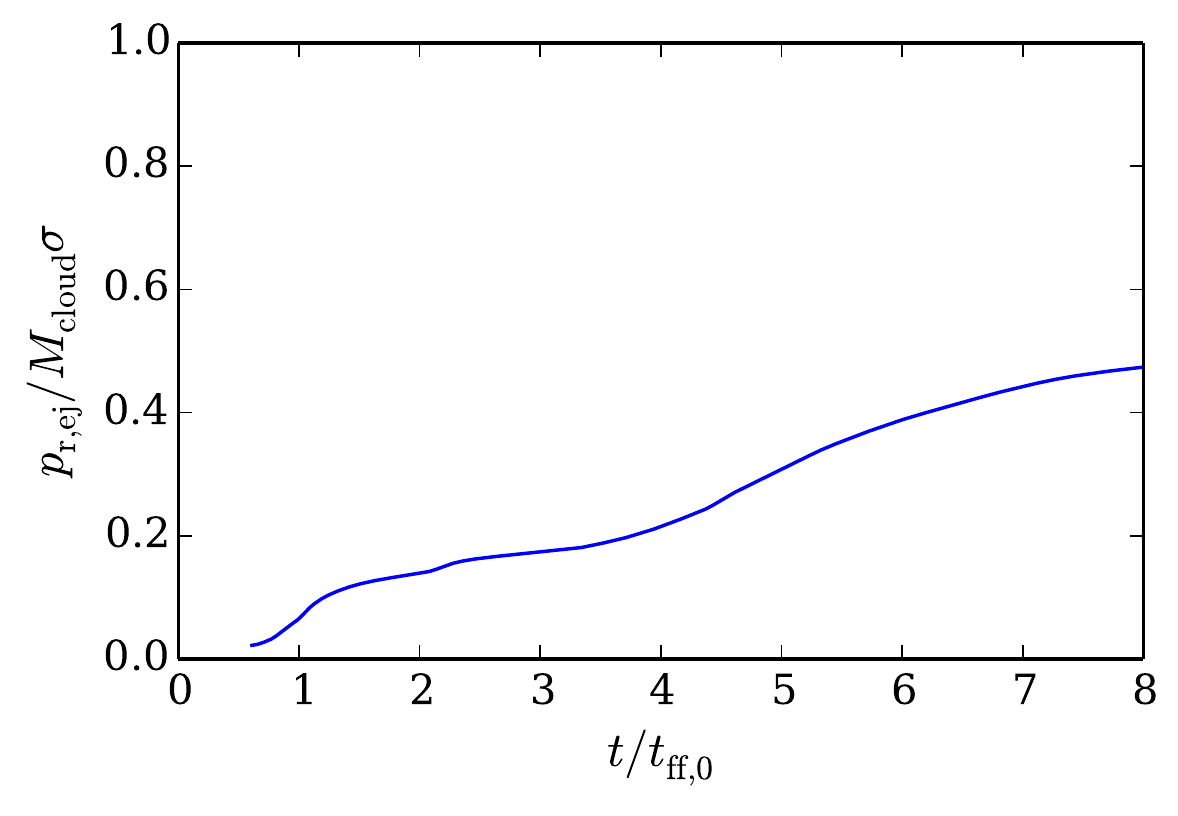}
\caption{Total ejected radial kinetic momentum in the fiducial model (K20).  
Momentum is measured in units of $p_\mathrm{turb, init}\equiv  \Mcloud \sigma$.
\label{evolution:fig:momejecthist}
}
\end{figure}

As discussed in Section~\ref{intro}, in addition to limiting the star 
formation efficiency in a given cloud, radiation feedback from clusters 
can also inject momentum into the surrounding \ac{ISM}.  Figure~\ref{evolution:fig:momejecthist} 
shows the total ejected kinetic momentum as a 
function of time for the fiducial model, which is 
$\sim 0.47 \Mcloud \sigma$ by the end of the run.
For the purposes of 
driving turbulence in the \ac{ISM} and therefore regulating future star formation
\citep[][hereafter \citetalias{Ostriker:2011}]{Ostriker:2011},
the most important quantity is the ratio of total kinetic momentum ejected to  
the total mass in stars formed, $p_*/M_*$.  For the fiducial model, 
$p_*/M_* = 17$ km s$^{-1}$; the ratio $p_*/(M_* \sigma)$ is equal to 
$0.78$.
This is consistent with the general expectation discussed in 
\citetalias{Ostriker:2011} that $p_*/M_*$ from \ac{IR} radiation feedback will be 
of order the velocity dispersion in the central star cluster that forms.

In Sections \ref{varyingopacity} and \ref{parameterstudy}, we discuss further 
how evolution and outcomes of the simulations vary for different model 
parameters.

\subsection{Radiation Structure in a Fiducial Model} \label{radstructure}

In this section, we again employ the fiducial model presented in Section~\ref{evolution} 
and now use it to explore the cloud's internal radiation 
structure as well as the competition between radiation and gravity.  
We wish to investigate the radiative structure after
stars have formed and significant feedback is underway, but before the
system reaches a limiting state, either by accreting the
bulk of the mass onto star clusters or ejecting it from the simulation
domain.  Thus, we have chosen to take snapshots of the variables at
time $t=3\tff$.  

\subsubsection{Radiation and Gravity in a Spherical System}
\label{sec:spherical}
Throughout this section, we compare our results to those of an isotropic 
dusty gas sphere surrounding a single, centrally-located massive and luminous 
cluster 
\citepalias[cf. Appendix A of][]{Ostriker:2011}
\citep[see also][]{Elmegreen:1983,Scoville:2001,Krumholz:2009a,Murray:2010}.  
For this system, in 
steady state the radiation flux is 
$F_{r,*}=L_*/(4 \pi r^2)=\Psi M_*/(4 \pi r^2)$, 
and the gravitational field from the cluster is $g_*=GM_*/r^2$.
Within any radial shell of mass $\delta M$, the ratio of the total radial 
radiation force $\kappa F_{r,*} \delta M /c$ 
to the total radial 
gravitational force $g_* \delta M$ from the cluster is therefore
\begin{eqnarray}
f_\mathrm{Edd,*} &\equiv& \frac{\kappa F_{r,*} }{cg_*} = \frac{\kappa\Psi }{4 \pi c G} \\
&=& 0.68 
\left(\frac{\kappa}{10 \mbox{ cm}^2 \mbox{ g}^{-1}}\right) 
\left(\frac{\Psi  }{1700 \mbox{ erg s}^{-1} \mbox{ g}^{-1}}\right). \notag
\label{eq:fEdd*}
\end{eqnarray}

Including the self-gravity from the gaseous sphere, the ratio of the outward 
force to the inward force on a radial shell at $r$ is 
\begin{equation}
f_\mathrm{Edd, sph}(r)=\frac{f_\mathrm{Edd,*}}{1+M_{\rm gas}(r)/M_*}, 
\label{eq:fEddrsph}
\end{equation}
where $M_{\rm gas}(r)$ is the mass of the 
gas sphere interior to $r$.  
For any gas mass $M_{\rm gas}(r)$ less than $M_\mathrm{max}=M_*(f_\mathrm{Edd,*}-1)$, 
the outward force on a shell at $r$ exceeds the inward force.
For the fiducial model, with $\kappa =20 \cmsg$, 
$f_\mathrm{Edd,*}=1.4$.
As a consequence, if the cloud were able to remain 
spherically symmetric while 
forming stars at its center, gas out to a radius 
for which $M_{\rm gas}(r)/M_*=0.4$ would 
have the local radiation force  exceeding gravity.  

If the net efficiency of star formation compared to the initial 
cloud mass is $\varepsilon_* = M_*/M_{\rm cloud}$, then the total remaining 
gas has $M_{\rm gas}(r_{\rm max})/M_*=(1-\varepsilon_*)/\varepsilon_*$. 
Substitution in Equation~\eqref{eq:fEddrsph} shows that 
the local Eddington ratio $f_\mathrm{Edd, sph}(r)$ would exceed unity 
at all radii in a spherical cloud when a fraction 
\begin{eqnarray}
\varepsilon_\mathrm{*,sph} &\equiv& f_\mathrm{Edd,*}^{-1} \\
&=& 1.5 \left(\frac{\kappa}{10 \mbox{ cm}^2 \mbox{ g}^{-1}}\right)^{-1} 
\left(\frac{\Psi  }{1700 \mbox{ erg s}^{-1} \mbox{ g}^{-1}}\right)^{-1}. \notag
\label{eq:eps*min}
\end{eqnarray}
of the original cloud has been 
converted to stars.\footnote{If all the remaining 
gas is collected in a single thin shell, 
as for the idealized system described in 
Appendix A of \citetalias{Ostriker:2011}, self-gravity is diluted so that 
the ratio of radiation to total gravitational forces 
is $f_\mathrm{Edd} = f_\mathrm{Edd,*}/(1+ 0.5M_\mathrm{shell}/M_*)$.   
Shell expulsion would commence when a fraction 
$\varepsilon_\mathrm{*,sh}=(2 f_\mathrm{Edd,*}-1)^{-1}$
of the gas is converted to stars; for the case $\kappa=20 \cmsg$ this would 
correspond to 56\%.} 
For an idealized spherical system with $\kappa=20 \cmsg$, 
74\% of the original cloud would have to be converted to stars for the 
remainder to be forced outward by radiation.  Of course, gas pressure 
forces can become important, so that gas originally 
at small radii where 
$f_\mathrm{Edd,sph}(r)\gg 1$ 
can sweep up and expel fluid elements that were originally 
at large radii; this reduces the actual \ac{SFE} 
$\varepsilon_\mathrm{*,final}$ below $\varepsilon_\mathrm{*,sph}$.  

The above discussion suggests that unless $\kappa$ is relatively large, 
radiation forces will not be able to expel a substantial portion of 
a cloud's mass.  In fact, larger values of $\kappa$ than would 
be realistic for \ac{IR} radiation at Milky Way dust abundances
may be required for any gas expulsion,
even in the limit of highly efficient star formation.   For the 
idealized spherical case, the (stellar-plus-gas) gravitational 
force exceeds the radiation force 
at the outer edge of the cloud even for $\varepsilon_*\rightarrow 1$ if 
$f_\mathrm{Edd,*} < 1$. 
Moreover, for gas very near any star cluster, the dominant forces 
are radiation and gravity, so only if $f_\mathrm{Edd,*} > 1$ can continued 
accretion be halted.  The condition $f_\mathrm{Edd,*} >1$ translates to 
$\kappa > \kappa_\mathrm{crit}$ where 
\begin{eqnarray}
\kappa_\mathrm{crit} &\equiv& \frac{4 \pi c G }{\Psi} \\
&=& 15 \cmsg \left(\frac{\Psi  }{1700 \mbox{ erg s}^{-1} \mbox{ g}^{-1}}\right)^{-1}. \notag
\label{eq:kappacrit} 
\end{eqnarray}
Although the true structure is non-spherical, this explains why we 
have selected the $\kappa=20 \cmsg$ model for detailed presentation of 
internal radiation structure: for lower-$\kappa$ models, radiation forces 
are not expected to be strong enough to prevent continuing accretion and 
expel substantial amounts of mass.

As we shall next discuss, the simple spherical analysis provides
useful guidance regarding the relative importance of radiation and
gravity, but the solution of the full \ac{RHD} equations for a turbulent
medium is more complex.  In particular, neither the radiation field
nor the gravitational field is spherically symmetric, and the gas 
density on which these fields act is highly nonuniform.  The radiation 
and gravitational forces---locally, averaged over spherical shells, 
or integrated over the whole cloud---therefore can differ substantially 
from estimates based on simple spherical models.

\subsubsection{Analysis of Radiation and Gravity in the Turbulent Cloud}
\label{sec:turbanalysis}
\begin{figure}
  \centering
  \epsscale{1}
  \plotone{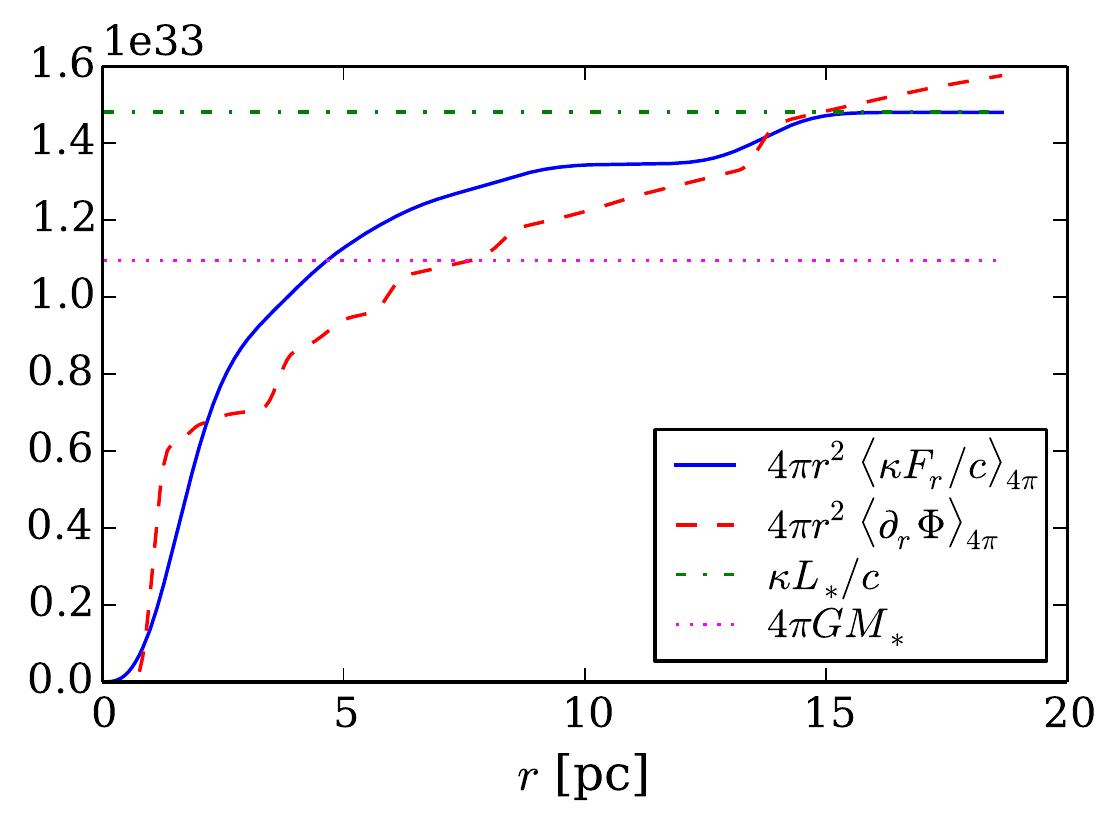}
\caption{Angle-averaged radial radiation force per unit mass, 
$\langle  \kappa F_r/c\rangle$ (solid), and the 
angle-averaged total gravitational force per
  unit mass, $\langle\partial_r \Phi \rangle$ (dash-dotted), 
both compensated by $4\pi r^2$, 
for the fiducial run (model K20) at time $t=3\tff$.  For
comparison, we plot $\kappa L_*/c$ (dashed) and $4\pi G M_*$  (dotted),
the radiation and gravitational forces that would be produced by a 
point-source cluster with the same total stellar mass as is present at 
$t=3\tff$ in the simulation.  
The  measured radiation force reaches 90\% of the point-source 
value at $\sim$10 pc from the center of mass, comparable to the
  radius of the original cloud.  
At $r>15 \pc$, $4\pi r^2\langle F_r \rangle$ asymptotes to $L_*$.
The measured gravitational force exceeds the stellar point-source value 
at large radii because a significant portion of the cloud's gas has neither 
been accreted nor expelled at this time.} \label{radstructure:fig:Fr}
\end{figure}

To quantify the internal cloud structure in a simple way, we compute
angle-averaged radial profiles, denoted by $\langle \cdot
\rangle_{4\pi}$.  To do this, we first compute the center of mass of the stars, 
$\mathbf{r}_\mathrm{CM}$, then linearly interpolate the Cartesian grid data onto an 
$N_r \times N_\phi \times N_\theta$ spherical grid with $N_r = N_\theta = N$ and $N_\phi
= 2N$ zones centered at $\mathbf{r}_\mathrm{CM}$, with $r \in [0,2 \Rcloud - r_\mathrm{CM}]$.\footnote{Note that $r_\mathrm{CM} \ll 2\Rcloud$ typically, so that $2\Rcloud-r_\mathrm{CM} \approx L_\mathrm{box}/2$.}  Finally, we average the interpolated data on this grid over all solid angles to obtain a
radial profile.  

Figure~\ref{radstructure:fig:Fr} compares, for $t=3\tff$, the
profiles of the two dominant competing effects in the problem: the
outward radial radiation force and the inward gravitational
force.  These two forces both scale with the mass (or, per unit volume, 
the density), which we omit
here, and both fall off as inverse square laws away from the center of
mass of the system; thus, we compensate by a factor of $4\pi r^2$ in each
profile.  For comparison, we also plot the 
radiation and stellar gravity forces/mass that would apply 
in the limit of a 
single point mass cluster at $\mathbf{r}_\mathrm{CM}$, which appear
here as constants ($\kappa L_*/c$ and $4\pi G M_*$, respectively) 
in our radially-compensated plot.  Far from the
center of mass of the system, the profile of the radiation force
approaches that of a point source, reaching 90\% of this value by
$r\sim 10$ pc, a distance comparable to the original cloud radius.
However, the gravitational force at this time still contains a
significant contribution from the remaining gas in the system 
(at this time, the total mass in gas is $3.4\times 10^5\Msun$, 
while the mass in stars is $6.6\times 10^5 \Msun$), so it
is slightly higher than the point-source cluster limit at large distance.
The mean specific radiation force exceeds the mean 
specific gravitational force for
most of the profile, indicating that in most zones 
the gas feels a net outward force at this point.  
Figure~\ref{evolution:fig:masshist} shows that the accretion rate 
onto the sink particles slows between $t \sim 2-3 \tff$ and essentially 
stops after $t \sim 3\tff$. We note that in the simple spherical model 
described above, Equation~\ref{eq:eps*min} suggests that 
radiation forces would exceed gravity forces everywhere only 
when $\varepsilon_* \rightarrow 0.7$ for $\kappa = 20 \cmsg$.  In fact, 
$\varepsilon_*$ is only $\sim 0.57$ at $t \sim 3\tff$, somewhat lower 
than this value.

\begin{figure}
\centering
\epsscale{1}
\plotone{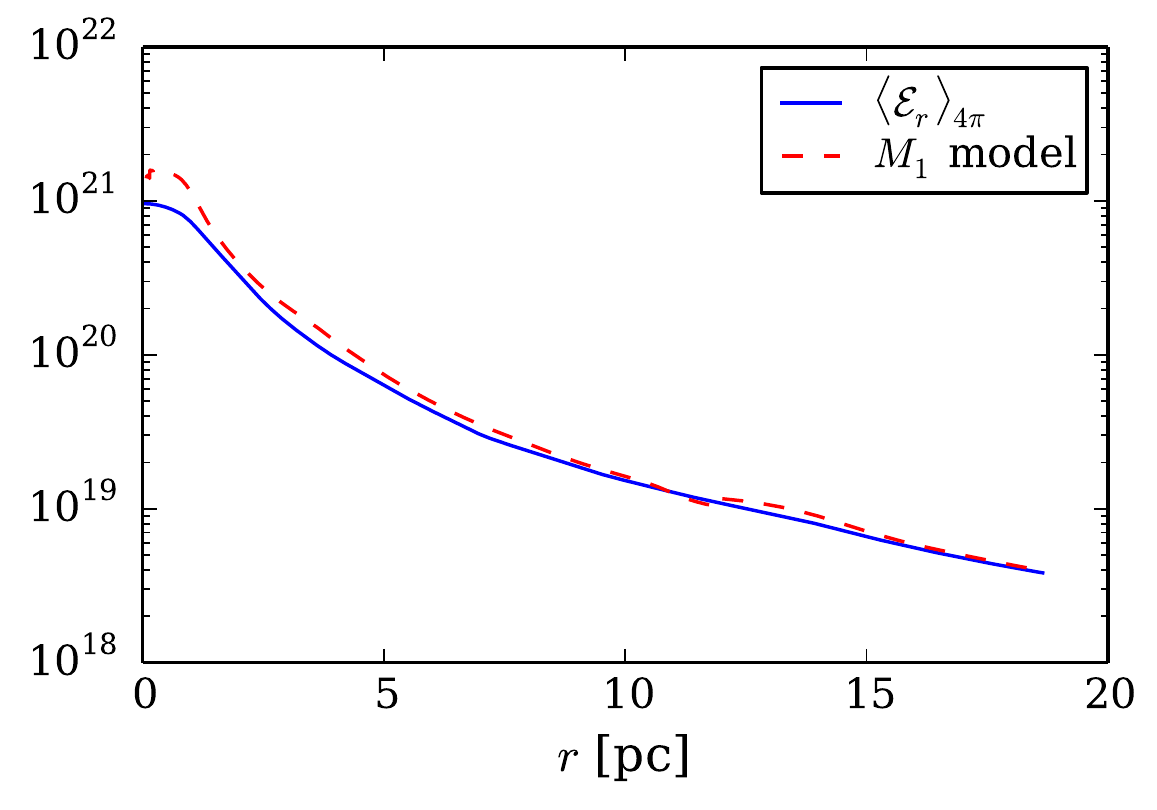}
\plotone{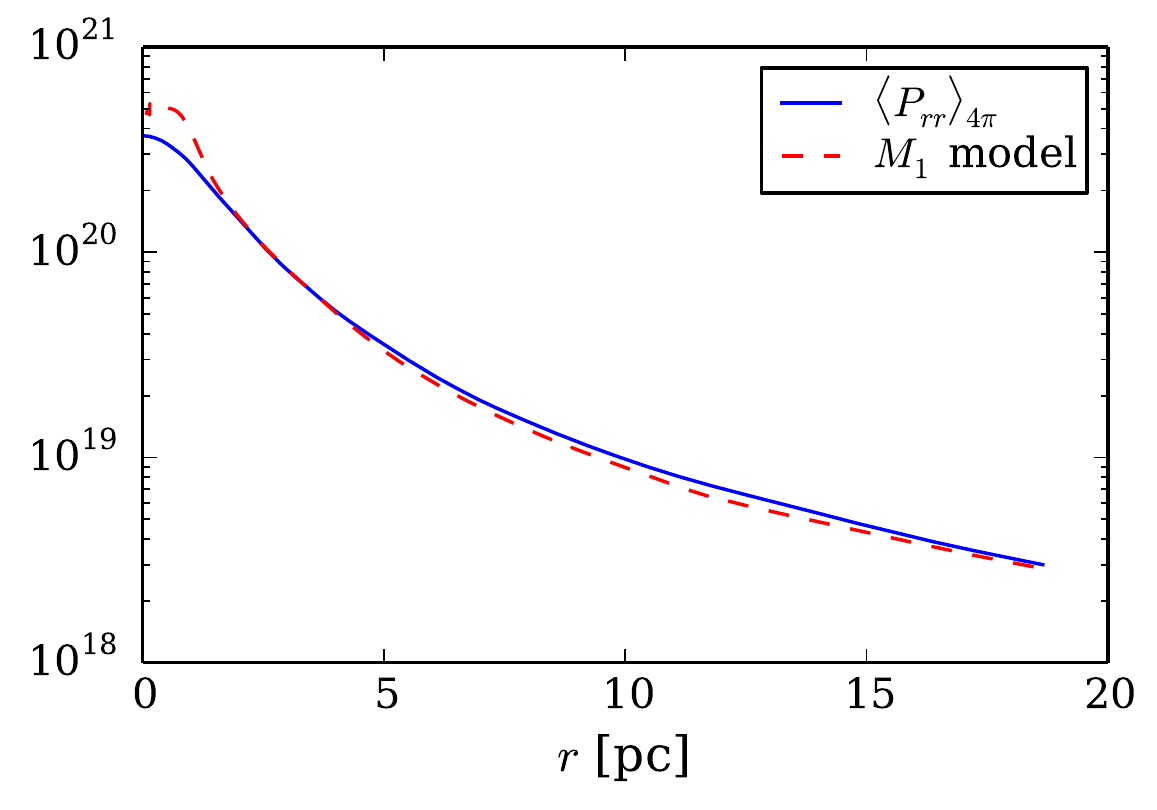}
\caption{Angle-averaged profiles of $\mathcal{E}$ (top) and $P_{rr}$
  (bottom) for model K20 at time $t=3\tff$ (solid curves).
  For comparison, we also plot the semi-analytic solutions (dashed) 
derived from the $M_1$ closure for a spherically symmetric system with 
density and radiation flux profiles given by angle-averages of 
the simulation data at this time.
  The measured data and semi-analytic solutions are comparable, 
showing that the radiation is in a quasi-steady state at this time.}
	\label{radstructure:fig:Er_Prr}
\end{figure}

In Figure~\ref{radstructure:fig:Er_Prr}, we plot the radiation energy
density, $\langle \mathcal{E} \rangle_{4\pi}$, and the radial component of
the radiation pressure tensor, $\langle P_{rr} \rangle_{4\pi}$, at
time $t=3\tff$.  For comparison, we also plot the
semi-analytic solutions for $\mathcal{E}$ and $P_{rr}$, respectively,
derived from the $M_1$ closure for a spherically symmetric system 
in steady state 
\citepalias[see Equations~110 and~111 of][]{Skinner:2013}.  
This requires the
solution of an \ac{ODE}\footnote{This \ac{ODE}
  has a regular singularity at $f=2\sqrt{3}/5\sim 0.7$ that 
poses numerical difficulties in its solution, resulting in 
a slight shift in the comparison profile near 
$\sim 12\pc$ for model K20 in Figure~\ref{radstructure:fig:Er_Prr}.} 
for the reduced flux, $f\equiv F/(c{\mathcal E})$, as
a function of the radial profiles $\langle \rho \rangle_{4\pi}$ and
$\langle F_r \rangle_{4\pi}$.  The angle-averaged profiles and
semi-analytic models are comparable for both $\mathcal{E}$ and
$P_{rr}$, indicating that the radiation field is, in an averaged sense,
close to quasi-steady state for the density distribution at this time.

\begin{figure}
  \centering
  \epsscale{1}
  \plotone{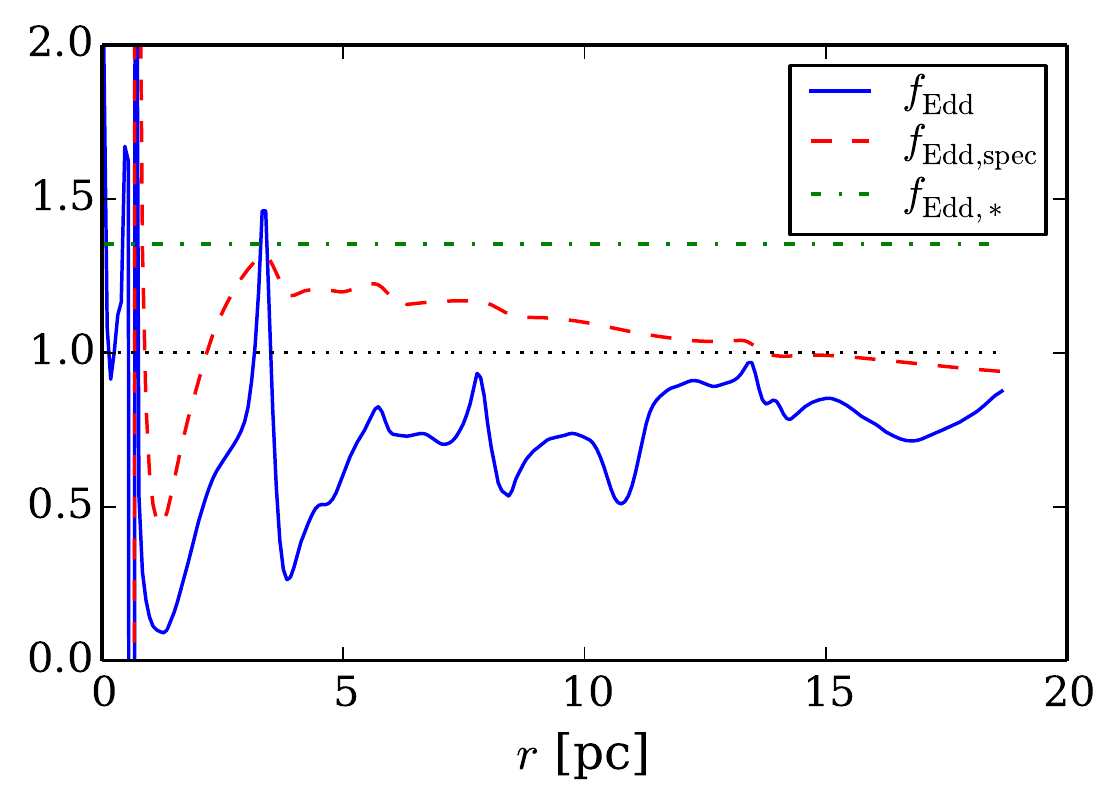}
\caption{Radial profile of the Eddington factor, $f_\mathrm{Edd}$
  (solid), averaged over local spherical shells (see
  Equation~\ref{radstructure:f_Edd_def}) for model K20 at $t=3\tff$.  The local Eddington ratio is 
  less than 1 (dotted) over most of the domain,
  indicating that the majority of the gas in the cloud feels a net
  inward force.  The ratio of angle-averaged \textit{specific}
  forces, $f_\mathrm{Edd, spec}\equiv \langle \kappa F_r/c\rangle_{4\pi}/\langle 
\partial_r \Phi \rangle_{4\pi}$ is also shown (dash-dotted) for comparison; 
this exceeds unity for most radii.
The dashed line shows $f_\mathrm{Edd,*}$ 
(see Equation~\ref{eq:fEdd*}), the Eddington ratio for a simple spherical
model with negligible gas self-gravity.
}
\label{radstructure:fig:f_Edd1}
\end{figure}

Figure~\ref{radstructure:fig:f_Edd1} shows (again for model K20 at $t=3\tff$) 
the radial profile of the Eddington factor, defined by
\begin{equation}
	f_\mathrm{Edd}(r) \equiv \frac{\langle \rho \kappa F_r/c \rangle_{4\pi}}{\langle \rho \partial_r \Phi \rangle_{4\pi}},  
\label{radstructure:f_Edd_def}
\end{equation}
where the angular averages in Equation~\eqref{radstructure:f_Edd_def} are
over local spherical shells.  The quantity $f_\mathrm{Edd}(r)$ measures
the ratio of the total radial radiation force acting on the gas
to the total gravitational force acting on that gas, 
within a shell at a given radius.  
For comparison, we show the ratio of angle-averaged \textit{specific} forces 
$f_\mathrm{Edd, spec}(r)$; i.e., 
with the density omitted from Equation~\eqref{radstructure:f_Edd_def}.
As previously seen in Figure~\ref{radstructure:fig:Fr}, where the 
angle-averaged specific radiation force 
exceeds that of gravity at most radii, 
$f_\mathrm{Edd, spec}(r)>1$ almost everywhere. In contrast, the ratio 
of angle-averaged total forces 
$f_\mathrm{Edd}(r)$ is below 1 
for most of the profile. For reference, we also show the quantity 
$f_\mathrm{Edd,*}$ defined in Equation~\eqref{eq:fEdd*}; this is 
well above both $f_\mathrm{Edd,spec}(r)$ and $f_\mathrm{Edd}(r)$ 
because self-gravity adds appreciably to the gravity of 
the stars at this time.  The difference between $f_\mathrm{Edd}$ and
$f_\mathrm{Edd,spec}$ suggests that the
radiation flux and gas density must be anti-correlated to
some extent, and highlights the importance of conducting fully 
three-dimensional simulations.  

\begin{figure}
  \centering
  \epsscale{1}
  \plotone{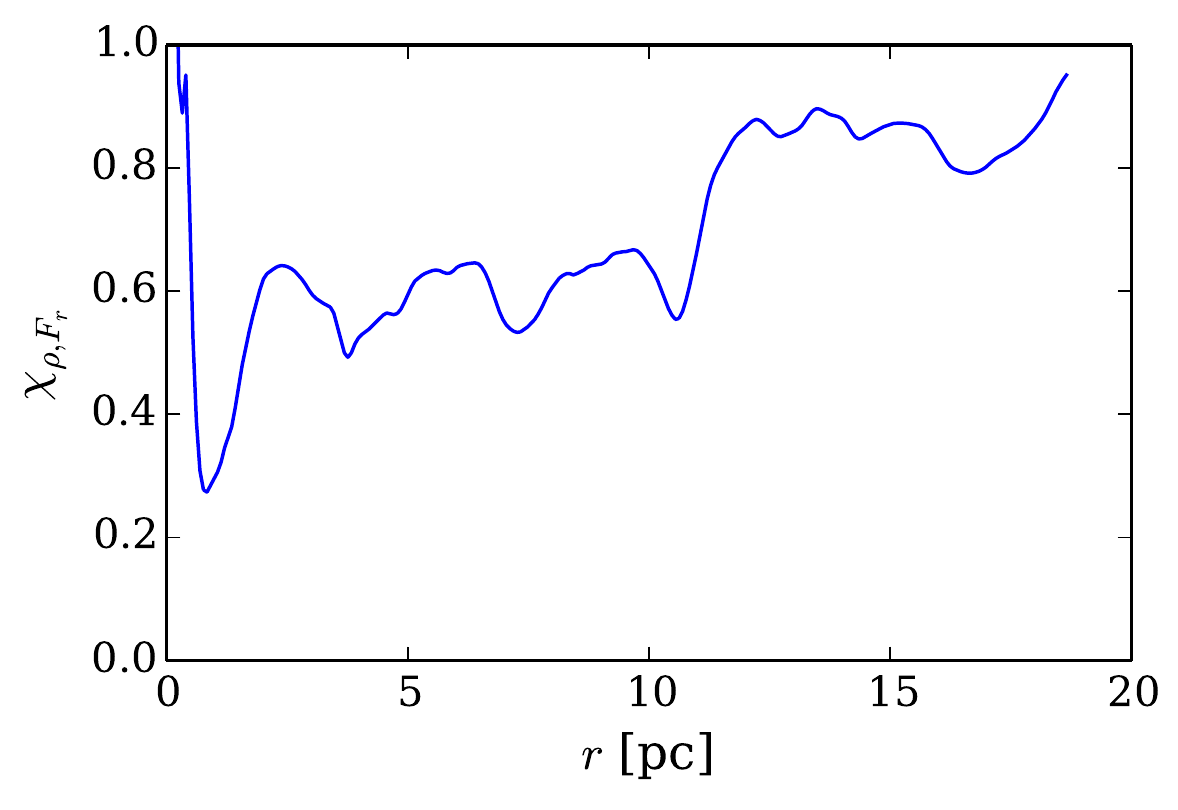}
\caption{Radial profile of $\chi_{\rho,F_r}$, the correlation fraction
  between $\rho$ and $F_r$ (see
  Equation~\ref{radstructure:d_Fr_corr_def}) for model K20 at $t=3\tff$.  
Values of $\chi_{\rho,F_r}$ less than unity show that
  $\rho$ and $F_r$ are somewhat anti-correlated over most of the
  domain.}
	\label{radstructure:fig:d_Fr_corr}
\end{figure}

In Figure~\ref{radstructure:fig:d_Fr_corr}, we plot the profile of $\chi_{\rho,F_r}$, the correlation fraction of $\rho$ and $F_r$, defined as
\begin{equation}
	\chi_{\rho,F_r} \equiv \frac{\langle \rho F_r \rangle_{4\pi}}{\langle \rho \rangle_{4\pi} \langle F_r \rangle_{4\pi}}.  
\label{radstructure:d_Fr_corr_def}
\end{equation}
When $\chi_{\rho,F_r}=1$, the gas density and radiation flux in a given 
radial shell  
have uncorrelated fluctuations, whereas $\chi_{\rho,F_r}<1$ implies that 
the fluctuations about the mean are anti-correlated.
Figure~\ref{radstructure:fig:d_Fr_corr}
indicates that there is some degree of anti-correlation between $\rho$
and $F_r$, which explains why $f_\mathrm{Edd}$ lies below
$f_\mathrm{Edd,spec}$.

\begin{figure}
  \centering
  \epsscale{1}
  \plotone{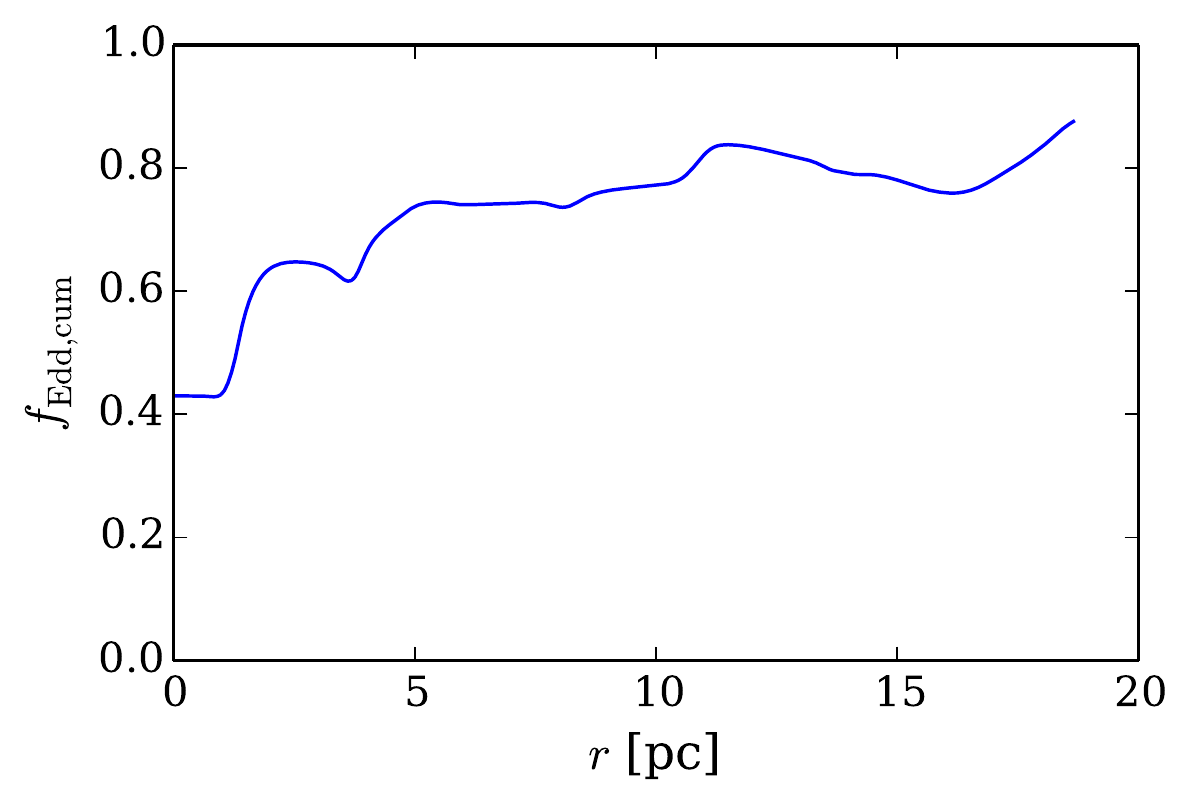}
\caption{Radial profile of the cumulative Eddington ratio,
  $f_\mathrm{Edd,cum}(r)$ (see
  Equation~\ref{radstructure:f_Edd_cum_def}), measured inward from
  the largest radius in the interpolation grid 
for model K20 at $t=3\tff$.
Since $f_\mathrm{Edd,cum}(r)<1$ over the entire domain, 
the gas feels a net inward force.}
	\label{radstructure:fig:f_Edd_cum}
\end{figure}

Figure~\ref{radstructure:fig:f_Edd_cum} shows the radial profile of $f_\mathrm{Edd,cum}$, the cumulative integral of $f_\mathrm{Edd}$ defined as
\begin{equation}
f_\mathrm{Edd,cum}(r) \equiv \frac{\int^{r_\mathrm{max}}_r \langle
  \rho \kappa F_r/c \rangle_{4\pi} \,4\pi r^2
  dr}{\int^{r_\mathrm{max}}_r \langle \rho \partial_r \Phi
  \rangle_{4\pi} \,4\pi r^2 dr} 
\label{radstructure:f_Edd_cum_def}.
\end{equation}
Note that in
Equation~\eqref{radstructure:f_Edd_cum_def}, the volume integrals are
cumulative, starting from the maximum radius of the spherical
interpolation grid, instead of only over local shells as in
Equation~\eqref{radstructure:f_Edd_def}.  Once again, this shows that
the net force that the majority of the gas feels is inward, since
$f_\mathrm{Edd,cum} < 1$ throughout the domain.  We note, however, 
that in spite of the \textit{net} inward force, and 
the \textit{overall} anti-correlation in the fluctuations of density and 
radiation flux, individual fluid elements can experience a 
radiation force (potentially aided by a pressure force) 
that exceeds the gravitational force.
This is why, as shown in Figures \ref{evolution:fig:masshist}
and \ref{evolution:fig:momejecthist}, mass and momentum can be ejected by 
the action of radiation forces.

\begin{figure}
  \centering
  \epsscale{1}
  \plotone{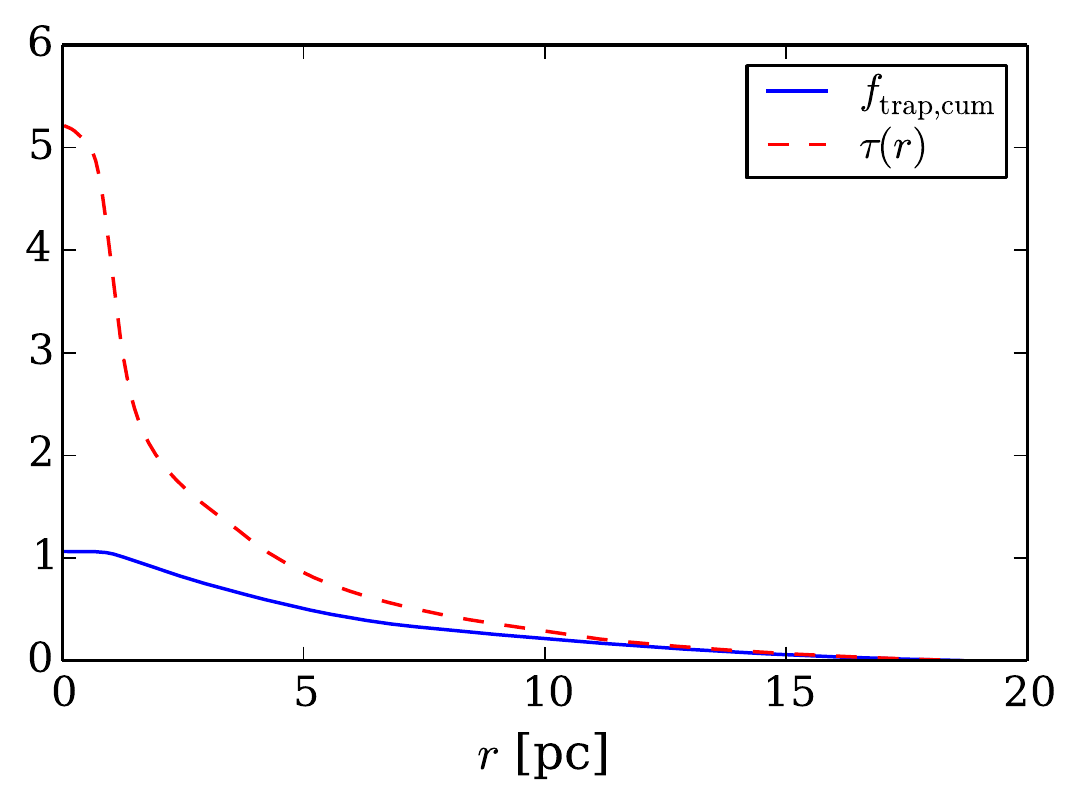}
\caption{Radial profile of the cumulative trapping factor,
  $f_\mathrm{trap,cum}(r)$ (see
  Equation~\ref{radstructure:f_trap_cum_def}), measured inward from
  the largest radius in the interpolation grid (solid), 
for model K20 at $t=3\tff$. Also shown (dashed) is the
  optical depth $\tau(r)$ (see Equation~\ref{radstructure:tau_r_def}), also
  measured radially-inward.  For a simple spherical system 
with a central point source,
  these would be identical; the discrepancy here is explained by the
  anti-correlation of $\rho$ and $F_r$ and the distribution of star formation. }
\label{radstructure:fig:f_trap_cum}
\end{figure}

In analytic models, effects of reprocessed 
\ac{IR} radiation are sometimes described in terms of a radiation 
``trapping factor'' \citep[e.g.,][]{Krumholz:2009a}.  This measures the 
enhancement in the radiation force due to reprocessed radiation energy trapped
in an opaque cloud.  The trapping factor is defined as a ratio of the 
radial radiation force on the gas to the force $L_*/c$ that would apply in 
the single-absorption limit.\footnote{Note that we do not include \ac{UV} radiation in these simulations,
so the trapping factor here only includes effects of \ac{IR} radiation.}  In
Figure~\ref{radstructure:fig:f_trap_cum}, we plot the cumulative
profile of the trapping factor defined by
\begin{equation}
	f_\mathrm{trap,cum} (r) \equiv \frac{\int^{r_\mathrm{max}}_r
          \langle \rho \kappa F_r/c \rangle_{4\pi} \,4\pi r^2
          dr}{L_*/c}, 
\label{radstructure:f_trap_cum_def}
\end{equation}

in comparison to the cumulative optical depth $\tau(r)$ defined by
\begin{equation}
	\tau(r) \equiv \int^{r_\mathrm{max}}_r \langle \rho \kappa
        \rangle_{4\pi} \,dr. 
\label{radstructure:tau_r_def}
\end{equation}
Both quantities are measured radially-inward from the maximum radius
of the interpolation grid. For a spherical, isotropic radiation field  
of a point source and a spherical gas density profile, 
$f_\mathrm{trap,cum}(r)$ and $\tau(r)$ would be identical. However,
Figure~\ref{radstructure:fig:f_trap_cum} shows 
that $f_\mathrm{trap,cum}(r)$ is substantially less than $\tau(r)$
(and is less than 1 over most of the domain).  Once again, much of 
this difference is explained 
by the anti-correlation between $\rho$ and $F_r$, as demonstrated in
Figure~\ref{radstructure:fig:d_Fr_corr}.  
In addition, the fact that sources are distributed rather than
concentrated in a single central point implies that the flux is not
radially directed at small $r$; the high density in this region
therefore leads to a greater contribution to $\tau$ 
(Equation~\ref{radstructure:tau_r_def})  than to $f_\mathrm{trap,cum}$ (Equation
\ref{radstructure:f_trap_cum_def}).  We note that for the present
models, the source distribution is in part due to the spatial
separation of multiple sink particles, and in part due to the finite
size of each radiation source ($\sim R_* = 1\pc$; see Equation
\ref{sourceprofile}) .

\subsection{Effects of Varying Opacity} \label{varyingopacity}

Here we present the results from the K series, in which we vary the
opacity while holding the cloud mass and radius fixed (see
Table~\ref{table:modelparams} for input parameter values).  As
explained in Section~\ref{models}, we use a wide
range of $\kappa$; this range contains values that substantially exceed the
realistic mean opacity of dust to \ac{IR}  
in star-forming \acp{GMC} for Solar neighborhood dust abundance, 
but it allows us to study
the physical dependence of the cloud outcomes on this principal
parameter.  In particular, the discussion leading to Equation~\ref{eq:kappacrit} 
suggests that the potential for reprocessed radiation to 
drive substantial mass loss from a cloud depends primarily on $\kappa$.
In high dust abundance systems
(either galaxies/ISM regions with high metallicity, or locally dust-enriched 
individual clouds), $\kappa$ may be 
in the regime where reprocessed radiation forces become quite important. 

The comparative properties for Series K are summarized in
Table~\ref{table:varyopacity}, where $\varepsilon_* \equiv M_*/M_\mathrm{tot}$ and 
$\varepsilon_\mathrm{gas}\equiv M_\mathrm{gas}/M_\mathrm{tot}$ are measured
over the entire computational domain, with the total mass defined as
$M_\mathrm{tot} \equiv M_* + M_\mathrm{gas} + M_\mathrm{ej}$; 
$f_\mathrm{Edd,cum}(r)$, $f_\mathrm{trap,cum}(r)$, and $\tau(r)$ are measured over 
the entire spherical interpolation domain for $r \rightarrow 0$
(see Equations \ref{radstructure:f_Edd_cum_def}, 
\ref{radstructure:f_trap_cum_def}, \ref{radstructure:tau_r_def}); and 
$\langle \chi_{\rho,F_r} \rangle_r$ is averaged over the spherical 
interpolation domain.  
The quantity 
$\varepsilon_\mathrm{ej}\equiv M_\mathrm{ej}/M_\mathrm{tot}$ is measured from the 
time integral of the mass flux through the surfaces of the computational box.  
We report results from analysis at time
$t=3\tff$, which is after significant star formation feedback has
begun, but before that feedback can destroy the cloud.

\begin{deluxetable*}{ccccccccccc}
  \tabletypesize{\footnotesize}
  \tablecaption{Intermediate Outcomes from Varying the Opacity}
  \tablewidth{0pt}
  \tablecolumns{11}
  \centering
  \tablehead{
    \colhead{\phantom{\quad}} &
    \colhead{Model} &
    \colhead{$f_\mathrm{Edd,*}$} &
    \colhead{$\varepsilon_*$} &
    \colhead{$\varepsilon_\mathrm{gas}$} &
    \colhead{$\varepsilon_\mathrm{ej}$} & 
    \colhead{$f_\mathrm{Edd,cum}$\tna} &
    \colhead{$\langle \chi_{\rho,F_r} \rangle_r$\tnb} &
    \colhead{$f_\mathrm{trap,cum}$\tna} &
    \colhead{$\tau$\tna} &
    \colhead{\phantom{\quad}}
    \vspace{0.1cm}
  }
  \startdata
	& K01        & 0.068 & 0.67 & 0.22 & 0.11 & 0.032 & 0.82 & 0.074 & 0.24 & \\
	& K05        & 0.34  & 0.66 & 0.23 & 0.11 & 0.12  & 0.77 & 0.25  & 1.1  & \\
	& K10        & 0.68  & 0.63 & 0.25 & 0.12 & 0.17  & 0.79 & 0.55  & 3.0  & \\
	& \bftab K20 & 1.4   & 0.57 & 0.30 & 0.13 & 0.43  & 0.81 & 1.1   & 5.2  & \\
	& K30        & 2.0   & 0.53 & 0.33 & 0.14 & 0.60  & 0.70 & 1.2   & 4.9  & \\
	& K40        & 2.7   & 0.50 & 0.33 & 0.16 & 0.99  & 0.63 & 1.1   & 4.1  & 
  \enddata
  \label{table:varyopacity}
\noindent
\tablenotetext{a}{Measured over the spherical interpolation grid, for 
$r \rightarrow 0$.}
\tablenotetext{b}{Measured over spherical shells, then radially averaged.}
\tablecomments{
Intermediate outcomes for the K series, in which the opacity is varied. 
All results are given at time $t=3 \tff$.
}
\end{deluxetable*}

As expected (cf. Equation~\ref{eq:eps*min}), 
the star formation efficiency $\varepsilon_*$ decreases
as $\kappa$ increases.  Correspondingly, both the fraction of
gas remaining in the computational domain, $\varepsilon_\mathrm{gas}$,
and the fraction ejected,\footnote{\label{footnote:eps_wind}It is not
  sufficient merely to note that the gas has left the computational
  domain through the outer boundary to conclude that it has been
  driven out as a wind, especially considering the ``diode'' outflow
  boundary condition we employ prevents gas from (re)entering through
  the same boundary.  In principle, some of the ejected gas might be able to return.
} 
$\varepsilon_\mathrm{ej}$, increase
somewhat with increasing opacity.  However, the differences in these 
intermediate efficiencies at $t=3 \tff$ are less than at late times, 
because the dynamical effect from radiation requires more time to 
develop fully.

In addition to the differences of the efficiencies with $\kappa$, 
Table~\ref{table:varyopacity} shows a number of other interesting 
effects.  First, the Eddington ratio integrated over the whole (spherical) 
domain, $f_\mathrm{Edd,cum}(r\rightarrow 0)$, 
increases with $\kappa$.  However, in all 
cases it remains less than unity.  Moreover, the correlation between 
gas and radiation flux, $\langle \chi_{\rho,F_r} \rangle_r$, decreases 
as $\kappa$ increases for the models with $f_\mathrm{Edd,*}>1$.  That is, 
cases with increasing potential for radiation to overwhelm gravity 
(larger $f_\mathrm{Edd,*}$) in part compensate for this with an increasing
anti-correlation of gas and radiation.  We further find that the 
cumulative trapping factor, $f_\mathrm{trap,cum}(r\rightarrow 0)$, 
can be far smaller 
than the integrated optical depth $\tau(r\rightarrow 0)$ from the center to the edge of 
the cloud; the ratio is a factor of 4 or 5 for the $f_\mathrm{Edd,*}>1$ models.
Thus, the estimate $\tau L/c$, sometimes used in subgrid models 
(within galaxy formation simulations) to 
represent the force from reprocessed radiation, can substantially 
overestimate the true radiation force.

\begin{figure}
  \centering
  \epsscale{1}
  \plotone{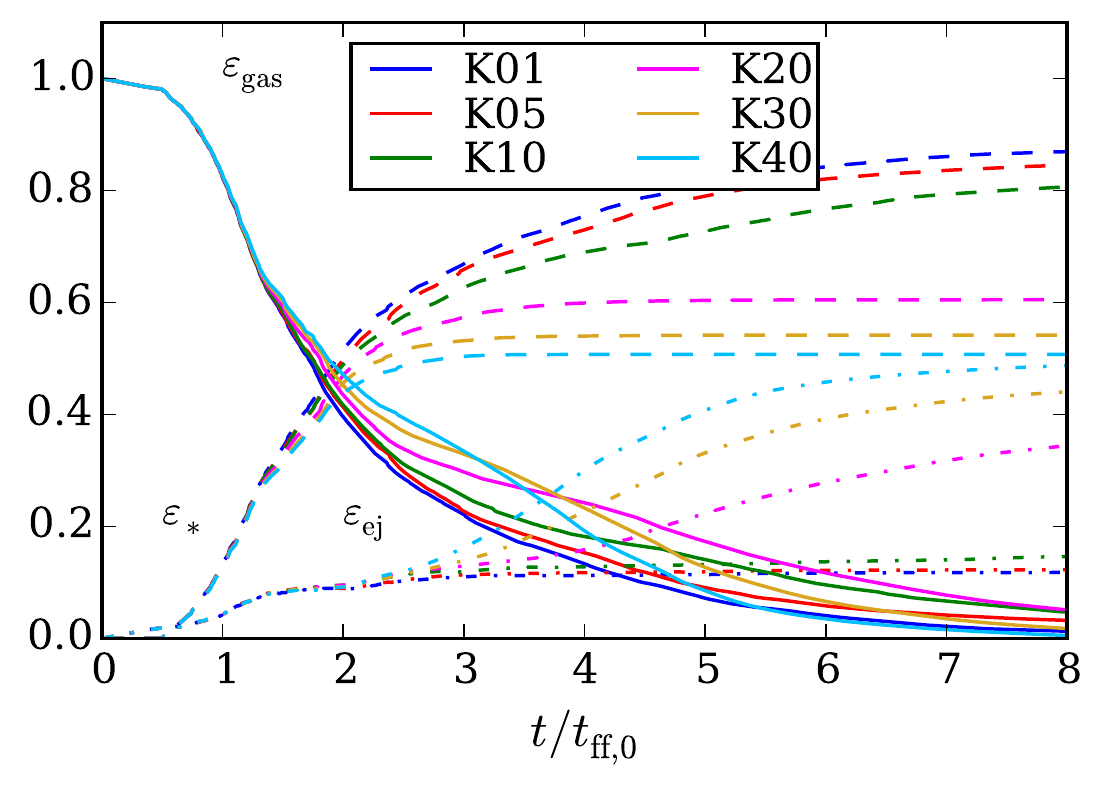}
	\caption{Time evolution of the mass fractions, 
$\varepsilon_* \equiv M_*/M_\mathrm{tot}$ (dashed),
$\varepsilon_\mathrm{gas} \equiv M_\mathrm{gas}/M_\mathrm{tot}$ (solid), 
and $\varepsilon_\mathrm{ej} \equiv M_\mathrm{ej}/M_\mathrm{tot}$ (dash-dotted), 
where $M_\mathrm{tot} \equiv M_* + M_\mathrm{gas} + M_\mathrm{ej}$ 
for the runs in the K series. Low-$\kappa$ models K01, K05, and K10 have 
$f_\mathrm{Edd,*}<1$, while high-$\kappa$ 
models K20, K30, K40 have $f_\mathrm{Edd,*}>1$.
There is a clear break in $\varepsilon_*$ and $\varepsilon_\mathrm{ej}$ between 
these groups. Time is in units of the initial free-fall time,
$t_\mathrm{ff}$, for each model.
}
\label{varyingopacity:fig:eps}
\end{figure}

\begin{figure}
  \centering
  \epsscale{1}
  \plotone{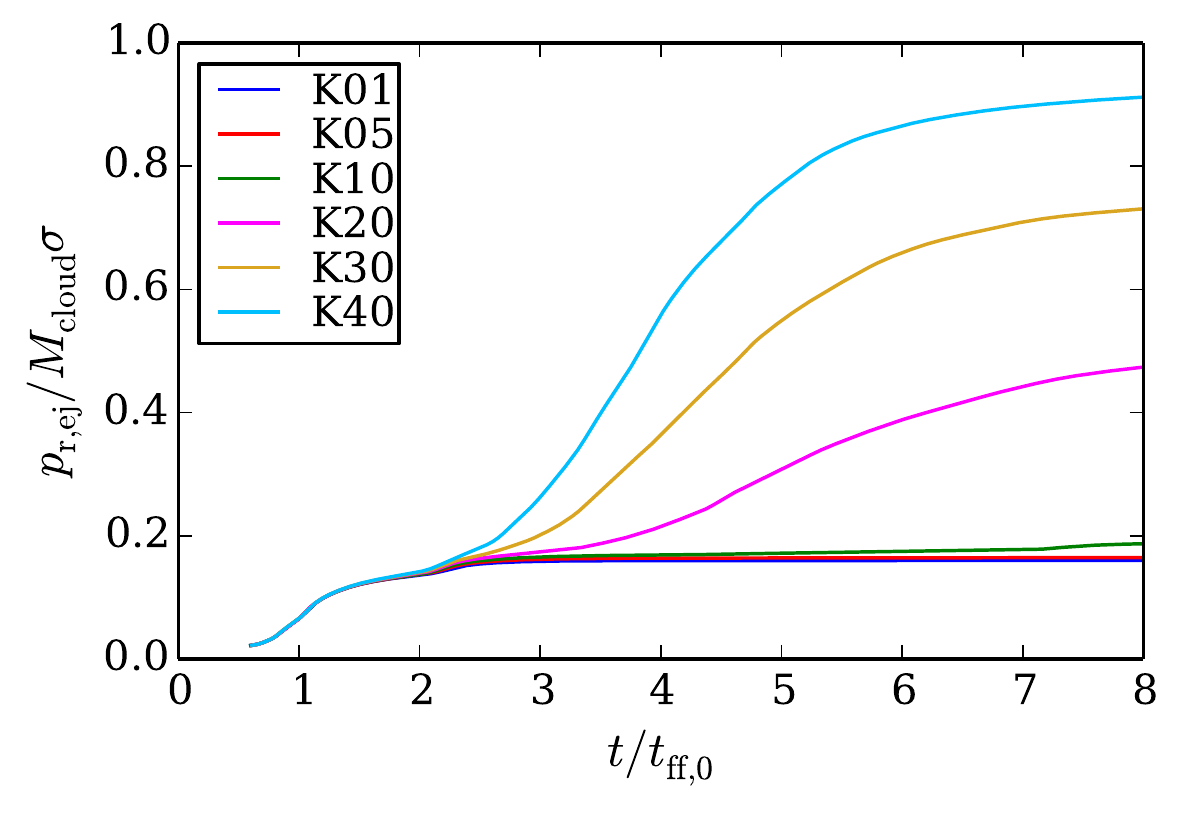}
	\caption{Time evolution of $p_\mathrm{r,ej}/p_\mathrm{turb,init}$, 
the time-integrated radial component of the kinetic momentum ejected from the
grid, in the center of mass frame of the star particles, in units of
the initial turbulent momentum $p_\mathrm{turb,init} \equiv \Mcloud\sigma$, for the runs in the K series.
Only models with $f_\mathrm{Edd,*}>1$ (K20, K30, K40) have substantial momentum 
loss driven by radiation.
}
\label{varyingopacity:fig:pr_ej2}
\end{figure}

\begin{figure}
  \centering
  \epsscale{1}
  \plotone{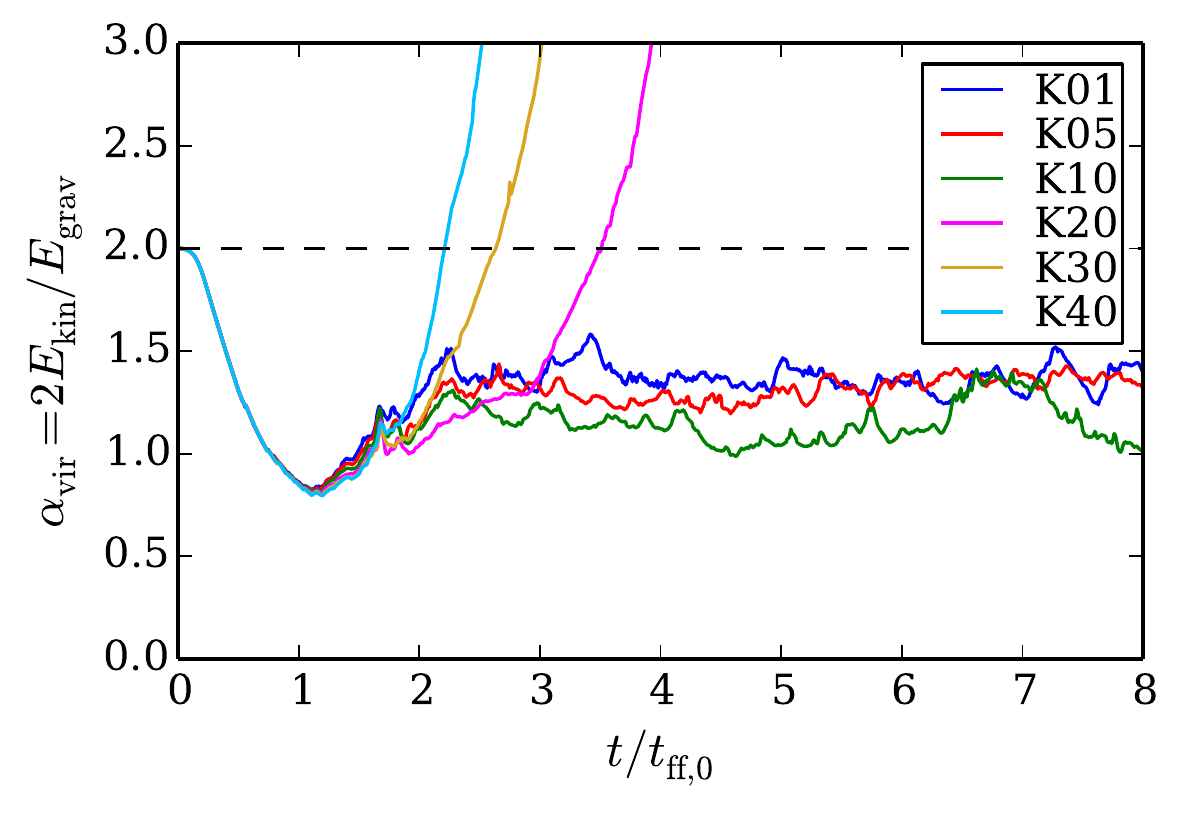}
\caption{Time evolution of $\alpha_\mathrm{vir}=2E_\mathrm{kin}/E_\mathrm{grav}$,
the total virial parameter for the gas on the grid, for the runs in the K 
series.  Note that the clouds with $\kappa > \kappa_\mathrm{crit}$ rapidly become unbound, 
within a few free-fall times, and do so sooner with increasing $\kappa$.}
	\label{varyingopacity:fig:alpha_vir}
\end{figure}

Figures~\ref{varyingopacity:fig:eps} -
\ref{varyingopacity:fig:alpha_vir} show the time evolution out to
$t=8\tff$ of several diagnostic variables.
Figure~\ref{varyingopacity:fig:eps} shows the time evolution of the
mass fractions, $\varepsilon_*$, $\varepsilon_\mathrm{gas}$, and
$\varepsilon_\mathrm{ej}$.  On the one hand, the time evolution of
$\varepsilon_\mathrm{gas}$ is similar for all values of $\kappa$, with
variations of up to 5-10\% over all runs in the K series.  By time
$t=8\tff$, less that 10\% of the original mass of the cloud remains,
with almost all of the gas in each run either accreted onto star
particles or expelled from the domain.
On the other hand, the evolutions of $\varepsilon_*$ and
$\varepsilon_\mathrm{ej}$ depend strongly on $\kappa$; as $\kappa$
increases, $\varepsilon_*$ decreases and $\varepsilon_\mathrm{ej}$
increases.  
Figure~\ref{varyingopacity:fig:eps} shows a clear break between 
models in the group with $\kappa = 1,$ 5, 10, compared to the group 
$\kappa = 20,$ 30, 40.  These groups have $f_\mathrm{Edd,*}<1$ and 
$f_\mathrm{Edd,*}>1$, respectively, so the break is consistent with 
general expectations discussed in Section~\ref{sec:spherical}  

There are additional more subtle effects as well.  First, note that runs 
K1 and K5 have almost identical mass ejection.  This 
is essentially the same as the mass that would be ejected in the absence 
of radiation, due to a small fraction ($\sim 10\%$) of the initial mass 
in the cloud being unbound.  For $\kappa=10$, the effects of radiation 
begins to be seen in reducing $\varepsilon_*$ and increasing 
$\varepsilon_\mathrm{ej}$ (starting at $\sim 3\tff$).  For the large-$\kappa$ group, the final value of 
$\varepsilon_*$ is reached by $t=4\tff$ in all cases.  However, gas ejection 
occurs more rapidly in run K40 than in the K20 and K10 runs.  For large $\kappa$, 
the final values of both $\varepsilon_*$ and 
$\varepsilon_\mathrm{ej}$ approach 0.5.

Figure~\ref{varyingopacity:fig:pr_ej2} shows the time evolution of
$p_\mathrm{r,ej}/(\Mcloud\sigma)$, the time-integrated
radial kinetic momentum ejected from the grid (in the center of mass frame of
the \ac{GMC}), in units of the characteristic 
initial turbulent momentum of the cloud 
$p_\mathrm{turb,init} \equiv \Mcloud\sigma$.
Similar to the results for the ejected mass, there is a clear break between
the low- and high-$\kappa$ models.  In all runs, $p_\mathrm{r,ej}$
increases similarly up to $\sim 2 \tff$, due to the initial turbulence 
expelling some of the gas from the grid.  After $t=2\tff$, the additional 
kinetic momentum ejected over the course
of the simulation is negligible for runs K1, K5, and K10, where
$f_\mathrm{Edd,*} < 1$, and increases dramatically with $\kappa$ for
runs K20, K30, and K40, where $f_\mathrm{Edd,*} > 1$. 

Finally, Figure~\ref{varyingopacity:fig:alpha_vir} shows the time
evolution of $\alpha_\mathrm{vir} \equiv 2
E_\mathrm{kin,gas}/E_\mathrm{grav,gas}$, the total virial parameter
for the gas on the grid.  The simulations begin with
$\alpha_\mathrm{vir}=2$, i.e., with the cloud in the just-bound state.
Then, after a decrease up to $t \sim \tff$ due to the decay of the
initial  turbulence, the virial parameter increases again
once star formation begins.  For runs K1, K5, and K10,
$\alpha_\mathrm{vir}$ settles into a roughly steady value between 1 and 1.5,
i.e., close to virial equilibrium.  In contrast, for
runs K20, K30, and K40, the virial parameters rapidly 
diverge, with larger-$\kappa$ models diverging
earlier, as the cloud is disrupted by radiation from 
the newly formed stars and the
unaccreted gas is ejected from the grid.  
Similar to the results shown in Figure~\ref{varyingopacity:fig:pr_ej2},
this suggests that a state
change occurs in systems with $f_\mathrm{Edd,*} \gtrsim 1$ so that the
gas quickly becomes unbound and is dispersed back into the diffuse \ac{ISM}.

\subsection{Cloud Evolution Outcomes: Parameter Study} \label{parameterstudy}

In this section, we examine the final outcomes for each series in our
parameter study.  Table~\ref{table:parameter_study} summarizes the
results from each series described in Section~\ref{models} with model
parameters given in Table~\ref{table:modelparams}.  From these data,
several general trends are evident.  First, we examine trends in the
final outcomes of the K series, in which the opacity $\kappa$ is
varied independently of the other simulation parameters.  As discussed
in Section~\ref{varyingopacity}, the star formation efficiency $\varepsilon_{*,final}$ 
decreases with increasing opacity, with a clear division between models K10 
and K20 near the critical value of
$f_\mathrm{Edd,*}=1$.  This division is also clearly seen in the values of
$p_{r,\mathrm{ej}}$, the total radial kinetic momentum ejected from the simulation grid.  
Table~\ref{table:parameter_study} includes results for $p_{r,\mathrm{ej}}$
normalized in three different ways, and for all choices the values 
from $\kappa \le 10$ models are all similar, while the values are much larger
for $\kappa \ge 20$.

\begin{deluxetable*}{ccccccccccc}
  \tabletypesize{\footnotesize}
  \tablecaption{Final Outcomes}
  \tablewidth{0pt}
  \tablecolumns{9}
  \centering
  \tablehead{
    \colhead{\phantom{\quad}} &
    \colhead{Model} &
    \colhead{$\varepsilon_\mathrm{*,sph}$} &
    \colhead{$\varepsilon_\mathrm{*,final}$} &
    \colhead{$\varepsilon_\mathrm{ej,final}$} &
    \colhead{$p_{r,\mathrm{ej}}$} &
    \colhead{$p_{r,\mathrm{ej}}$} &
    \colhead{$p_{r,\mathrm{ej}}$} &
    \colhead{$p_{r,\mathrm{ej}}/M_*$} &
    \colhead{$\alpha_\mathrm{vir,4}$\tna} &
    \colhead{\phantom{\quad}} \\
    \colhead{} &
    \colhead{} &
    \colhead{} &
    \colhead{} &
    \colhead{} &
    \colhead{$\overline{\Mcloud \sigma}$} &
    \colhead{$\overline{M_* \sigma}$} &
    \colhead{$\overline{M_\mathrm{ej} v_\mathrm{esc}}$} &
    \colhead{($\mathrm{km}\mbox{ s}^{-1}$)} &
    \colhead{} &
    \colhead{}
    \vspace{0.1cm}
  }  
  \startdata
	& K1         & 15    & 0.87 & 0.12 & 0.16 & 0.18 & 1.4 & 3.9 & 1.3 & \\
	& K5         &  3.0  & 0.85 & 0.12 & 0.16 & 0.19 & 1.4 & 4.1 & 1.3 & \\
	& K10        &  1.5  & 0.81 & 0.15 & 0.19 & 0.23 & 1.3 & 4.9 & 1.1 & \\
	& \bftab K20 &  0.75 & 0.61 & 0.34 & 0.47 & 0.78 & 1.4 & 17  & 3.3 & \\
	& K30        &  0.50 & 0.54 & 0.44 & 0.73 & 1.3  & 1.7 & 29  & ---\tnb & \\
	& K40        &  0.38 & 0.51 & 0.49 & 0.91 & 1.8  & 1.9 & 38  & ---\tnb & \\
	\vspace{-0.2cm} \\
	\tableline
	\vspace{-0.2cm} \\
	&  R7.1       & 0.75 & 0.70 & 0.25 & 0.60 & 0.87 & 2.5  & 22 & 1.9 & \\
	&  \bftab R10 & 0.75 & 0.61 & 0.34 & 0.47 & 0.78 & 1.4  & 17 & 3.3 & \\
	&  R14.1      & 0.75 & 0.53 & 0.39 & 0.33 & 0.62 & 0.86 & 11 & 5.2 & \\
	\vspace{-0.2cm} \\
	\tableline
	\vspace{-0.2cm} \\
	&  M0.5      & 0.75 & 0.50 & 0.44 & 0.41 & 0.82 & 0.95 & 12 & 2.8 & \\
	&  \bftab M1 & 0.75 & 0.61 & 0.34 & 0.47 & 0.78 & 1.4  & 17 & 3.3 & \\
	&  M2        & 0.75 & 0.71 & 0.23 & 0.52 & 0.73 & 2.3  & 22 & 1.6 & \\
	\vspace{-0.2cm} \\
	\tableline
	\vspace{-0.2cm} \\
	&  R5M0.25      & 0.75 & 0.58 & 0.36 & 0.60 & 1.0  & 1.7 & 16 & 1.5 & \\
	&  R7.1M0.5     & 0.75 & 0.62 & 0.33 & 0.54 & 0.86 & 1.7 & 16 & 4.0 & \\
	&  \bftab R10M1 & 0.75 & 0.61 & 0.34 & 0.47 & 0.78 & 1.4 & 17 & 3.3 & \\
	&  R14.1M2      & 0.75 & 0.61 & 0.36 & 0.41 & 0.68 & 1.2 & 17 & 4.8 & \\
	&  R20M4        & 0.75 & 0.66 & 0.27 & 0.35 & 0.53 & 1.3 & 16 & 1.7 & 
  \enddata
  \label{table:parameter_study}
\noindent
\tablenotetext{a}{The virial parameter is given at time $t=4\tff$.}
\tablenotetext{b}{In runs K30 and K40, the cloud has become unbound by $t=4\tff$ and the gas density is at the floor value, such that 
$\alpha_\mathrm{vir,4}$ is not longer meaningful.}
\tablecomments{All results are given at time $t=8\tff$, except for 
$\alpha_\mathrm{vir,4}$.}
\end{deluxetable*}

Second, we examine the simulation outcomes in the R, M, and RM series
as a function of the cloud surface density and Mach number.  The opacity 
is $\kappa = 20$ for all of these models, so that $f_{\rm Edd,*}=1.4$.  On the
one hand, the star formation efficiencies are all comparable
throughout the RM series, where the surface density and opacity are
held constant as the cloud radius and mass are varied.  On the other
hand, as the cloud radius and mass are independently varied in the R
and M series, respectively, the cloud surface density changes, and 
there is a much stronger variation in the star formation
efficiencies in these series. 
 This suggests that---all else being
equal---clouds with higher surface density eject a
smaller fraction of their initial mass.  
Interestingly, the trend for $\varepsilon_\mathrm{*,final}$ with $\tau$ is the opposite 
in varying-$\Sigma$ models from varying-$\kappa$ models.
That is, an increase of $\kappa$ for fixed cloud mass and radius decreases $\varepsilon_\mathrm{*,final}$,
whereas an increase of mass or decrease of radius for fixed $\kappa$ increases $\varepsilon_\mathrm{*,final}$.  
Also, we note that the pairs (M0.5,M1) and 
(R5M0.25, R10M1) have the same matched values of the initial velocity 
dispersion (16 and 23 $\kms$, see Table~\ref{table:modelparams}), 
but $\varepsilon_\mathrm{*,final}$ varies more strongly for the M series than for the RM 
series.  That is, surface density appears to be more important than 
the potential well depth for the final star formation efficiency of a cloud.

We further note that although $f_\mathrm{Edd,*}>1$ distinguishes models in which 
$\varepsilon_*$ is reduced by radiation from those in which radiation does not 
affect $\varepsilon_*$, the values $\varepsilon_\mathrm{*, final}$ in 
Table~\ref{table:parameter_study} are {\it not} consistent with Equation~\eqref{eq:eps*min}.  
In particular, the R, M, and RM series all have 
$f_\mathrm{Edd,*}=1.4$ and $\varepsilon_\mathrm{*,sph}=0.74$, while 
Table~\ref{table:parameter_study} shows a range of values for 
$\varepsilon_\mathrm{*, final}$.  Similarly, $\varepsilon_\mathrm{*,sph}$ decreases 
from $0.74$ to $0.37$ from K20 to K40, but the decrease in 
$\varepsilon_\mathrm{*, final}$ is fractionally smaller.  This shows that a simple
``Eddington-type'' spherical model is inadequate for quantitatively 
predicting the net \ac{SFE} in a cloud, as controlled by radiation feedback.

In all series, we see that $p_{r,\mathrm{ej}}/M_*$, the kinetic momentum
ejected per stellar mass formed, is comparable to the initial velocity
dispersion $\sigma$ of the clouds, to within a factor of 2 for all
runs with $f_\mathrm{Edd,*}>1$.  The values are between 10 and
40$\kms$.  
Interestingly, just as for $\varepsilon_\mathrm{*,final}$, the ratio
$p_{r,\mathrm{ej}}/M_*$ appears to depend more strongly on the cloud's
surface density than its potential well depth; all the models in the
RM series have essentially the same value of $p_{r,\mathrm{ej}}/M_*$
even though they have varying potential well depth (i.e., varying
$\sigma$), while $p_{r,\mathrm{ej}}/M_*$ increases toward higher
$\Sigma$ models in the R and M series.  

In principle, it is possible that some fraction of the gas that is
ejected from the simulation grid might in reality ultimately be able
to re-collapse.  As a proxy for whether or not the gas can re-collapse,
we compute the escape velocity
\begin{equation}
	v_\mathrm{esc} \equiv \left( \frac{2G\Mcloud}{R_\mathrm{box}} \right)^{1/2},
\end{equation}
where $R_\mathrm{box} = L_\mathrm{box}/2 = 2\Rcloud$ is the distance from the center to the
edge of the computational box.  In Table~\ref{table:parameter_study}, we compare $p_{r,\mathrm{ej}}$ to the product of the total ejected gas mass and this 
escape speed.   We see once again that this ratio is
correlated with $\Sigma$, the initial surface density of the cloud,
and only for the lowest value of $\Sigma$ considered (0.33$\gcms$ 
in models M0.5 and R14.1)
is this ratio less than 1.  For all other models, the
ratio $p_{r,\mathrm{ej}}/M_\mathrm{ej} v_\mathrm{esc}$ is greater than
1, indicating that the bulk of the ejected gas is likely unbound.

\section{Summary and Discussion}  \label{conclusions}

In this work, we present the results of a set of numerical \ac{RHD}
simulations of star-forming, turbulent \acp{GMC}.  We focus on the
dynamical effects of reprocessed radiation that originates in massive
star clusters, in particular, on the ability of radiation forces to
limit further gravitational collapse and destroy the \ac{GMC} by driving
gas outward.  
Our models investigate conditions and processes similar to those experienced
and created by forming \acp{SSC}.
Our simulations are conducted using the
\textit{Hyperion} extension of the \textit{Athena} code;
\textit{Hyperion} solves the two-moment \ac{RHD} equations using the
$M_1$ closure and the \ac{RSLA} (see SO13).  To represent the
interaction of \ac{IR} radiation with the dusty \ac{ISM}, we adopt the radiative
equilibrium condition, such that the absorption and emission of
radiation by the fluid are assumed to balance identically everywhere.
Star clusters are treated in an idealized manner, with regions of
collapsing gas creating sink particles that become local radiation
sources.  We adopt spatially-uniform opacity and an isothermal
equation of state for simplicity, but consider models with a wide
range of opacity, $\kappa = 1-40\cmsg$, to test the dependence on this
important parameter.  We also explore a range of cloud masses and
sizes similar to those observed in starburst regions, with initial cloud 
surface density $\Sigma = 0.3-1.3 \gcms$ (or
$1600-6000\Surf$).  All models are initiated with a
turbulent velocity dispersion, $\sigma$, such that kinetic and
gravitational energy are in equipartition.  Table~\ref{table:modelparams}
summarizes the input model parameters.  While idealized in several
respects, to our knowledge this is the first study that has used
self-consistent, three-dimensional \ac{RHD} simulations to investigate the
dynamics of \ac{IR} radiation feedback in cluster-forming turbulent \acp{GMC}.

In all of our models, evolution to a final state occurs over several
free-fall times (see, e.g., Figure~\ref{varyingopacity:fig:eps}).  By
$\sim 1 \tff$, gravitational collapse leads to the formation of the
first star particles, and by $\sim 2-4 \tff$, they gain most of the
mass they will accrete in their lifetimes.  The evolution of gas mass
is similar in all models, with the difference that in models with low
opacity, essentially all the gas is consumed by star formation, while
in models with high opacity, accretion is halted by radiative
feedback.  For runs
with $\kappa =1, 5, 10 \cmsg$ ($f_\mathrm{Edd,*}<1$),
accretion onto star particles slows after $\sim 2\tff$; the
surrounding gas maintains a quasi-steady value of the virial parameter
$\sim 1-1.5$ over the next $\sim 6 \tff$
(Fig. \ref{varyingopacity:fig:alpha_vir}), and little additional mass
is ejected from the simulation box.  For the runs with 
$\kappa = 20, 30, 40 \cmsg$ ($f_\mathrm{Edd,*}>1$), the virial parameter 
diverges after $\sim 2-3 \tff$, and most
of the remaining gas is ejected from the box.

The most important parameter in determining $\varepsilon_\mathrm{*,final}$, the net
efficiency of star formation over the lifetime of a cloud, is
$f_\mathrm{Edd,*}\equiv \kappa \Psi/(4 \pi cG)$, where $\Psi$ is the
mean luminosity-to-mass ratio. When $f_\mathrm{Edd,*}<1$, for $\kappa
< 15 \cmsg$ (taking $\Psi = 1700 \mbox{ erg s}^{-1} \mbox{ g}^{-1}$;
see Equation~\ref{eq:fEdd*}), star formation efficiencies
are high (exceeding 80\%); almost all the mass ejected from the box (at $t< \tff$) is
due to the initial turbulence rather than radiation feedback.
When $f_\mathrm{Edd,*} > 1$, efficiencies
are lower and decrease as $f_\mathrm{Edd,*}$
increases, as shown for the K series in Table~\ref{table:parameter_study}.
Of secondary importance is the cloud surface density $\Sigma$; models 
with lower $\Sigma$ have lower $\varepsilon_\mathrm{*,final}$ 
(see the R and M series in 
Table~\ref{table:parameter_study}).
\footnote{We note that the relative sensitivity to $\kappa$ and relative 
insensitivity to $\Sigma$ for the dynamical response of a turbulent, 
self-gravitating cloud to reprocessed radiation 
is the opposite of the dynamical response to direct \ac{UV} radiation from embedded
clusters, 
which strongly depends on $\Sigma$ \citep[Raskutti, Ostriker, \& Skinner 2015,
in preparation; see also][]{Fall:2010,Thompson:2014}}
Models with high $\kappa$ and low $\Sigma$ have  $\varepsilon_\mathrm{*,final}\sim 0.5$. 

The value of $f_\mathrm{Edd,*}$ is also the discriminant between
models that have high vs. low values for the ratio of ejected momentum
to stellar mass formed, $p_{r,\mathrm{ej}}/M_*$ (see
Table~\ref{table:parameter_study}).  
However, we note that the
``high'' values ($p_{r,\mathrm{ej}}/M_* \sim 10 - 40 \kms$) are still
quite low compared to the momentum/stellar mass from other feedback
sources \citep[cf.][]{Ostriker:2011}; 
for supernovae, this ratio is $\sim 3000 \kms$
\citep[see, e.g.,][]{Kim:2015}.  
For $f_\mathrm{Edd,*}>1$, we find that the ejected momentum,
$p_{r,\mathrm{ej}}$ is of order $\sigma M_*$.  We also find that 
$p_{r,\mathrm{ej}}/M_*$ increases with the cloud surface density 
$\Sigma$.  On average, gas that is ejected 
has speed a few times the escape speed of the system.

Our simulations indicate that a large value of $\kappa$ ($> 10 \cmsg$) 
would be required for reprocessed radiation to eject a
significant proportion of the mass in a cloud.  This value is perhaps
unphysically large for \ac{IR} opacities 
at Solar neighborhood abundance even for warm dust 
\citep[see][for temperature-dependent Rosseland mean opacity]{Semenov:2003}, 
with the implication that 
reprocessed radiation would not substantially limit star formation in 
typical Milky Way \acp{GMC}. 
However, for galaxies with metallicities higher than Solar, 
or in dust-enriched regions, such large opacities are possible.
The minimum
$\kappa$ for which feedback limits star formation in our turbulent
simulations is consistent with the prediction of an extremely simple
model: an isotropic spherical cloud surrounding a central stellar
cluster, in which gravity and radiation are the only forces.  
For this
highly idealized situation, Equation~\eqref{eq:kappacrit} shows that
the critical opacity for which radiation forces can begin to exceed
gravity immediately outside the cluster (preventing further accretion)
is $\kappa_\mathrm{crit}=15 \cmsg (\Psi/1700 \mbox{ erg s}^{-1} \mbox{ g}^{-1})^{-1}$.  
In spite of
having a similar critical transition, however, the simple spherical
model does not describe the detailed functional dependence of the
numerical models on $\kappa$; for example, unlike Equation
(\ref{eq:eps*min}), the K series does not show 
$\varepsilon_\mathrm{*,final}\propto \kappa^{-1}$ at $\kappa > \kappa_\mathrm{crit}$ 
(when $f_\mathrm{Edd,*}>1$).

One difference between the simple spherical model and our turbulent
simulations is the anti-correlation between the gas density, $\rho$,
and the radial flux, $F_r$ (see Figures~\ref{radstructure:fig:f_Edd1}
and~\ref{radstructure:fig:d_Fr_corr}).  Thus, whereas 
$f_{\rm Edd,sph}(r) \propto f_\mathrm{Edd,*} \propto \kappa$ for the spherical
model (Equation~\ref{eq:fEddrsph}), the cumulative
$f_\mathrm{Edd,cum}(r)$ (see Equation~\ref{radstructure:f_Edd_cum_def})
measured in our simulations is not linear in $\kappa$ (see
Table~\ref{table:varyopacity}).  In our fiducial model, Figure
\ref{radstructure:fig:f_Edd1} shows that while $f_{\rm Edd}(r)<1$ at
most radii, $f_{\rm Edd,spec}>1$ at most radii; this is a consequence
of anti-correlation (see Equation~\ref{radstructure:f_Edd_def} 
and \ref{radstructure:d_Fr_corr_def}).  For
the K series, the anti-correlation between density and radial flux
increases at higher values of $\kappa$, and $f_\mathrm{Edd,cum}(r\rightarrow 0)$
remains less than unity for all values of $\kappa$.  Nevertheless,
even when the cumulative Eddington factor is less than unity, some
fluid elements become unbound, reducing $\varepsilon_\mathrm{*,final}$ at high enough 
$\kappa$.

\cite{Krumholz:2012} found, in their 2D \ac{RHD} simulations of 
turbulent disks using
\ac{FLD}, that the radiation flux is strongly anti-correlated with the
matter distribution. However, the \ac{FLD} approximation may
contribute in part to this result; it is well-known that radiation in this
approximation can easily ``leak'' around dense structures without
creating shadows behind them \citep[see, e.g.,][]{Hayes:2003}.
This question was investigated by \cite{Davis:2014}, who 
demonstrated flux-density
anti-correlations in turbulent \ac{RHD} simulations 
using both \ac{FLD} and more accurate \ac{VET} methods. They
show that the anti-correlation is much stronger using the \ac{FLD}
method compared to their \ac{VET} method, since the former does not
account for the relative insensitivity of the flux to strong density
contrasts in filamentary structures when the optical depths across
their widths are of order unity or less.  
Because the $M_1$ approximation we have adopted does not 
resolve angular variations in 
the intensity of radiation and can have difficulty capturing
the true radiation field in some situations, it will be important to
check the results we have obtained with other more accurate (but more
computationally expensive) \ac{RHD} methods such as \ac{VET}.

Motivated by analytic spherical models of cluster formation in dusty clouds
\citep{Krumholz:2009a,Murray:2010}, current galaxy formation simulations 
\citep[e.g.,][]{Hopkins:2011,Hopkins:2014,Agertz:2013} 
have adopted simple subgrid treatments of
the force arising from reprocessed \ac{IR} radiation.  In these treatments, the
single-scattering \ac{UV} force, $L_*/c$, is boosted via photon trapping by 
a factor $\tau_\mathrm{IR}$, the
mean optical depth to \ac{IR} through a cloud.  In fact, our comparisons between 
$f_\mathrm{trap,cum}$ and $\tau$ (defined in Equations 
\ref{radstructure:f_trap_cum_def} and \ref{radstructure:tau_r_def})
as shown in Figure~\ref{radstructure:fig:f_trap_cum} and Table~\ref{table:varyopacity}
indicate that $\tau L_*/c$ 
 may overestimate the true reprocessed radiation force by a factor of 
$\sim 4-5$.  First, the force may be reduced due to
anti-correlation of the flux and density, as noted above.  
Second, the assumption of a single,
centrally-embedded cluster may be too na\"ive.  A  massive 
cloud may contain several clusters 
in a distributed configuration, especially at early stages before these 
can merge.  Radiation forces on gas within a cloud with distributed sources 
is subject to cancellation, and only far from the center of mass of
the distribution would the radiation field approach
that of a single concentrated source.  Meanwhile, the largest contribution 
to the optical depth may be from the dense central region of the cloud.  

In addition to accounting for radiation-matter anti-correlation and
distributed radiation sources, it is also important for models that
apply \ac{IR} feedback via a subgrid model to ensure that the inward
gravitational forces are consistent with the imposed outward radiation
forces.  This requires that gravity be spatially and temporally well-resolved 
within any clouds where radiation forces are applied.  If
gravity is softened at small scales, then collapse may not occur as
rapidly as it realistically should, which would give imposed radiation
forces an unphysical advantage.  A situation of this kind might help
explain why \cite{Hopkins:2011} concluded that reprocessed radiation
could play a dominant role in regulating star formation (for their
``HiZ'' model), even though their \ac{SPH} simulations adopted a value
$\kappa=5 \cmsg$ that we found leads to a negligible reduction in a cloud's
star formation efficiency -- whether for fully turbulent simulations
or an idealized spherical system.  If limited resolution vitiates the direct 
action of small-scale gravity in star-forming clouds, then to avoid a
gravity/radiation imbalance it would be necessary to incorporate 
effects of that gravity
as part of a subgrid model for effects of radiation. More generally, the limited resolution of
galaxy formation simulations precludes direct simulation of many 
small-scale physical processes, but as feedback is crucial to the
control of star formation, application of subgrid treatments is
unavoidable.  Spatially resolved direct \ac{RHD}/\ac{MHD} simulations can be
used to identify the most important feedback processes, and to design
and calibrate subgrid treatments that capture these key effects.

In the present work, we have focused on the effects of reprocessed \ac{IR}
radiation, with our simulations suggesting that this form of feedback
is unlikely to significantly reduce star formation within \acp{GMC} unless 
the dust abundance and opacity are higher than expected for Solar metallicity
conditions.
Alternatively, since the opacity $\kappa$ and luminosity-to-mass ratio
$\Psi$ only appear as a product, a top-heavy stellar mass function could 
increase $f_\mathrm{Edd,*}$, potentially leading to a reduction in the 
\ac{SFE}.  \citet{Turner:2015} present intriguing evidence of both 
dust self-enrichment and a boosted $\Psi$ in Cloud D within NGC 5253. 
Even so, this cloud has estimated SFE $\sim 0.6$, similar to the values of 
$\varepsilon_\mathrm{*, final}$ we obtain in our high $f_\mathrm{Edd,*}$ and 
$\Sigma\sim 0.33 \gram\ \pcc$ models.

Finally, we note that the direct \ac{UV} radiation from young, hot clusters has an
advantage over \ac{IR} in that optical depths become large even when clouds
have much lower column densities. Initial study 
(Raskutti, Ostriker, \& Skinner 2015, in preparation)
of the effects of 
(non-ionizing) \ac{UV}, using \ac{RHD} models similar to those of this paper, 
suggests that this direct radiation may be effective in limiting the \ac{SFE} within
low surface density \acp{GMC}.

\acrodef{BVP}{boundary value problem}
\acrodef{CFL}{Courant-Friedrichs-Lewy}
\acrodef{CTU}{corner transport upwind}
\acrodef{DFT}{discrete Fourier transform}
\acrodef{EOS}{equation of state}
\acrodef{FFT}{fast Fourier transform}
\acrodef{FLD}{flux-limited diffusion}
\acrodef{GMC}{giant molecular cloud}
\acrodef{GMRES}[GMRES]{generalized minimal-residual}
\acrodef{HLL}[HLL]{Harten-Lax-van Leer}
\acrodef{HWHM}{half-width at half-maximum}
\acrodef{IMF}{initial mass function}
\acrodef{IR}{infrared}
\acrodef{ISM}{interstellar medium}
\acrodef{IVP}{initial value problem}
\acrodef{MHD}{magnetohydrodynamics}
\acrodef{MPI}[MPI]{Message-Passing Interface}
\acrodef{MUSCL}{monotone upwind method for scalar conservation laws}
\acrodef{ODE}{ordinary differential equation}
\acrodef{OTVET}{optically thin variable Eddington tensor}
\acrodef{PM}{particle mesh}
\acrodef{PDE}{partial differential equation}
\acrodef{RHD}{radiation hydrodynamics}
\acrodef{RMS}{root mean-square}
\acrodef{RSLA}{reduced speed of light approximation}
\acrodef{SSC}{super star cluster}
\acrodef{SFE}{star formation efficiency}
\acrodef{SFR}{star formation rate}
\acrodef{SN}{supernova}
\acrodefplural{SN}[SNe]{supernovae}
\acrodef{SPH}{smooth particle hydrodynamics}
\acrodef{TSC}{triangular-shaped cloud}
\acrodef{UV}{ultraviolet}
\acrodef{VET}{variable Eddington tensor}
\acrodef{VL}{van Leer}

\acknowledgments

We thank the referee for providing a helpful report.  We also thank
Taysun Kimm, Chris Matzner, and Jim Stone for helpful comments and
suggestions on the manuscript.  This work was supported by Grants
No. AST-1312006 and PHY-1144374 from the National Science
Foundation. MAS is supported by the Max-Planck/Princeton Center for
Plasma Physics.  Part of this project was conducted during a visit to
the KITP at U.C. Santa Barbara, which is supported by the National
Science Foundation under Grant No. PHY-1125915. Simulations were
performed on the computational resources supported by the PICSciE
TIGRESS High Performance Computing Center at Princeton University.

\bibliography{apj-jour,references}

\begin{thebibliography}{74}
\expandafter\ifx\csname natexlab\endcsname\relax\def\natexlab#1{#1}\fi

\bibitem[{{Agertz} {et~al.}(2013){Agertz}, {Kravtsov}, {Leitner}, \&
  {Gnedin}}]{Agertz:2013}
{Agertz}, O., {Kravtsov}, A.~V., {Leitner}, S.~N., \& {Gnedin}, N.~Y. 2013,
  \apj, 770, 25

\bibitem[{{Bertoldi} \& {McKee}(1992)}]{Bertoldi:1992}
{Bertoldi}, F., \& {McKee}, C.~F. 1992, \apj, 395, 140

\bibitem[{{Binette} {et~al.}(1997){Binette}, {Wilson}, {Raga}, \&
  {Storchi-Bergmann}}]{Binette:1997}
{Binette}, L., {Wilson}, A.~S., {Raga}, A., \& {Storchi-Bergmann}, T. 1997,
  \aap, 327, 909

\bibitem[{{Dale} {et~al.}(2012){Dale}, {Ercolano}, \& {Bonnell}}]{Dale:2012}
{Dale}, J.~E., {Ercolano}, B., \& {Bonnell}, I.~A. 2012, \mnras, 424, 377

\bibitem[{{Dale} {et~al.}(2013){Dale}, {Ercolano}, \& {Bonnell}}]{Dale:2013a}
---. 2013, \mnras, 430, 234

\bibitem[{{Davis} {et~al.}(2014){Davis}, {Jiang}, {Stone}, \&
  {Murray}}]{Davis:2014}
{Davis}, S.~W., {Jiang}, Y.-F., {Stone}, J.~M., \& {Murray}, N. 2014, ArXiv
  e-prints

\bibitem[{{Dobbs} {et~al.}(2014){Dobbs}, {Krumholz}, {Ballesteros-Paredes},
  {Bolatto}, {Fukui}, {Heyer}, {Low}, {Ostriker}, \&
  {V{\'a}zquez-Semadeni}}]{Dobbs:2014}
{Dobbs}, C.~L., {et~al.} 2014, Protostars and Planets VI, 3

\bibitem[{{Dopita} {et~al.}(2002){Dopita}, {Groves}, {Sutherland}, {Binette},
  \& {Cecil}}]{Dopita:2002}
{Dopita}, M.~A., {Groves}, B.~A., {Sutherland}, R.~S., {Binette}, L., \&
  {Cecil}, G. 2002, \apj, 572, 753

\bibitem[{{Draine}(2011{\natexlab{a}})}]{Draine:2011}
{Draine}, B.~T. 2011{\natexlab{a}}, \apj, 732, 100

\bibitem[{{Draine}(2011{\natexlab{b}})}]{Draine:2011a}
---. 2011{\natexlab{b}}, Physics of the Interstellar and Intergalactic Medium
  (Princeton: Princeton University Press)

\bibitem[{{Elmegreen}(1983)}]{Elmegreen:1983}
{Elmegreen}, B.~G. 1983, \mnras, 203, 1011

\bibitem[{{Fall} {et~al.}(2010){Fall}, {Krumholz}, \& {Matzner}}]{Fall:2010}
{Fall}, S.~M., {Krumholz}, M.~R., \& {Matzner}, C.~D. 2010, \apjl, 710, L142

\bibitem[{{Federrath} \& {Klessen}(2012)}]{Federrath:2012}
{Federrath}, C., \& {Klessen}, R.~S. 2012, \apj, 761, 156

\bibitem[{{Geen} {et~al.}(2015){Geen}, {Rosdahl}, {Blaizot}, {Devriendt}, \&
  {Slyz}}]{Geen:2015}
{Geen}, S., {Rosdahl}, J., {Blaizot}, J., {Devriendt}, J., \& {Slyz}, A. 2015,
  \mnras, 448, 3248

\bibitem[{{Gnedin} \& {Abel}(2001)}]{Gnedin:2001}
{Gnedin}, N.~Y., \& {Abel}, T. 2001, \na, 6, 437

\bibitem[{{Gong} \& {Ostriker}(2009)}]{Gong:2009}
{Gong}, H., \& {Ostriker}, E.~C. 2009, \apj, 699, 230

\bibitem[{{Gong} \& {Ostriker}(2011)}]{Gong:2011}
---. 2011, \apj, 729, 120

\bibitem[{{Gong} \& {Ostriker}(2013)}]{Gong:2013}
---. 2013, \apjs, 204, 8

\bibitem[{{Gonz{\'a}lez} {et~al.}(2007){Gonz{\'a}lez}, {Audit}, \&
  {Huynh}}]{Gonzalez:2007}
{Gonz{\'a}lez}, M., {Audit}, E., \& {Huynh}, P. 2007, \aap, 464, 429

\bibitem[{{Harper-Clark} \& {Murray}(2009)}]{Harper-Clark:2009}
{Harper-Clark}, E., \& {Murray}, N. 2009, \apj, 693, 1696

\bibitem[{{Hayes} \& {Norman}(2003)}]{Hayes:2003}
{Hayes}, J.~C., \& {Norman}, M.~L. 2003, \apjs, 147, 197

\bibitem[{{Hockney} \& {Eastwood}(1988)}]{Hockney:1988}
{Hockney}, R.~W., \& {Eastwood}, J.~W. 1988, Computer simulation using
  particles (Bristol: Hilger)

\bibitem[{{Hopkins} {et~al.}(2014){Hopkins}, {Kere{\v s}}, {O{\~n}orbe},
  {Faucher-Gigu{\`e}re}, {Quataert}, {Murray}, \& {Bullock}}]{Hopkins:2014}
{Hopkins}, P.~F., {Kere{\v s}}, D., {O{\~n}orbe}, J., {Faucher-Gigu{\`e}re},
  C.-A., {Quataert}, E., {Murray}, N., \& {Bullock}, J.~S. 2014, \mnras, 445,
  581

\bibitem[{{Hopkins} {et~al.}(2011){Hopkins}, {Quataert}, \&
  {Murray}}]{Hopkins:2011}
{Hopkins}, P.~F., {Quataert}, E., \& {Murray}, N. 2011, \mnras, 417, 950

\bibitem[{{Iffrig} \& {Hennebelle}(2015)}]{Iffrig:2015}
{Iffrig}, O., \& {Hennebelle}, P. 2015, \aap, 576, A95

\bibitem[{{Johnson} \& {Kobulnicky}(2003)}]{Johnson:2003}
{Johnson}, K.~E., \& {Kobulnicky}, H.~A. 2003, \apj, 597, 923

\bibitem[{{Johnson} {et~al.}(2001){Johnson}, {Kobulnicky}, {Massey}, \&
  {Conti}}]{Johnson:2001}
{Johnson}, K.~E., {Kobulnicky}, H.~A., {Massey}, P., \& {Conti}, P.~S. 2001,
  \apj, 559, 864

\bibitem[{{Johnson} {et~al.}(2015){Johnson}, {Leroy}, {Indebetouw}, {Brogan},
  {Whitmore}, {Hibbard}, {Sheth}, \& {Evans}}]{Johnson:2015}
{Johnson}, K.~E., {Leroy}, A.~K., {Indebetouw}, R., {Brogan}, C.~L.,
  {Whitmore}, B.~C., {Hibbard}, J., {Sheth}, K., \& {Evans}, A.~S. 2015, \apj,
  806, 35

\bibitem[{{Kawamura} {et~al.}(2009){Kawamura}, {Mizuno}, {Minamidani},
  {Filipovi{\'c}}, {Staveley-Smith}, {Kim}, {Mizuno}, {Onishi}, {Mizuno}, \&
  {Fukui}}]{Kawamura:2009}
{Kawamura}, A., {et~al.} 2009, \apjs, 184, 1

\bibitem[{{Kepley} {et~al.}(2014){Kepley}, {Reines}, {Johnson}, \&
  {Walker}}]{Kepley:2014}
{Kepley}, A.~A., {Reines}, A.~E., {Johnson}, K.~E., \& {Walker}, L.~M. 2014,
  \aj, 147, 43

\bibitem[{{Kim} \& {Ostriker}(2015)}]{Kim:2015}
{Kim}, C.-G., \& {Ostriker}, E.~C. 2015, \apj, 802, 99

\bibitem[{{Kobulnicky} \& {Johnson}(1999)}]{Kobulnicky:1999}
{Kobulnicky}, H.~A., \& {Johnson}, K.~E. 1999, \apj, 527, 154

\bibitem[{{Krumholz} \& {Matzner}(2009)}]{Krumholz:2009a}
{Krumholz}, M.~R., \& {Matzner}, C.~D. 2009, \apj, 703, 1352

\bibitem[{{Krumholz} {et~al.}(2004){Krumholz}, {McKee}, \&
  {Klein}}]{Krumholz:2004}
{Krumholz}, M.~R., {McKee}, C.~F., \& {Klein}, R.~I. 2004, \apj, 611, 399

\bibitem[{{Krumholz} \& {Thompson}(2012)}]{Krumholz:2012}
{Krumholz}, M.~R., \& {Thompson}, T.~A. 2012, \apj, 760, 155

\bibitem[{{Krumholz} {et~al.}(2014){Krumholz}, {Bate}, {Arce}, {Dale},
  {Gutermuth}, {Klein}, {Li}, {Nakamura}, \& {Zhang}}]{Krumholz:2014}
{Krumholz}, M.~R., {et~al.} 2014, ArXiv e-prints

\bibitem[{{Larson}(1969)}]{Larson:1969}
{Larson}, R.~B. 1969, \mnras, 145, 271

\bibitem[{{Leisawitz} {et~al.}(1989){Leisawitz}, {Bash}, \&
  {Thaddeus}}]{Leisawitz:1989}
{Leisawitz}, D., {Bash}, F.~N., \& {Thaddeus}, P. 1989, \apjs, 70, 731

\bibitem[{{Leitherer} {et~al.}(1999){Leitherer}, {Schaerer}, {Goldader},
  {Gonz{\'a}lez Delgado}, {Robert}, {Kune}, {de Mello}, {Devost}, \&
  {Heckman}}]{Leitherer:1999}
{Leitherer}, C., {et~al.} 1999, \apjs, 123, 3

\bibitem[{{Leroy} {et~al.}(2015){Leroy}, {Bolatto}, {Ostriker}, {Rosolowsky},
  {Walter}, {Warren}, {Donovan Meyer}, {Hodge}, {Meier}, {Ott}, {Sandstrom},
  {Schruba}, {Veilleux}, \& {Zwaan}}]{Leroy:2015}
{Leroy}, A.~K., {et~al.} 2015, \apj, 801, 25

\bibitem[{{Levermore} \& {Pomraning}(1981)}]{Levermore:1981}
{Levermore}, C.~D., \& {Pomraning}, G.~C. 1981, \apj, 248, 321

\bibitem[{{Lopez} {et~al.}(2011){Lopez}, {Krumholz}, {Bolatto}, {Prochaska}, \&
  {Ramirez-Ruiz}}]{Lopez:2011}
{Lopez}, L.~A., {Krumholz}, M.~R., {Bolatto}, A.~D., {Prochaska}, J.~X., \&
  {Ramirez-Ruiz}, E. 2011, \apj, 731, 91

\bibitem[{{Lopez} {et~al.}(2014){Lopez}, {Krumholz}, {Bolatto}, {Prochaska},
  {Ramirez-Ruiz}, \& {Castro}}]{Lopez:2014}
{Lopez}, L.~A., {Krumholz}, M.~R., {Bolatto}, A.~D., {Prochaska}, J.~X.,
  {Ramirez-Ruiz}, E., \& {Castro}, D. 2014, \apj, 795, 121

\bibitem[{{Martizzi} {et~al.}(2015){Martizzi}, {Faucher-Gigu{\`e}re}, \&
  {Quataert}}]{Martizzi:2015}
{Martizzi}, D., {Faucher-Gigu{\`e}re}, C.-A., \& {Quataert}, E. 2015, \mnras,
  450, 504

\bibitem[{{Matzner}(2002)}]{Matzner:2002}
{Matzner}, C.~D. 2002, \apj, 566, 302

\bibitem[{{McKee} \& {Ostriker}(2007)}]{McKee:2007}
{McKee}, C.~F., \& {Ostriker}, E.~C. 2007, Annual Review of Astronomy and
  Astrophysics, 45, 565

\bibitem[{{Miura} {et~al.}(2012){Miura}, {Kohno}, {Tosaki}, {Espada}, {Hwang},
  {Kuno}, {Okumura}, {Hirota}, {Muraoka}, {Onodera}, {Minamidani}, {Komugi},
  {Nakanishi}, {Sawada}, {Kaneko}, \& {Kawabe}}]{Miura:2012}
{Miura}, R.~E., {et~al.} 2012, \apj, 761, 37

\bibitem[{{Murray}(2009)}]{Murray:2009}
{Murray}, N. 2009, \apj, 691, 946

\bibitem[{{Murray}(2011)}]{Murray:2011}
---. 2011, \apj, 729, 133

\bibitem[{{Murray} {et~al.}(2010){Murray}, {Quataert}, \&
  {Thompson}}]{Murray:2010}
{Murray}, N., {Quataert}, E., \& {Thompson}, T.~A. 2010, \apj, 709, 191

\bibitem[{{Ostriker} \& {Shetty}(2011)}]{Ostriker:2011}
{Ostriker}, E.~C., \& {Shetty}, R. 2011, \apj, 731, 41

\bibitem[{{Pellegrini} {et~al.}(2007){Pellegrini}, {Baldwin}, {Brogan},
  {Hanson}, {Abel}, {Ferland}, {Nemala}, {Shaw}, \&
  {Troland}}]{Pellegrini:2007}
{Pellegrini}, E.~W., {et~al.} 2007, \apj, 658, 1119

\bibitem[{{Penston}(1969)}]{Penston:1969}
{Penston}, M.~V. 1969, \mnras, 144, 425

\bibitem[{{Reines} {et~al.}(2008){Reines}, {Johnson}, \& {Hunt}}]{Reines:2008}
{Reines}, A.~E., {Johnson}, K.~E., \& {Hunt}, L.~K. 2008, \aj, 136, 1415

\bibitem[{{Rogers} \& {Pittard}(2013)}]{Rogers:2013}
{Rogers}, H., \& {Pittard}, J.~M. 2013, \mnras, 431, 1337

\bibitem[{{Rosdahl} {et~al.}(2013){Rosdahl}, {Blaizot}, {Aubert}, {Stranex}, \&
  {Teyssier}}]{Rosdahl:2013}
{Rosdahl}, J., {Blaizot}, J., {Aubert}, D., {Stranex}, T., \& {Teyssier}, R.
  2013, \mnras, 436, 2188

\bibitem[{{Scoville} {et~al.}(2001){Scoville}, {Polletta}, {Ewald}, {Stolovy},
  {Thompson}, \& {Rieke}}]{Scoville:2001}
{Scoville}, N.~Z., {Polletta}, M., {Ewald}, S., {Stolovy}, S.~R., {Thompson},
  R., \& {Rieke}, M. 2001, \aj, 122, 3017

\bibitem[{{Semenov} {et~al.}(2003){Semenov}, {Henning}, {Helling}, {Ilgner}, \&
  {Sedlmayr}}]{Semenov:2003}
{Semenov}, D., {Henning}, T., {Helling}, C., {Ilgner}, M., \& {Sedlmayr}, E.
  2003, \aap, 410, 611

\bibitem[{{Skinner} \& {Ostriker}(2013)}]{Skinner:2013}
{Skinner}, M.~A., \& {Ostriker}, E.~C. 2013, \apjs, 206, 21

\bibitem[{{Stone} \& {Gardiner}(2009)}]{Stone:2009}
{Stone}, J.~M., \& {Gardiner}, T. 2009, \na, 14, 139

\bibitem[{{Stone} {et~al.}(2008){Stone}, {Gardiner}, {Teuben}, {Hawley}, \&
  {Simon}}]{Stone:2008}
{Stone}, J.~M., {Gardiner}, T.~A., {Teuben}, P., {Hawley}, J.~F., \& {Simon},
  J.~B. 2008, \apjs, 178, 137

\bibitem[{{Stone} {et~al.}(1998){Stone}, {Ostriker}, \& {Gammie}}]{Stone:1998}
{Stone}, J.~M., {Ostriker}, E.~C., \& {Gammie}, C.~F. 1998, \apjl, 508, L99

\bibitem[{{Thompson} \& {Krumholz}(2014)}]{Thompson:2014}
{Thompson}, T.~A., \& {Krumholz}, M.~R. 2014, ArXiv e-prints

\bibitem[{{Thompson} {et~al.}(2005){Thompson}, {Quataert}, \&
  {Murray}}]{Thompson:2005}
{Thompson}, T.~A., {Quataert}, E., \& {Murray}, N. 2005, \apj, 630, 167

\bibitem[{{Townsley} {et~al.}(2003){Townsley}, {Feigelson}, {Montmerle},
  {Broos}, {Chu}, \& {Garmire}}]{Townsley:2003}
{Townsley}, L.~K., {Feigelson}, E.~D., {Montmerle}, T., {Broos}, P.~S., {Chu},
  Y.-H., \& {Garmire}, G.~P. 2003, \apj, 593, 874

\bibitem[{{Truelove} {et~al.}(1997){Truelove}, {Klein}, {McKee}, {Holliman},
  {Howell}, \& {Greenough}}]{Truelove:1997}
{Truelove}, J.~K., {Klein}, R.~I., {McKee}, C.~F., {Holliman}, II, J.~H.,
  {Howell}, L.~H., \& {Greenough}, J.~A. 1997, \apjl, 489, L179

\bibitem[{{Tsai} {et~al.}(2009){Tsai}, {Turner}, {Beck}, {Meier}, \&
  {Ho}}]{Tsai:2009}
{Tsai}, C.-W., {Turner}, J.~L., {Beck}, S.~C., {Meier}, D.~S., \& {Ho},
  P.~T.~P. 2009, \aj, 137, 4655

\bibitem[{{Turner} {et~al.}(2015){Turner}, {Beck}, {Benford}, {Consiglio},
  {Ho}, {Kov{\'a}cs}, {Meier}, \& {Zhao}}]{Turner:2015}
{Turner}, J.~L., {Beck}, S.~C., {Benford}, D.~J., {Consiglio}, S.~M., {Ho},
  P.~T.~P., {Kov{\'a}cs}, A., {Meier}, D.~S., \& {Zhao}, J.-H. 2015, \nat, 519,
  331

\bibitem[{{Turner} {et~al.}(2000){Turner}, {Beck}, \& {Ho}}]{Turner:2000}
{Turner}, J.~L., {Beck}, S.~C., \& {Ho}, P.~T.~P. 2000, \apjl, 532, L109

\bibitem[{{Turner} {et~al.}(1998){Turner}, {Ho}, \& {Beck}}]{Turner:1998}
{Turner}, J.~L., {Ho}, P.~T.~P., \& {Beck}, S.~C. 1998, \aj, 116, 1212

\bibitem[{{Walch} \& {Naab}(2014)}]{Walch:2014}
{Walch}, S.~K., \& {Naab}, T. 2014, ArXiv e-prints

\bibitem[{{Walch} {et~al.}(2012){Walch}, {Whitworth}, {Bisbas}, {W{\"u}nsch},
  \& {Hubber}}]{Walch:2012}
{Walch}, S.~K., {Whitworth}, A.~P., {Bisbas}, T., {W{\"u}nsch}, R., \&
  {Hubber}, D. 2012, \mnras, 427, 625

\bibitem[{{Whitmore} {et~al.}(2014){Whitmore}, {Brogan}, {Chandar}, {Evans},
  {Hibbard}, {Johnson}, {Leroy}, {Privon}, {Remijan}, \&
  {Sheth}}]{Whitmore:2014}
{Whitmore}, B.~C., {et~al.} 2014, \apj, 795, 156

\bibitem[{{Yeh} \& {Matzner}(2012)}]{Yeh:2012}
{Yeh}, S.~C.~C., \& {Matzner}, C.~D. 2012, \apj, 757, 108

\end{thebibliography}
\clearpage

\appendix

\section{Fourier Transform Poisson Solver with Open Boundary Conditions}

Equation~\eqref{gravity:phi} may be rewritten as a discrete
convolution over the simulation domain $[0,L_x] \times [0,L_y] \times
[0,L_z]$, divided into $N_x N_y N_z$ equal zones.  Letting $(a,b,c)$
and $(a',b',c')$ represent zone-center integer indices, we may define
the Green function kernel
$\mathcal{G}(x_a,y_b,z_c;x_{a'},y_{b'},z_{c'}) =
\mathcal{G}(|x_a-x_{a'}|,|y_b-y_{b'}|,|z_c-z_{c'}|)=
\mathcal{G}(|a-a'|\Delta x,|b-b'|\Delta y,|c-c'|\Delta z)
$ 
as a symmetric function on an extended domain 
$[-L_x,L_x] \times [-L_y,L_y] \times[-L_z,L_z]$ or equivalently
$[-N_x,N_x-1] \times [-N_y,N_y-1] \times[-N_z,N_z-1]$.  
We also extend $\rho(x_{a},y_{b},z_{c})$ over the larger
domain, setting the value to zero for $a<0$, $b<0$, or $c<0$.
Equation~\eqref{gravity:phi} then becomes
\begin{eqnarray}
	\Phi(x_a,y_b,z_c) &=& G \sum_{a'=-N_x}^{N_x-1}
        \sum_{b'=-N_y}^{N_y-1} \sum_{c'=-N_z}^{N_z-1} \notag \\ 
        && \times \mathcal{G}(x_a,y_b,z_c;x_{a'},y_{b'},z_{c'}) \notag\\ 
        && \times \rho(x_{a'},y_{b'},z_{c'}) \,\Delta x \Delta y \Delta z. 
\label{gravity:extendphidiscrete}
\end{eqnarray}

Taking both $\mathcal{G}_{ijk}$ and $\rho_{ijk}$ to be $2N_x$-,
$2N_y$-, and $2N_z$-periodic sequences in the indices $i$, $j$, and
$k$, respectively, and using the discrete analog of the Fourier
Convolution Theorem, it follows from
Equation~\eqref{gravity:extendphidiscrete} that
\begin{eqnarray}
\Phi_{ijk} &=& \frac{G}{(2N_x)(2N_y)(2N_z)} \sum_{l=0}^{2N_x-1}
\sum_{m=0}^{2N_y-1} \sum_{n=0}^{2N_z-1} \notag \\ 
&& \times\hat{\mathcal{G}}_{lmn} \hat{\rho}_{lmn} \notag \\ 
&& \times\exp\left[-2\pi \imath \left(\frac{il}{2N_x} + \frac{jm}{2N_y} +
  \frac{kn}{2N_z}
  \right)\right], 
\label{gravity:extendphidiscreteconvolution}
\end{eqnarray}
where $\hat{\mathcal{G}}_{lmn}$ and $\hat{\rho}_{lmn}$ are the
respective \acp{DFT} of the sequences $\mathcal{G}_{ijk}$ and
$\rho_{ijk}$.  This method is computationally efficient since the
\acp{DFT} can be computed via \acp{FFT}.

In this work, we compute $\hat{\mathcal{G}}_{lmn}$ directly via the
\ac{DFT} of the periodic sequence
\begin{eqnarray}
\mathcal{G}_{ijk} &=& -\left\{[(i \bmod 2N_x)\Delta x]^2 \right.\notag \\ 
&& + [(j \bmod 2N_y)\Delta y]^2 \notag \\ 
&& \left. + [(k \bmod 2N_z)\Delta z]^2
        \right\}^{-1/2},
\label{gravity:greensfunction}
\end{eqnarray}
where we set $\mathcal{G}_{000} = 0$ to avoid dividing by zero.
The \ac{DFT} is computed once and stored.  An alternative to this approach
is to use the \ac{DFT} obtained from the finite-difference approximation to
the Laplace equation,
\begin{eqnarray}
	\hat{\mathcal{G}}_{lmn} &=& 2\pi \left\{ \frac{\cos(\pi l/N_x) - 1}{\Delta x^2} \right. \notag \\ && +\frac{\cos(\pi m/N_y) - 1}{\Delta y^2} \notag \\ && \left. + \frac{\cos(\pi n/N_z) - 1}{\Delta z^2} \right\}^{-1}, \label{gravity:indirectkernel}
\end{eqnarray}
(see the Appendix of \citealt{Gong:2013}), where once again we set $\hat{\mathcal{G}}_{000} = 0$.  The approach we adopt here is more accurate for long-range forces compared to the result in Equation~\eqref{gravity:indirectkernel}, although it does not have a closed-form analytic expression, and therefore must be stored in an array of size $8N_x N_y N_z$.

\end{document}